\documentstyle[preprint,prd,aps,floats,natbib]{revtex}
\tighten			
\def\fur{f\"{u}r}			
\def\Universitat{Universit\"{a}t}	
\def\Brugmann{{\text{Br\"{u}gmann}}}

\def\Cecile{{\text{\text{C\'{e}cile}}}}
\def\Gomez{{\text{G\'{o}mez}}}

\def\Hamade{{\text{Hamad\'{e}}}}
\def\Masso{{\text{Mass\'{o}}}}

\def\OMurchadha{{\text{\'{O}~Murchadha}}}

\def\code#1{{\sc #1}}				
\def\defn#1{``#1''}				
\def\xdefn#1{#1}				
\def\nop#1{\hbox{}}				
\def\cf{cf.\hbox{}}
\def\eg{e.g.\hbox{}}

\def\enmasse{{\it en masse\/}}
\def\etal{{\it et~al.\/}}
\def\ie{i.e.\hbox{}}

\def\Cecile{\text{C\'{e}cile}}
\def\Gomez{\text{G\'{o}mez}}
\def\Hamade{\text{Hamad\'{e}}}
\def\Masso{\text{Mass\'{o}}}
\def\OMurchadha{\text{\`{O}~Murchadha}}

\def\absp{{\text{absp}}}
\def\abst{{\text{abst}}}
\def\areal{{\text{areal}}}
\def\bh{{\text{bh}}}
\def\bg{{\text{bg}}}

\def\constant{{\text{constant}}}

\def\Gaussian{{\text{Gaussian}}}

\def\init{{\text{init}}}

\def\max{{\text{max}}}
\def\min{{\text{min}}}
\def\MS{{\text{MS}}}

\def\rel{{\text{rel}}}
\def\relp{{\text{relp}}}
\def\relt{{\text{relt}}}

\def\total{{\text{total}}}
\def\ulp{{\text{ulp}}}
\def\ab{_{ab}}
\def\ij{_{ij}}
\def\upij{^{ij}}
\def\dfrac#1#2{{\displaystyle \frac{#1}{#2}}}
\def\tfrac#1#2{{\textstyle \frac{#1}{#2}}}

\def\dhalf{\dfrac{1}{2}}
\def\thalf{\tfrac{1}{2}}

\def\tquarter{\tfrac{1}{4}}
\def\csmash#1{\hbox to 0em{\hss{#1}\hss}}	
\def\lsmash#1{\hbox to 0em{{#1}\hss}}		
\def\rsmash#1{\hbox to 0em{\hss{#1}}}		
\def\cwso#1#2{{\setbox0=\hbox{#2}\hbox to\wd0{\hss{#1}\hss}}}
\def\lwso#1#2{{\setbox0=\hbox{#2}\hbox to\wd0{{#1}\hss}}}
\def\rwso#1#2{{\setbox0=\hbox{#2}\hbox to\wd0{\hss{#1}}}}
\def\three{{}^{(3)}}				
\def\four{{}^{(4)}}				
\def\gtsim{\gtrsim}
\def\ltsim{\lesssim}
\def\del{\nabla}
\def\diag{\mathop{\rm diag}\nolimits}		
\def\Lie{\hbox{\pounds}}
\def\llangle{\mathopen{\langle\!\langle}}  
\def\rrangle{\mathclose{\rangle\!\rangle}} 
\def\llfloor{\mathopen{\lfloor\!\!\lfloor}}  
\def\rrfloor{\mathclose{\rfloor\!\!\rfloor}} 

\def\magnitude#1{\llangle #1 \rrangle}
\def\iwr{{\text{\tt w}}}		
\def\wr{w}				
\def\J{{\sf J}}
\def\K{{\sf K}}
\def\u{{\sf u}}
\def\U#1{\hbox{\underline{{\mathsurround=0em$#1$}}}}	
\def\eqn#1{$(\text{#1})$}
\def\eqref#1{\eqn{\protect\ref{#1}}}
\def\jtciteprefix{}
\def\jtcitenameprefix#1{}
\def\jtcitenamesuffix{}
\def\jtcitesprefix{}
\def\jtcitesnameprefix#1{}
\def\jtcitesnamesuffix{}
\def\jtcitenumericvsname#1#2{#2}
\def\jtcite{\citet}
\def\supersqueezetable{\let\tabbodyfont\tiny}
\begin{document}
\bibliographystyle{jtnatbib}	
\def\newblock{}			
\preprint{UWThPh-1999-38}
\title{A $3+1$ Computational Scheme for Dynamic			\\
       Spherically Symmetric Black Hole Spacetimes		\\
       -- II: Time Evolution}
\author{Jonathan Thornburg}
\address{Institut \fur{} Theoretische Physik, \Universitat{} Wien	\\
	 Boltzmanngasse~5, A-1090 Wien, Austria
\thanks{E-mail \hbox{\tt <jthorn@galileo.thp.univie.ac.at>}}}
\date{\today}
\maketitle
%
%
\begin{abstract}
This is the second in a series of papers describing a $3+1$ computational
scheme for the numerical simulation of dynamic black hole spacetimes.
In this paper we focus on the problem of numerically time-evolving a
given black-hole--containing initial data slice in spherical symmetry.
We avoid singularities via the \defn{black-hole exclusion} or
\defn{horizon boundary condition} technique, where the slices meet the
black hole's singularity, but on each slice a spatial neighbourhood of
the singularity is excluded from the domain of the numerical computations.

We first discuss some of the key design choices which arise with the
black hole exclusion technique:  Where should the computational domain's
inner boundary be placed?  Should a free or a constrained evolution
scheme be used?  How should the coordinates be chosen?

We then give a detailed description of our numerical evolution scheme.
We focus on a standard test problem, the time evolution of an asymptotically
flat spherically symmetric spacetime containing a black hole surrounded
by a massless scalar field, but our methods should also extend to more
general black hole spacetimes.  We assume that the black hole is
already present on the initial slice.  We use a free evolution, with
Eddington-Finkelstein--like coordinates and the inner boundary placed
at a fixed (areal) coordinate radius well inside the horizon.

Our numerical scheme is based on the method of lines (MOL), where the
spacetime PDEs are first discretized in space only, yielding a system
of coupled ODEs for the time evolution of the field variables along
the spatial-grid-point world lines.  These ODEs are then time-integrated
by standard finite difference methods.  In contrast to the more common
``space and time together'' finite differencing schemes, we find that
MOL schemes are considerably simpler to implement, and make it much
easier to construct stable higher order differencing schemes.  Our
MOL scheme uses 4th~order finite differencing in space and time, with
5~and 6~point spatial molecules and a 4th~order Runge-Kutta time
integrator.  The spatial grid is smoothly nonuniform, but not adaptive.

We present several sample numerical evolutions of black holes
accreting scalar field shells, showing that this scheme is stable,
very accurate, and can evolve ``forever''.  As an example of the
typical accuracy of our scheme, for a grid resolution of
$\Delta r/r \approx 3\%$ near the horizon, the errors in
$g\ij$\,($K\ij$) components at $t = 100m$ in a dynamic evolution
are ${\ltsim}\, 10^{-5}$\,($3 \,{\times}\, 10^{-7}$) in most of
the grid, while the energy constraint is preserved to
${\approx}\, 3 \,{\times}\, 10^{-5}$ of its individual
partial-derivative terms.

When the black hole accretes a relatively thin and massive scalar
shell, for a short time 3~distinct apparent horizons are present,
and the apparent horizons move at highly superluminal speeds;
this has important implications for how a black-hole--exclusion
computational scheme should handle the inner boundary.

Our other results for the scalar field phenomenology are generally
consistent with past work, except that we see a very different late-time
decay of the field near the horizon.  We suspect this is a numerical
artifact.
\end{abstract}
%
%
\draft				
%
%
\section{Introduction}
\label{sect-introduction}

Dynamic black hole spacetimes are interesting both as laboratories in
which to study the basic physics and phenomenology of strong-field
gravitation, and as probable very strong astrophysical sources of
gravitational radiation.  These systems are strongly relativistic,
time-dependent, and often highly asymmetric, and so are hard to study
in detail except by numerical methods.

This is the second in a series of papers describing a $3\,{+}\,1$ computational
scheme for the numerical simulation of dynamic black hole spacetimes.
In the first paper in this series
(\jtciteprefix\jtcite{Thornburg-1999-sssf-initial-data}, hereinafter
paper~I), we described our numerical initial data construction, and
presented some example initial data slices.  In this paper we discuss
our numerical time-evolution scheme and related topics, and present
several example spacetimes computed using this evolution scheme.

Black hole spacetimes necessarily contain singularities, which a $3\,{+}\,1$
computational scheme must somehow avoid in order for the $3\,{+}\,1$ equations
to remain well-defined and numerically tractable.  Traditionally this
avoidance has been done by using \defn{freezing} slicings, where the
lapse function is forced to drop to a very small value in the spatial
vicinity of a singularity-to-be, causing proper time to cease to advance
there and thus ``freezing'' the evolution before it reaches the singularity.
Unfortunately, by advancing at very different proper-time rates in
different parts of space, such slices necessarily ``stretch''
\footnote{
	 The term \defn{grid stretching} is often to
	 describe this, but we prefer \defn{slice stretching},
	 as the effect is actually inherent in the
	 {\em continuum\/} freezing-slicing $3\,{+}\,1$
	 equations, and is unrelated to the numerical
	 grid.
	 }
{}, causing the $3\,{+}\,1$ field variables to develop increasingly steep
gradients and non-smooth features, usually near to or just inside the
horizon.  This then leads to a variety of numerical problems (see, for
example,
\jtcitesprefix\jtcite{Smarr-1984-in-GR10,Shapiro-Teukolsky-1986-in-Centrella}).

To avoid these problems,
\jtcitenameprefix{Unruh}
\jtciteprefix\jtcite{Unruh-1984-BHE-suggestion}
\jtcitenamesuffix{}
suggested evolving only the region of spacetime outside the black
holes.  This \defn{black hole exclusion}, or \defn{black hole excision},
or \defn{horizon boundary condition} technique
\footnote{
	 We prefer the first two terms as focusing
	 attention on the underlying process --
	 the exclusion of the black hole from the
	 computational domain -- rather than on
	 the particular boundary condition used for
	 implementation.  Indeed, in our computational
	 scheme the boundary in question is actually
	 well inside the horizon, so ``horizon
	 boundary condition'' would be something
	 of a misnomer.
	 }
{} has been developed by a number of researchers
(\eg{}~\jtcitesprefix\jtcite{Thornburg-MSc,Thornburg-1987-2BH-initial-data,
Thornburg-1991-BHE-talk,Seidel-Suen-1992-BHE,Thornburg-PhD,
Scheel-Shapiro-Teukolsky-1995a-BHE-Brans-Dicke,ADMSS-1995-BHE,
Scheel-Shapiro-Teukolsky-1995b-BHE-Brans-Dicke,Marsa-PhD,
Marsa-Choptuik-1996-sssf,SBCST-1997-3D-xzy-BHE-evolution,
2BH-grand-challenge-alliance-1998-moving-BH}).
This technique involves choosing slices which intersect the singularity,
but on each slice excluding (excising) some spatial neighbourhood of
the singularity from the domain of the numerical computations.  With
singularities thus avoided, the slicing may then be (and generally is)
chosen to avoid or minimize slice stretching throughout the computational
domain.  The boundary of the excluded region, the \defn{inner boundary}
of the computational domain, is typically placed on or somewhat inside
the apparent horizon.

Using this technique, a well-designed evolution scheme can evolve an
initial slice ``forever'' without encountering coordinate singularities,
thus attaining the ``Holy Grail of numerical relativity''
(\jtciteprefix\jtcite[page~76]{Shapiro-Teukolsky-1986-in-Centrella}),
``\dots{} a code that simultaneously avoids
  singularities, handles black holes, maintains
  high accuracy, and runs forever.''

In this paper we first discuss some general aspects of the black hole
exclusion technique, focusing on the treatment of the inner boundary
and on how the coordinates should be chosen.  We then give a detailed
description of our black-hole--exclusion numerical evolution scheme
for dynamic black hole spacetimes.  Finally, we present several sample
spacetimes computed using this evolution scheme.

Our goal is a computational scheme which can be generalized to a wide
range of black hole spacetimes, so as much as possible we first present
and develop our algorithms for the generic case of a spacetime with no
symmetries and arbitrary matter fields present, and only later specialize
to a specific testbed system.  However, the numerical results discussed
here all pertain to a relatively simple testbed system: an asymptotically
flat spherically symmetric spacetime containing a black hole surrounded
by a (massless, minimally coupled) scalar field.
\footnote{
	 We have previously developed and implemented a
	 similar scheme for the evolution of a vacuum
	 axisymmetric spacetime containing a single black
	 hole (\jtciteprefix\jtcite{Thornburg-PhD}), but for
	 that system we were unable to obtain time evolutions
	 free of finite differencing instabilities.  We
	 believe these instabilities were primarily due
	 to the handling of the $z$~axis in the polar
	 spherical coordinate system, and were not
	 intrinsic to the black hole exclusion technique,
	 but we haven't proven this.
	 }

The spherically symmetric scalar field and similar systems have been
studied by a number of past researchers, including (among others)
an extensive analytical study by
\jtcitesnameprefix{Christoudolou}
\jtcitesprefix\jtcite{Christoudolou-1986a-sssf,Christoudolou-1986b-sssf,
Christoudolou-1987a-sssf,Christoudolou-1987b-sssf,
Christoudolou-1991-sssf,Christoudolou-1993-sssf}
\jtcitesnamesuffix{},
other valuable analytical work by
\jtciteprefix\jtcite{Gomez-Winicour-1992-sssf-2+2-asymptotics},
$3\,{+}\,1$ studies by
\jtcitesprefix\jtcite{Choptuik-PhD,Choptuik-1991-consistency,
Seidel-Suen-1992-BHE,Choptuik-1993-self-similarity,
Bernstein-Bartnik-1995-sssf,
Scheel-Shapiro-Teukolsky-1995a-BHE-Brans-Dicke,
Scheel-Shapiro-Teukolsky-1995b-BHE-Brans-Dicke,ADMSS-1995-BHE,
Marsa-PhD,Marsa-Choptuik-1996-sssf},
$2+2$ studies by
\jtcitesprefix\jtcite{Goldwirth-Piran-1987-sssf-2+2,
Goldwirth-Ori-Piran-1989-in-Frontiers,
Gomez-Winicour-1992-sssf-2+2-numerical-methods,
Hamade-Stewart-1996-sssf-2+2},
a hybrid $3\,{+}\,1$ and $2+2$ study by
\jtciteprefix\jtcite{GLPW-1996-sssf-3+1-and-2+2},
and an interesting comparison of $3\,{+}\,1$ and $2+2$ methods by
\jtciteprefix\jtcite{Choptuik-Goldwirth-Piran-1992-sssf-cmp-3+1-vs-2+2}.
\footnote{
	 There has also been a wide variety of work
	 (both analytical and numerical) by many authors
	 investigating the self-similar behavior
	 discovered by
\jtcitenameprefix{Choptuik}
\jtciteprefix\jtcite{Choptuik-1993-self-similarity}
\jtcitesnamesuffix{}.
	 }
{}  Of these, our work is closest in spirit to that of
\jtcitesprefix\jtcite{Marsa-PhD,Marsa-Choptuik-1996-sssf},
but we use slightly different spatial coordinates, a very different
inner-boundary treatment, and very different numerical methods.

In the remainder of this paper,
we first summarize our notation
	(section~\ref{sect-notation}),
then discuss the tradeoffs of where the inner boundary should be
placed in a black-hole--excluded evolution scheme
	(section~\ref{sect-location-of-inner-bndry}),
whether a free or a constrained evolution scheme should be used
	(section~\ref{sect-free-vs-constrained-evolution}),
and various general criteria for choosing the coordinates
	(section~\ref{sect-general-conds-for-coords}).
We then discuss our formulation of the continuum $3\,{+}\,1$ geometry and
scalar field equations, our coordinate conditions, and our boundary
conditions for the evolution and coordinate equations
	(section~\ref{sect-continuum-physics}).
We then discuss our numerical methods
	(section~\ref{sect-numerical-methods})
and our diagnostics for assessing the numerical results
	(section~\ref{sect-diagnostics}).
We then present and analyze a number of sample numerical evolutions
	(section~\ref{sect-sample-evolutions}).
We end the main body of the paper with some conclusions and directions
for further research
	(section~\ref{sect-conclusions}).
Finally, in the appendices we tabulate the detailed evolution equations
for the spherically symmetric scalar field system
	(appendix~\ref{app-sssf-equations}),
give a brief introduction to the method of lines
	(appendix~\ref{app-MOL-intro}),
summarize our methodology for convergence testing and briefly review
the technique of Richardson extrapolation
	(appendix~\ref{app-convergence-tests+Richardson-extrap}),
and describe our techniques for trying to estimate the effects of
floating-point roundoff errors by artificially adding a tiny amount
of noise to the field variables during an evolution
	(appendix~\ref{app-adding-noise}).
%
%
\section{Notation}
\label{sect-notation}

We generally follow the sign and notation conventions of
\jtcitenameprefix{Misner, Thorne, and Wheeler}
\jtciteprefix\jtcite{MTW}
\jtcitenamesuffix{},
with a $(-,+,+,+)$ spacetime metric signature, $G = c = 1$ units,
and all masses and coordinate distances also taken as dimensionless.
We assume the usual Einstein summation convention for all repeated
indices, and we use the Penrose abstract-index notation as described
by (\eg{}) \jtciteprefix\jtcite{Wald}.  However, for pedagogical convenience
we often blur the distinction between a tensor as an abstract geometrical
object and the vector or matrix of a tensor's coordinate components.

We use the standard $3\,{+}\,1$ formalism of \jtciteprefix\jtcite{ADM-1962}
(see \jtcitesprefix\jtcite{York-1979-in-Yellow,York-1983-in-Red} for
recent reviews).  For our spherically-symmetric--scalar-field test
system, we use coordinates $(t,r,\theta,\varphi)$, with $\theta$ and
$\varphi$ the usual spherical-symmetry coordinates.  However, in general
we leave $t$ and $r$ arbitrary, \ie{} unless otherwise noted we make
no assumptions about the specific choice of the lapse function or the
radial component of the shift vector.

The distinction between 3- and 4-tensors is usually clear from context,
but where ambiguity might arise we use prefixes $\three$ and $\four$
respectively, as in~$\three\! R$ and $\four\! R$.  Any tensor without
a prefix is by default a 3-tensor.  $\Lie_v$ denotes the Lie derivative
operator along the 4-vector field $v^a$.

We use $abcd$ for spacetime (4-)\,indices, and $\partial_a$
denotes the spacetime coordinate partial derivative operator
$\partial / \partial x^a$.  $g\ab$ denotes the spacetime metric
and $\del_a$ the associated 4-covariant derivative operator.

We use $ijkl$ for spatial \hbox{(3-)\,indices}.  $\partial_i$ denotes
the spatial coordinate partial derivative operator
$\partial / \partial x^i$.  $g\ij$ denotes the 3-metric of a slice,
$\del_i$ the associated 3-covariant derivative operator, and $g$
the determinant of the matrix of $g\ij$'s coordinate components.
$\alpha$ and $\beta^i$ denote the $3\,{+}\,1$ lapse function and shift
vector respectively.  $n^a$ denotes the (timelike) future pointing
unit normal to the slices.
$K\ij \equiv \thalf \Lie_n g\ij \equiv - \del_i n_j$ denotes the
3-extrinsic curvature of a slice, and $K \equiv K_i{}^i$ its trace.
$\rho \equiv n^a n^b T\ab$ and $j^i = - n_a T^{ai}$ denote the locally
measured energy and 3-momentum densities respectively.  $T\ij$ denotes
the spatial part of stress-energy tensor, and $T = T_i{}^i$ its trace.

$\diag[ \cdots ]$ denotes the diagonal matrix with the specified
diagonal elements.  $x(\delta)y$ denotes the arithmetic progression
$x$, $x + \delta$, $x + 2\delta$, $x + 3\delta$, \dots, $y$.
$\Gaussian(x{=}A, \sigma{=}B)$ denotes the Gaussian function
$\exp(-\thalf z^2)$, where $z \equiv (x - A) / B$.  For a symmetric
rank~2 covariant tensor $T\ij$, we define the pointwise \defn{magnitude}
tensor norm $\magnitude{T\ij} \equiv \sqrt{T\ij T\upij}$.

When discussing finite difference molecules, we often refer to a
molecule as itself being a discrete operator, the actual application
to a grid function being implicit.  We denote a molecule's
\defn{application point}, the grid point where the molecule's result
is defined, by underlining the corresponding molecule coefficient.
We define the \defn{radius} of a molecule in a given direction to be
the maximum distance in grid points away from the application point
in that direction, where the molecule still has a nonzero coefficient.
We define a \defn{centered} molecule as one having the same radius
in both directions.  For any coordinate $x$, $\Delta x$ denotes a
uniform-grid finite difference grid spacing in~$x$.  Thus, for example,
we describe the usual 3~point centered 2nd~order 2nd~derivative molecule
by
\begin{equation}
\partial_{xx} =
	\frac{1}{(\Delta x)^2}
	\Bigl[
	\begin{array}{c@{\quad}c@{\quad}c}
	+1	 & \U{-2} & +1		
	\end{array}
	\Bigr]
		+ O \bigl( (\Delta x)^2 \bigr)
					\, \text{,}	
\end{equation}
and say that it has radius~1 in both directions.

Consider a family of molecules all approximating the same continuum
(differential) operator, one molecule in the family being centered
(with radia $r_0$ in both directions), and the remaining molecules
being off-centered.  If an off-centered molecule has radia $r_-$
and $r_+$ in the $-$ and $+$ directions respectively, we formalize
the intuitive notion of ``off-centering distance'' by defining its
\defn{offset} to be (the positive integer) $r_0 - r_-$ if $r_- < r_0$,
or (the negative integer) $r_+ - r_0$ if $r_+ < r_0$ (we assume that
exactly one of these cases holds).  Finally, we define the offset
of a centered molecule to be $0$.  Thus, for example, we say that
the 4~point off-centered 2nd~order 2nd~derivative molecule
\begin{equation}
\partial_{xx} =
	\frac{1}{(\Delta x)^2}
	\Bigl[
	\begin{array}{c@{\quad}c@{\quad}c@{\quad}c}
	\U{+2}	& -5	& +4	& -1	
	\end{array}
	\Bigr]
		+ O \bigl( (\Delta x)^2 \bigr)
					\, \text{,}	
\end{equation}
which has radia $r_- = 0$ and $r_+ = 3$, has offset $+1$.
%
%
\section{The Location of the Inner Boundary}
\label{sect-location-of-inner-bndry}

In $3\,{+}\,1$ numerical relativity we generally don't know the spacetime
we're simulating in advance, but rather we construct it ``on the fly''
during a numerical evolution.  In using the black hole exclusion
technique, then, the precise region of spacetime to be excluded from
the numerical evolution must in general be chosen {\em dynamically\/}
in each slice.
\footnote{
\jtcitenameprefix{Choptuik}
\jtciteprefix\jtcite{Choptuik-1992-remark-multipass-time-evolution}
\jtcitesnamesuffix{}
	 has suggested that a multipass evolution
	 algorithm might be useful here, with the
	 future behavior of the black holes being
	 approximately known from a previous
	 lower-resolution evolution of the same
	 spacetime.  This would ease the ``lack of
	 advance knowledge'' problem considerably,
	 but an ``on the fly'' scheme would still
	 be needed for the initial evolution of a
	 previously unknown spacetime.
	 }
{}  In this section we consider how this choice -- specified in
practice by the inner boundary location -- should be made.

We assume throughout this discussion, and indeed everywhere in this
work, that a black hole or holes is already present in our initial
data, and hence that every slice contains a black hole.  (This
restriction could probably be weakened, but we haven't investigated
this.)
%
%
\subsection{Event or Apparent Horizon?}
\label{sect-event-or-apparent-horizon}

The excluded region must clearly satisfy three conditions:
\begin{enumerate}
\item	\label{condition-all-singularities-in-excluded-region}
	There must be no singularities in the non-excluded
	\defn{exterior} region of spacetime (the computational
	domain), \ie{} any singularities must be contained within
	the excluded region of spacetime.
\item	\label{condition-no-singularities-close-to-exterior-region}
	Furthermore, to prevent numerical difficulties there must
	be no singularities ``close'' to the exterior region, \ie{}
	any singularities must be at least some (strictly positive)
	distance away from the computational domain, this distance
	being uniformly bounded from below during an evolution.
\item	\label{condition-exterior-region-evolution-well-posed}
	The evolution equations on the exterior region of spacetime
	must be well-posed (causal), \ie{} no future pointing null
	geodesic may cross from the excluded region to the exterior
	region.  In other words, everywhere on the inner boundary
	the light cones must point exclusively {\em out\/} of the
	computational domain and {\em into\/} the excluded region.
\footnote{
\label{footnote-hydrodynamics-causality-analogy}
	 Using a common analogy with hydrodynamics,
	 we might imagine trying to simulate fluid
	 flow in the exterior region, and requiring
	 that fluid should be flowing into the
	 excluded region at a supersonic speed.
	 }
\end{enumerate}

In terms of these conditions alone, the ideal choice for the
inner boundary location would be the event horizon.
Condition~\ref{condition-exterior-region-evolution-well-posed}
would then hold by definition, and
condition~\ref{condition-all-singularities-in-excluded-region} would be
precisely the well-known cosmic censorship hypothesis, which is already
widely assumed in~$3\,{+}\,1$ numerical evolutions.  Unfortunately, the event
horizon can't be located {\em during\/} a numerical evolution, as it's
defined (\jtcitesprefix\jtcite{Hawking-1973-in-Les-Houches,Hawking-Ellis})
in an inherently acausal manner: it can only be determined if the
entire future development of the slice is known.  (As discussed by
\jtcitesprefix\jtcite{ABBLMSSSW-1995-numerical-event-horizons,
LMSSW-1996-numerical-event-horizons},
in practice the event horizon may be located to good accuracy given
only the usual numerically generated approximate development to a
nearly stationary state, but the fundamental acausality still remains.)

Instead, we assume that our spacetime and slicing are such that each
slice contains one or more apparent horizons, one of which we use as
a reference point for placing the inner boundary.  We first consider
the simplest case, where the inner boundary coincides with the
outermost apparent horizon.  (For pedagogical convenience, in this
and the next subsection we assume there's only a single black hole
present, but this restriction isn't in any way essential to our
arguments.)

Condition~\ref{condition-exterior-region-evolution-well-posed} then
follows from the definition of an apparent horizon, which, in the words
of
\jtcitenameprefix{Hawking and Ellis}
\jtciteprefix\jtcite[page~323]{Hawking-Ellis}
\jtcitesnamesuffix{},
``\dots{} moves outward at least as fast as
  light; and moves out faster than light if
  any matter or radiation falls through it.''
\footnote{
	 \jtciteprefix\jtcite{Hawking-Ellis}
	 \jtcitenumericvsname{doesn't}{don't} prove
	 this assertion, but it's easily demonstrated
	 (\jtciteprefix\jtcite{Unruh-Wald-1993-proof-horizon-causality})
	 by considering the expansion of the congruence
	 of outgoing null geodesics orthogonal to an
	 apparent horizon.  By definition, this expansion
	 is zero on the apparent horizon.  But the
	 Raychaudhuri equation (more precisely its
	 analog for null geodesics) implies that the
	 derivative of this expansion along the congruence
	 is negative semidefinite, so the congruence
	 must be nonexpanding on future slices.  This
	 congruence must therefore lie within the
	 trapped region in these (future) slices,
	 and hence lie on or within their (outermost)
	 apparent horizons.
	 }
{}  \jtcitesprefix\jtcite{Israel-1986-PRL-AH-confinement,
Israel-1986-CJP-AH-confinement} \jtcitenumericvsname{have}{has}
also shown that under suitable technical conditions, an apparent
horizon on an initial slice can always be extended in time in a
``rigid'' (locally area-preserving) manner so as to act
``\dots{} as a wall that permanently seals off
  its interior contents from causal influence
  on the environment.''
However, in general this extension differs from the time development
of the apparent horizon.  It would be interesting to investigate this
extension's use in black-hole--exclusion inner boundary placement, but
so far as we know this hasn't yet been done.

With the inner boundary coinciding with the apparent horizon,
condition~\ref{condition-all-singularities-in-excluded-region} is
equivalent to requiring all singularities to lie within (outermost)
apparent horizons.  In terms of generic black hole spacetimes, this
\defn{apparent cosmic censorship} assumption is quite strong, in
fact considerably stronger than ``standard'' cosmic censorship.  In
particular, \jtciteprefix\jtcite{Wald-Iyer-1991-Schw-slicing-with-no-AH}
have shown that even in Schwarzschild spacetime there exist (angularly
anisotropic) slices which approach arbitrarily close to the $r = 0$
singularity, yet contain no apparent horizons.  (The related work
of \jtciteprefix\jtcite{AHST-1992-initial-data-cosmic-censorship} is
also relevant here.)

In practice, we simply follow the usual course in numerical relativity
of assuming all necessary properties of spacetime -- in our case including
apparent cosmic censorship (in fact the somewhat stronger
condition~\ref{condition-no-singularities-close-to-exterior-region})
and the light cones pointing entirely inwards on the inner boundary
-- and on this basis proceeding with the numerical computations,
verifying our assumptions after the fact.
%
%
\subsection{On or Inside the Apparent Horizon?}
\label{sect-on-or-inside-apparent-horizon}

Our analysis of the previous subsection takes the inner boundary to
be located coincident with an apparent horizon,
\footnote{
	 If a multidimensional Cartesian-topology grid
	 is used, then the inner boundary necessarily
	 has an irregular ``staircase'' shape.  However,
	 if this is chosen to approximate the apparent
	 horizon to within a single grid spacing, then
	 for present purposes we ignore the non-smoothness,
	 and still refer to such an inner boundary as
	 ``coincident'' with the apparent horizon.
 	 }
{} typically the outermost one, and this location has been used
successfully by
\jtcitesprefix\jtcite{Thornburg-PhD,Marsa-PhD,Marsa-Choptuik-1996-sssf,
SBCST-1997-3D-xzy-BHE-evolution}.
Alternatively, the inner boundary may be placed somewhat {\em inside\/}
an apparent horizon (again typically the outermost one), as has been
done successfully by
\jtcitesprefix\jtcite{Seidel-Suen-1992-BHE,
Scheel-Shapiro-Teukolsky-1995a-BHE-Brans-Dicke,ADMSS-1995-BHE}.
These latter authors all chose to leave a macroscopic distance
(\ie{} one large compared to the grid spacing) between the inner
boundary and the nearest apparent horizon, but still another option
would be to reduce this distance to only a few grid points, \ie{} to
place the inner boundary a few grid points inside an apparent horizon.
(We don't know of any numerical tests of this last option.)

Placing the inner boundary coincident with an apparent horizon has
several advantages:
\begin{itemize}
\item	As discussed in section~\ref{sect-event-or-apparent-horizon},
	it ensures the well-posedness (causality) of the exterior
	region evolution (at least so long as each slice does indeed
	contain an apparent horizon).  (In contrast, if the inner
	boundary is somewhat inside an apparent horizon, then the
	causality of the exterior region evolution depends on the
	details of the spacetime, the slicing, and the spatial
	coordinate choice.)

\item	It avoids ``wasting'' any grid points by placing them inside
	an apparent horizon, where they'd be unable to causally
	influence the exterior evolution.  (In practice, though,
	this inefficiency doesn't seem to be serious.)

\item	It's relatively easy to implement, particularly if the
	coordinates and finite difference grid use a polar spherical
	topology, as then the coordinates and grid may be chosen
	so the relevant apparent horizon is (\ie{} coincides with)
	a coordinate and grid sphere.

\item	The condition that the initial data have an apparent horizon
	coinciding with the inner boundary, provides a natural inner
	boundary condition for the York initial data algorithm, and
	also makes easy to use this algorithm to construct
	multiple--black-hole initial data
	(\jtcitesprefix\jtcite{Thornburg-MSc,Thornburg-1987-2BH-initial-data}).

\item	The condition that the relevant apparent horizon remain
	coincident with the inner boundary during the evolution,
	provides a natural inner boundary condition for the radial
	component of the shift vector.  In contrast, if the inner
	boundary is inside the horizon then enforcing such a condition
	is more difficult, as the corresponding ``boundary condition''
	still needs to be enforced at the horizon, despite this no
	longer being a boundary of the numerical grid.
\end{itemize}

However, placing the inner boundary somewhat inside an apparent horizon
(whether by a macroscopic distance or only by a few grid points), also
has a number of advantages:
\begin{itemize}
\item	It helps to prevent any numerical inaccuracies at the boundary
	from affecting the exterior evolution.
\jtcitesprefix\jtcite{Thornburg-1991-BHE-talk,Seidel-Suen-1992-BHE,
Thornburg-PhD,Scheel-Shapiro-Teukolsky-1995a-BHE-Brans-Dicke,
ADMSS-1995-BHE}
	all cite this reason for placing the inner boundary inside
	the horizon in their black-hole--excluded computational
	scheme.

\item	A black-hole--excluded evolution often requires that the
	apparent horizon or horizons be explicitly located, \ie{}
	that the apparent horizon equation be numerically solved,
	at each time step.  Like the exterior evolution, it's desirable
	for the apparent-horizon--finding process to be unaffected
	by any computational errors at the inner boundary.

\item	Because the apparent horizon equation can't be solved
	exactly in closed form, numerical apparent-horizon finders
	invariably use iterative algorithms.  (See, \eg{},
\jtcitesprefix\jtcite{BCSST-1996-horizon-finding,
Thornburg-1996-horizon-finding,ACLMSS-1998-horizon-finding}
	for recent discussions of numerical apparent-horizon--finding
	algorithms.)  These algorithms require the $3+1$ field
	variables ($g\ij$ and $K\ij$) to be defined throughout a
	spatial {\em neighbourhood\/} of the apparent horizon being
	located.  Placing the inner boundary somewhat inside an
	apparent horizon naturally provides this.  In contrast,
	placing the inner boundary coincident with an apparent
	horizon would require some form of extrapolation to
	continue the field variables inside the ``inner boundary''
	for the horizon finder's use.  This seems inelegant, and
	might also be numerically ill-conditioned.

\item	In a dynamic spacetime where we don't know the apparent
	horizon's motion in advance, the coordinate position of the
	inner boundary and/or the value of the shift vector there
	must in general be dynamically adjusted to \defn{track} the
	apparent horizon's motion, \ie{} to maintain the apparent
	horizon at a desired (possibly time-dependent) coordinate
	position as the evolution proceeds.  As discussed by
\jtciteprefix\jtcite[sections 3.1.5--3.1.7 and~4.5]{Thornburg-PhD},
	depending on the algorithms used, this \defn{horizon tracking}
	may be somewhat imprecise, \ie{} at times the inner boundary
	may drift somewhat in its position relative to the apparent
	horizon.  If this is the case, and the drift happens to move
	the inner boundary outwards, then the numerical evolution may
	cease to be well-posed (\ie{} it may violate the causality
	condition~\ref{condition-exterior-region-evolution-well-posed}
	of section~\ref{sect-event-or-apparent-horizon}) unless there's
	a ``buffer zone'' between the apparent horizon and the inner
	boundary.
\end{itemize}

Considerable further research is needed to investigate these issues,
and to help clarify the relative merits of the different possible
inner boundary placements.
%
%
\subsection{Multiple Apparent Horizons}
\label{sect-multiple-apparent-horizons}

The problem of placing the inner boundary is more complicated when
a slice may contain multiple apparent horizons, either enclosing
distinct volumes or topologically nested inside one other.  In terms
of our discussion in section~\ref{sect-event-or-apparent-horizon},
it's generally preferable to place the inner boundary near (on or
somewhat inside) the {\em outermost\/} apparent horizon within any
given black hole.  This minimizes the chances of any singularities
being too close to the exterior region (and also minimizes the number
of grid points ``wasted'' by being inside the outermost apparent
horizon and thus causally isolated from the exterior evolution),
while still ensuring that the exterior region evolution remains
well-posed.

In practice, multiple apparent horizons are often present only for a
limited time interval in an evolution.  As an example of the general
type of phenomenology to be expected, consider the 400.pqw1 evolution
discussed in
section~\ref{sect-sample-evolutions/BH-growth+apparent-horizon-motion},
where the initial slice contains a black hole surrounded by a relatively
thin and massive scalar field shell.
Figure~\ref{fig-400.pqw1-h+m(h)[t=0-50+19-20]} shows the apparent
horizon positions as a function of coordinate time~$t$ for this evolution.
Notice that initially there is only one apparent horizon, then at
$t \approx 19.13$ two new apparent horizons form outside the original
one, then at $t \approx 19.71$ the original (inner) and middle
apparent horizons merge and disappear, leaving only the (new) outer
apparent horizon at later times.

Because the original apparent horizon disappears at $t\approx 19.71$,
a black-hole--exclusion evolution scheme which had previously been
using it as a reference for placing the inner boundary, must somehow
switch to using the new outer apparent horizon at later times.
[Similarly, in a black-hole--collision simulation, when the two black
holes coalesce and their previous (separate) apparent horizons disappear,
a black-hole--exclusion evolution scheme must change from using their
separate apparent horizons as inner-boundary reference positions before
the coalescence, to using the resulting single black hole's new apparent
horizon afterwards (assuming it has one).]

There are a number of ways such a transition might be made.  For example:
\begin{itemize}
\item	One possibility would be to smoothly move the inner boundary
	from the inner apparent horizon out to the outer one during
	the time interval ($19.13 \ltsim t \ltsim 19.71$ in our example)
	where the slices contain both.  Unfortunately, this requires
	a substantial inner-boundary motion, possibly at a speed faster
	than the coordinate speed of light, or even faster than the
	grid CFL limit.  (For our example, a coordinate distance
	$\Delta r \approx 3.0$ must be covered in a coordinate time
	$\Delta t \ltsim 0.6$, so some part of the motion must necessarily
	be at a speed in excess of 5 times the coordinate speed of light.)
	Such rapid motion may well cause numerical problems, both near
	the inner boundary and possibly (if the coordinate and grid
	conditions don't sufficiently attenuate the inner boundary's
	motion) also further out in the grid.  At the very least,
	such highly \defn{superluminal} inner-boundary motion would
	likely require some type of causal differencing
(\jtciteprefix\jtcite{Alcubierre-PhD,
Alcubierre-Schutz-1994-causal-differencing}).
\item	To avoid individual grid points having to move so far and so
	fast, one could instead simply drop (delete) from the grid a
	suitable number of the innermost grid points at some suitable
	time.  This would have the effect of discontinuously moving
	the inner boundary outwards, while leaving the spatial coordinate
	system and the remainder of the grid unchanged.
\item	A variant of this latter possibility would be to again drop
	some of the innermost grid points, but simultaneously
	recoordinatize the slice and/or (discontinuously) change the
	grid points' spatial coordinate positions, so as to maintain
	some desired horizon-tracking condition, such as there being
	a grid point or layer of grid points precisely on the horizon.
	Note that for the black-hole--coalescence case the computational
	domain changes topology as well as size and shape, so {\em some\/}
	discontinuous change in the grid is essential.
\end{itemize}

As with the initial placement of the inner boundary, the treatment
of such multiple-apparent-horizon slices is greatly complicated if
spacetime -- and in particular the behavior of the apparent horizons
-- isn't known in advance, but rather is computed only ``on the fly''
during a numerical evolution.  (For example, in the context of
figure~\ref{fig-400.pqw1-h+m(h)[t=0-50+19-20]}, however the transition
from the original to the new outermost apparent horizon is made,
it's probably desirable to do it somewhat {\em before\/} the original
apparent horizon disappears.  But how can this impending disappearance
be detected in advance?)

Again, much further research is needed to properly investigate these
and other possibilities for how such multiple--apparent-horizon
spacetimes and slicings should be treated.
%
%
\subsection{Our Present Scheme}
\label{sect-present-inner-boundary-scheme}

In our present computational scheme we take a very simplistic approach
to these issues: we place the inner boundary at a fixed time-independent
spatial coordinate position (radius) $r = r_\min$, chosen to be well
inside the horizon's coordinate position at all times.  Our code
explicitly computes the light cones on the inner boundary, and monitors
them to ensure the exterior region evolution remains causal.
\footnote{
	 In detail, the check is that the outgoing light-cone
	 speed $c_+$ defined by~\eqref{eqn-light-cone},
	 satisfies $c_+ \leq 0$ on the inner boundary
	 at all times.  This condition has always been
	 satisfied throughout all our evolutions.
	 }

This inner boundary placement has proved adequate for our present
purposes, but clearly wouldn't suffice for more general spacetimes,
particularly (non--spherically-symmetric) ones containing moving
black holes.
%
%
\section{Free versus Constrained Evolution}
\label{sect-free-vs-constrained-evolution}

$3\,{+}\,1$ evolution schemes are customarily classified into \defn{free},
\defn{partially constrained}, or \defn{fully constrained} evolution
schemes, depending on whether none, some, or all of the constraints
are used in evolving the field variables.

It was once thought (\eg{}~\jtcitesprefix\jtcite{Piran-1983-in-Red,
Stewart-1984-numerical-relativity-survey}) that the finite differenced
constraint and evolution equations are inherently inconsistent,
rendering free and constrained evolution schemes fundamentally different
in meaning, and calling into question whether either can truly yield
good approximations to solutions of the continuum $3\,{+}\,1$ equations.
However, it later became clear from the work of
\jtcitesnameprefix{Choptuik}
\jtcitesprefix\jtcite{Choptuik-PhD,Choptuik-1991-consistency}
\jtcitesnamesuffix{}
that there is in fact no such inconsistency, \ie{} that properly
constructed free and constrained evolution schemes are both valid,
and (if they use the same order of finite differencing) yield results
which are within an $O(1)$ factor of each other in accuracy.

With one (important) exception discussed below, the choice of free
versus constrained evolution is thus purely a pragmatic one, to be
made on grounds of finite differencing stability, easy of implementation,
and $O(1)$ factors in accuracy and efficiency.

There is one serious problem with constrained evolution schemes, though:
They generally entail elliptic or parabolic (constraint-incorporating)
updating equations for the $g\ij$ and $K\ij$ components.  These equations
require boundary conditions, and while outer boundary conditions pose
no (conceptual) problem, with the black-hole--exclusion technique
there's no obvious way to obtain inner boundary conditions during an
evolution.  Further research to develop constrained black-hole--exclusion
evolution schemes would be very desirable.

In this work we have chosen a free evolution, both because of the
inner-boundary-condition problem, and because free evolution schemes
are easier to implement and easier to generalize efficiently to
multiple spatial dimensions.
%
%
\section{General Conditions for the Coordinates}
\label{sect-general-conds-for-coords}

How should the coordinates be chosen in a black-hole--exclusion
evolution?  To permit easy interpretation and accurate numerical
solution of the $3\,{+}\,1$ equations, we suggest requiring the coordinates
to satisfy the following general ``well-behavedness'' criteria
throughout the computational domain (including near to and on both
the event and apparent horizons):
\begin{enumerate}
\item	\label{item-first-coordinate-criterion}
	The slices must be spacelike and asymptotically flat.
\item	The coordinates must cover all the ``interesting'' regions of
	spacetime (certainly all of spacetime outside the black hole)
	to the future of the initial slice.
\item	The coordinates must be nonsingular (smooth) and nondegenerate
	(spacetime-filling)
\item	Similarly, the light cones should be non-degenerate, and in
	particular they should not be ``infinitely closed up'' or
	``infinitely wide open'' as they are (for example) near
	the horizon in the Schwarzschild slicing of Schwarzschild
	spacetime.
\item	\label{item-last-coordinate-criterion}
	\label{item-all-field-variables-smooth}
	All the $3\,{+}\,1$ field variables should be smooth functions
	of the coordinates.
\footnote{
	 This implicitly assumes that the underlying
	 spacetime geometry (\eg{}~the 4-Riemann tensor)
	 is itself smooth.  Note that if the type of
	 self-similar behavior discovered by
\jtcitenameprefix{Choptuik}
\jtciteprefix\jtcite{Choptuik-1993-self-similarity}
\jtcitesnamesuffix{}
	 is present, this requirement means we must
	 exclude at least a neighbourhood of any
	 self-similarity points.
	 }
\end{enumerate}

To allow well-behaved long-time evolutions, we also require these
conditions to continue to hold in the limit $t \to \infty$.
(Requiring condition~\ref{item-all-field-variables-smooth} in this
limit specifically rules out the type of sharp gradients and non-smooth
features which arise from slice stretching.)  Finally, we'd like our
coordinate conditions to be readily generalizable to generic
multiple--black-hole spacetimes.

Our conditions for the coordinates are very similar to those of
\jtciteprefix\jtcite[page~76]{Shapiro-Teukolsky-1986-in-Centrella},
\jtcitenumericvsname{which describes}{who describe} the
``Holy Grail of numerical relativity'' as
``\dots{} a code that simultaneously avoids
  singularities, handles black holes, maintains
  high accuracy, and runs forever.''

Schwarzschild spacetime provides a particularly useful test case
for considering coordinate choice.  As a number of previous researchers
have observed
(\eg{}~\jtcitesprefix\jtcite{Marsa-PhD,Marsa-Choptuik-1996-sssf}),
(ingoing) Eddington-Finkelstein coordinates $(t,r,\theta,\varphi)$\,
\footnote{
	 Recall (\jtciteprefix\jtcite[box~31.2]{MTW})
	 that these coordinates are defined by taking
	 $r$ to be an areal radial coordinate, $\theta$
	 and $\varphi$ the standard angular coordinates,
	 and choosing the slicing so that $t+r$ is an
	 (ingoing) null coordinate.  For reference, we
	 list some of the $3\,{+}\,1$ field variables
	 for Schwarzschild spacetime in these
	 coordinates in appendix~A of paper~I;
	 \jtciteprefix\jtcite[appendix~2]{Thornburg-PhD}
	 gives a more extensive list.
	 }
{} are ideally suited for use with the black-hole--exclusion
technique.  Figure~\ref{fig-Schw/EF-slicing}
%
%
\addtocounter{footnote}{1}
\footnotetext{
\label{footnote-x(delta)y}
	     Recall that $x(\delta)y$ denotes the
	     arithmetic progression $x$, $x + \delta$,
	     $x + 2\delta$, $x + 3\delta$, \dots, $y$.
	     }
{} shows the Eddington-Finkelstein slicing of Schwarzschild spacetime,
plotted in both Kruskal-Szekeres and Eddington-Finkelstein coordinates.
These coordinates satisfy all of our
criteria~\hbox{\ref{item-first-coordinate-criterion}
--\ref{item-last-coordinate-criterion}} described above.  Note that
the Eddington-Finkelstein slices are {\em not\/} maximal: $K$ is
nonzero, and in fact spatially variable, throughout the slices.

For example, the light cones shown in
figure~\hbox{\ref{fig-Schw/EF-slicing}(b)} are well-behaved near to
and on the horizon, whereas a similar plot in Schwarzschild coordinates
(\jtcite[figure~32.1(a)]{MTW}) shows pathological behavior there.
More generally, in these coordinates the 4-metric (and hence all the
$3\,{+}\,1$ field variables) remains well-behaved everywhere in spacetime
-- including near to and on the horizon -- except near the singularity
itself.

In terms of our discussion in
sections~\ref{sect-on-or-inside-apparent-horizon}
and~\ref{sect-present-inner-boundary-scheme}, it's instructive
to check the causality of the exterior region evolution for the
combination of Eddington-Finkelstein coordinates (slices) and an
$r = \constant$ inner boundary position.  For this case it's
easy to see that the exterior region evolution does in fact remain
causal for any (stationary) inner boundary position inside the
horizon.  This can be seen as the slopes of the outgoing legs of the
light cones plotted in figure~\hbox{\ref{fig-Schw/EF-slicing}(b)}:
these point inwards for $r < 2m$ and outwards for $r > 2m$.

For Kerr spacetime, Kerr coordinates
\footnote{
	 Recall (\jtciteprefix\jtcite[box~33.2]{MTW})
	 that these generalize Eddington-Finkelstein
	 coordinates for Schwarzschild spacetime: the
	 Kerr slicing in Kerr spacetime is defined by
	 taking the usual (areal $r$), and requiring
	 that $t+r$ be an (ingoing) null coordinate.
	 \jtciteprefix\jtcite[appendix~2]{Thornburg-PhD}
	 lists some of the $3\,{+}\,1$ field variables
	 for Kerr spacetime in these coordinates.
	 }
{} are similarly well suited for use with the black-hole--exclusion
technique, and similarly satisfy all of our
criteria~\hbox{\ref{item-first-coordinate-criterion}
--\ref{item-last-coordinate-criterion}} described above.  The
Kerr-coordinates 4-metric is similarly well-behaved everywhere outside
the ring singularity, including near to and on the (event) horizon.

For Kerr slicing and a stationary $r = \constant$ inner boundary
position, the exterior region evolution remains causal provided the
inner boundary remains between the inner and outer horizons
(\jtciteprefix\jtcite[section~3.1.5]{Thornburg-PhD}).  This restriction
isn't a problem so long as the black hole has non-maximal spin, which
we require in any case for the validity of our computational scheme.
[An maximal-spin (extremal) Kerr spacetime is infinitesimally close
to a spacetime with no event horizon and a naked singularity, rendering
our fundamental black-hole--exclusion technique invalid.]
%
%
\section{Continuum Physics}
\label{sect-continuum-physics}

We now discuss the basic continuum physics of our testbed system and
our computational scheme.
%
%
\subsection{$3\,{+}\,1$ Geometry and Scalar Field Equations}

%
%
\subsubsection{Generic Spacetime}
\label{sect-continuum-physics/generic-spacetime}

We first briefly review our formulation of the continuum $3\,{+}\,1$ equations
in their generic form, making no assumptions about spacetime's symmetries.

As is customary for the $3\,{+}\,1$ formalism, we assume that spacetime is
globally hyperbolic, and we introduce global spacetime coordinates
$x^a = (t,x^i)$, where $t$ is a global time coordinate.  Given the
spacetime metric $\four g\ab$, we then define the usual 3-metric $g\ij$,
\hbox{(3-)\,lapse} function $\alpha$, and \hbox{(3-)\,shift} vector
$\beta^i$, so the spacetime line element becomes
\begin{equation}
\four ds^2 \equiv \four g\ab \, dx^a \, dx^b
	= - (\alpha^2 - \beta_i \beta^i) \, dt^2
	  + 2 \beta_i \, dt \, dx^i
	  + g\ij \, dx^i \, dx^j
					\, \text{.}	
\end{equation}
Given $g\ij$, we define the usual 3-covariant derivative operator
$\del_i$ and \hbox{(3-)\,Ricci} tensor and scalar $R\ij$ and
$R \equiv R_i{}^i$.

Taking $n^a$ to be the future pointing timelike unit normal to
the slices, we define the usual \hbox{(3-)\,extrinsic} curvature
$K\ij \equiv \thalf \Lie_n g\ij \equiv - \del_i n_j$.  Given the
spacetime stress-energy tensor $T\ab \equiv \four T\ab$, we define
the usual $3\,{+}\,1$ scalar field variables $T\ij \equiv \four T\ij$,
$\rho \equiv n^a n^b T\ab$, and $j^i \equiv - n_a T^{ai}$.

The $3\,{+}\,1$ evolution and constraint equations are then
(\jtcitesprefix\jtcite{York-1979-in-Yellow,York-1983-in-Red})
\begin{mathletters}
						\label{eqn-evolution-generic}
\begin{eqnarray}
\partial_t g\ij
	& = &	{}
		- 2 \alpha K\ij
		+ \U{\Lie_\beta g\ij}
									\\
\partial_t K\ij
	& = &	{}
		- \del_i \del_j \alpha
		+ \alpha \left( R\ij - 2 g^{kl} K_{ik} K_{jl} + K K\ij \right)
		+ \U{\Lie_\beta K\ij}			\nonumber	\\
	&   &	\qquad
		{}
		+
		4 \pi \alpha \left(
			     \vphantom{T^i} (T - \rho) g\ij - 2 T\ij
			     \right)
									\\
\U{\Lie_\beta g\ij}
	& = &	\U{\del_i \beta_j} + \U{\del_j \beta_i}			\\
	& = &	\U{\beta^k \partial_k g\ij}
		+ \U{g_{ik} \partial_j \beta^k}
		+ \U{g_{jk} \partial_i \beta^k}
									\\
\U{\Lie_\beta K\ij}
	& = &	\U{\beta^k \del_k K\ij}
		+ \U{K_{ik} \del_j \beta^k}
		+ \U{K_{jk} \del_i \beta^k}				\\
	& = &	\U{\beta^k \partial_k K\ij}
		+ \U{K_{ik} \partial_j \beta^k}
		+ \U{K_{jk} \partial_i \beta^k}
\end{eqnarray}
\end{mathletters}
and
\begin{mathletters}
						\label{eqn-constraints-generic}
\begin{eqnarray}
C \equiv
	\Bigl( R - K\ij K\upij + K^2 \Bigr)
	-
	\Bigl( 16 \pi \rho \Bigr)
		& = &	0
					\label{eqn-energy-constraint-generic}
									\\
C^i \equiv
	\Bigl( \del_j K\upij - \del^i K \Bigr)
	-
	\Bigl( 8 \pi j^i \Bigr)
		& = &	0
\end{eqnarray}
\end{mathletters}
respectively, where the shift vector Lie derivative terms are
\underline{underlined} for reasons discussed in
section~\ref{sect-spatial-FD}.  We use a free evolution scheme, \ie{}
after constructing initial data (discussed in detail in paper~I) we
use the constraints only as diagnostics of our computational scheme's
accuracy.

We take the scalar field $\phi$ to satisfy the \hbox{4-scalar} wave
equation $\del_a \del^a \phi = 0$, and to have the usual stress-energy
tensor
\begin{equation}
4 \pi T\ab = (\partial_a \phi) (\partial_b \phi)
	     - \thalf g_{ab} (\partial_c \phi) (\partial^c \phi)
					\, \text{.}	
\end{equation}

We then define the $3\,{+}\,1$ scalar field variables
\begin{mathletters}
							\label{eqn-Pi-Q-defn}
\begin{eqnarray}
P_i	& = &	\del_i \phi						\\
Q	& = &	\frac{1}{\alpha}
		\left( \partial_t \phi - \beta^i \del_i \phi \right)
					\, \text{,}	
\end{eqnarray}
\end{mathletters}
so that
\begin{mathletters}
						\label{eqn-partial-phi=fn(P+Q)}
\begin{eqnarray}
\partial_t \phi	& = &	\alpha Q + \beta^k P_k				\\
\del_i \phi	& = &	P_i
					\, \text{.}	
\end{eqnarray}
\end{mathletters}
These latter definitions~\eqref{eqn-Pi-Q-defn}
and~\eqref{eqn-partial-phi=fn(P+Q)} are similar, but not identical,
to those of \jtcitesprefix\jtcite{Choptuik-PhD,Choptuik-1991-consistency,
Choptuik-1993-self-similarity,Marsa-PhD,Marsa-Choptuik-1996-sssf}.

Straightforward but tedious calculations then give the scalar field
evolution equations as
\begin{mathletters}
						\label{eqn-PiQ-evolution}
\begin{eqnarray}
\partial_t P_i
	& = &	\del_i \left( \alpha Q + \beta^k P_k \right)		\\
\partial_t Q
	& = &	\del_i \left( \alpha P^i \right)
		+ \alpha K Q
		+ \beta^i \del_i Q
\end{eqnarray}
\end{mathletters}
and the $3\,{+}\,1$ scalar field variables as
\begin{mathletters}
\begin{eqnarray}
4 \pi \rho	& = &	\thalf \left( P_k P^k + Q^2 \right)		\\
4 \pi j_i	& = &	- P_i Q						\\
4 \pi T\ij	& = &	P_i P_j
			+ \thalf g\ij \left( - P_k P^k + Q^2 \right)	\\
4 \pi T		& = &	- \thalf P_k P^k + \tfrac{3}{2} Q^2
					\, \text{.}	
\end{eqnarray}
\end{mathletters}

For the black hole exclusion technique, we assume that each slice
contains one or more apparent horizons.  As discussed by (\eg{})
\jtciteprefix\jtcite{York-1989-in-Frontiers}, an apparent horizon satisfies
the equation
\begin{equation}
H \equiv
	\del_i n^i + K_{ij} n^i n^j - K
		= 0
					\, \text{,}	
						\label{eqn-horizon-generic}
\end{equation}
where we take $n^i$ to be the outward-pointing unit normal to the
horizon, and for future use we define the \defn{horizon function} $H$
as the left hand side of this equation.
%
%
\subsubsection{Spherical Symmetry}
\label{sect-continuum-physics/spherical-symmetry}

We now assume that spacetime is spherically symmetric, and that the
spatial coordinates are $x^i = (r,\theta,\varphi)$, with the usual polar
spherical topology.  We take the $3\,{+}\,1$ field tensors to have the
coordinate components
\begin{mathletters}
\begin{eqnarray}
g\ij
	& \equiv &
		\diag
		\left[
		\begin{array}{c@{\quad}c@{\quad}c}
		A	& B	& B \sin^2 \theta	
		\end{array}
		\right]
									\\
K\ij
	& \equiv &
		\diag
		\left[
		\begin{array}{c@{\quad}c@{\quad}c}
		X	& Y	& Y \sin^2 \theta	
		\end{array}
		\right]
									\\
\beta^i
	& \equiv &
		\left[
		\begin{array}{c@{\quad}c@{\quad}c}
		\beta	& 0	& 0	
		\end{array}
		\right]
									\\
P_i
	& \equiv &
		\left[
		\begin{array}{c@{\quad}c@{\quad}c}
		P	& 0	& 0	
		\end{array}
		\right]
					\, \text{.}	
\end{eqnarray}
\end{mathletters}
Notice that we do {\em not\/} factor out either a conformal factor
or any $r^2$ factors from the 3-metric components.
\footnote{
	 Although this somewhat simplifies the $3\,{+}\,1$
	 equations, it may degrade the accuracy of our
	 computational scheme.  In particular, it's at
	 least plausible that factoring out $r^2$~factors
	 might give improved accuracy at large~$r$.
	 It might be interesting to try a side-by-side
	 accuracy comparison between otherwise-identical
	 factored and unfactored schemes, but we haven't
	 investigated this.
	 }

With these assumptions it's then straightforward, though tedious,
to express all the other $3\,{+}\,1$ variables in terms of the
\defn{state variables} $A$, $B$, $X$, $Y$, $P$, and $Q$, and their
spatial (1st and 2nd) derivatives.  For reference, we list the
resulting equations for the constraints and most of our diagnostics
in appendix~B of paper~I; we list the equations for our remaining
diagnostics in section~\ref{sect-diagnostics/energy-constraint} of
this paper, and those for the evolution equations in
appendix~\ref{app-sssf-equations}.

Note that although spherical symmetry means there can be only one
black hole in spacetime, there may be multiple apparent horizons.
We use $h$ to denote the or an apparent horizon's coordinate position
(radius).  We describe our methods for numerically locating apparent
horizons in section~B1 of paper~I.

Given our functional form for the metric, it's easy to see that the
range of causal spatial velocities within the light cone is given by
\begin{equation}
c_- \equiv
{} - \beta - \frac{\alpha}{\sqrt{A}}
	\leq
	\frac{\partial r}{\partial t}
	\leq
		{} - \beta + \frac{\alpha}{\sqrt{A}}
		\equiv c_+
					\, \text{.}	
							\label{eqn-light-cone}
\end{equation}
%
%
\subsection{Coordinate Conditions}
\label{sect-coord-conds}

Although we have experimented with a number of different slicings
and spatial coordinate conditions (some of these are described in
\jtciteprefix\jtcite{Thornburg-PhD}), in this paper we discuss only a
single, relatively simple, slicing and spatial coordinate condition:

As suggested by
\jtcitesprefix\jtcite{Choptuik-1993-generalized-EF-slicing-suggestion,
Marsa-PhD,Marsa-Choptuik-1996-sssf}, we choose the lapse function
$\alpha$ by following the Eddington-Finkelstein requirement that
$t+r$ be an (ingoing) null coordinate.  This gives the
\defn{generalized Eddington-Finkelstein} slicing condition
\begin{equation}
\frac{\alpha}{\sqrt{A}} + \beta = 1
					\, \text{.}	
					\label{eqn-generalized-EF-lapse}
\end{equation}

We choose the shift vector $\beta^i$ to keep the areas of constant-$r$
surfaces (coordinate spheres) temporally constant during the evolution,
\ie{} $\partial_t g_{\theta\theta} = 0$, giving the \defn{constant area}
spatial coordinate condition
\begin{equation}
{}
- 2 \alpha Y + \beta \partial_r B
	= 0
					\, \text{.}	
						\label{eqn-constant-area-shift}
\end{equation}
This shift vector condition generalizes the usual areal radial
coordinate: if the radial coordinate on the initial slice is areal
($g_{\theta\theta} = r^2$), then the constant-area coordinate
condition will preserve this throughout the evolution.

[In practice we normally do choose our initial data to have an areal
radial coordinate, but this is solely for convenience in data analysis;
our computational scheme doesn't require it.  Indeed, in our computational
scheme the time evolution commutes with 3-coordinate-transformations,
so a different spatial coordinate choice on the initial slice would
leave the slicing and the spatial-coordinate world lines invariant,
merely relabelling the latter.  Such invariance is a useful property
of a computational scheme (\jtciteprefix\jtcite[section~2.6]{Thornburg-PhD}).]

These coordinate conditions are easy to implement and very cheap to
to compute, requiring only a few algebraic operations at each grid
point to compute $\alpha$ and $\beta$.  Moreover, they satisfy all
of our general coordinate-choice criteria described in
section~\ref{sect-general-conds-for-coords}.  Unfortunately,
these coordinates implicitly require the existence of a ``radial''
coordinate, so they don't generalize well to multiple-black-hole
spacetimes.  Moreover, the constant area spatial coordinate condition
can only constraint a single degree of freedom of the shift vector at
each event, which is insufficient to fully specify the coordinates in
a nonspherical spacetime.  We will discuss more general and generalizable
coordinate conditions in a future paper.

In anticipation of future generalization to different coordinate
conditions, we specifically do {\em not\/} assume any particular
coordinate conditions in deriving any of our equations outside this
section, \ie{} elsewhere in this paper we take the lapse function and
shift vector as generic.  That is, for example, despite the constant-area
shift vector condition~\eqref{eqn-constant-area-shift}, we still
numerically evolve $B \equiv g_{\theta\theta}$ in the same manner
as the other state variables.
%
%
\subsection{Boundary Conditions}
\label{sect-BCs}

Conceptually, our computational domain should be all of spacetime
exterior to the inner boundary $r = r_\min$.  However, as is customary
in~$3\,{+}\,1$ numerical evolutions of asymptotically flat spacetimes, we
further truncate our computational domain at a finite outer boundary
radius $r = r_\max$.
%
%
\subsubsection{Evolution Equations}
\label{sect-BCs/evolution-eqns}

As discussed in section~\ref{sect-location-of-inner-bndry},
provided $c_+ \leq 0$ on the inner boundary, the evolution equations
need no (continuum) boundary conditions there.  (We discuss the
finite difference treatment of the inner boundary in
section~\ref{sect-spatial-FD}.)

We impose approximate outgoing radiation boundary conditions at
the outer boundary.  This is most elegantly done by matching the
numerically computed field variables to a tensor--spherical-harmonic
decomposition of the gravitational field, as discussed by
\jtcitesprefix\jtcite{Thorne-1980-RMP-tensor-spherical-harmonics,
Abrahams-Evans-1988-gravitational-radiation-matching,
Abrahams-1989-in-Frontiers,
Abrahams-Evans-1990-gravitational-radiation-matching,
Abrahams-Shapiro-Teukolsky-1995-gravitational-radiation-matching}.
\footnote{
	 See also the gravitational-radiation--theory survey of
	 \jtciteprefix\jtcite{Pirani-1962-gravitational-radiation-survey}.
	 }
{}  However, such boundary conditions are quite complicated to derive
and implement.  Instead, here we use a much simpler approximation,
the simple pseudo-scalar Sommerfeld-type outgoing radiation condition
of \jtcitesprefix\jtcite{Trautman-1958a-radiation-BCs,
Trautman-1958b-radiation-BCs,Fock-1959-radiation-BCs}.
\footnote{
	 \jtciteprefix\jtcite{Givoli-1991-radiation-BC-review}
	 gives an excellent survey of numerical methods
	 for non-reflecting boundary conditions in various
	 areas of computational physics.  As well,
	 \jtciteprefix\jtcite{Israeli-Orszag-1981-radiation-BCs}
	 have suggested a \defn{sponge filter} technique
	 which may be useful in improving the accuracy of
	 such outer boundary conditions;
\jtcitesprefix\jtcite{Choptuik-PhD,Marsa-PhD,Marsa-Choptuik-1996-sssf}
	 have used this technique successfully in numerical
	 relativity computations similar to ours.
	 }
{}  This condition has been used successfully in~$3$-dimensional
$(x,y,z)$ numerical relativity calculations by
\jtciteprefix\jtcite{Nakamura-Oohara-1989-3D-evolution-in-Frontiers}.

For our formulation of outgoing radiation boundary conditions (an
adaptation of that of \jtciteprefix\jtcite{Thornburg-PhD}), we let
$u$ denote any of the state-vector field variables $A$, $B$, $X$,
$Y$, $P$, or $Q$, and $u_\bg$ denote the same field variable at the
corresponding coordinate position in some \defn{background} slice
to be chosen later.  We first factor out $u$ and $u_\bg$'s asymptotic
coordinate dependence at large radia by defining a constant parameter
$m$ such that that $u = O(r^m)$ at large radia in a flat spacetime.

We then define $\delta u \equiv u - u_\bg$ and take $\delta u / r^m$
to (approximately) satisfy a simple flat-spacetime scalar outgoing
radiation boundary condition
\begin{equation}
\frac{\delta u}{r^m}
	\approx
		\frac{f(r - c_+t)}{r^n}
				\qquad
				\text{(at $r = r_\max$)}
				\label{eqn-scaled-delta-u-outgoing-radiation}
\end{equation}
for some (unspecified) function $f$, where $c_+$ is the outgoing
light-cone speed defined by~\eqref{eqn-light-cone}, and $n$ is a
constant parameter chosen based on the expected asymptotic fall-off
of outward-propagating perturbations in~$\delta u$.

Taking the time derivative of~\eqref{eqn-scaled-delta-u-outgoing-radiation},
assuming the background field variables to be time-independent, and
simplifying, we obtain the approximate scalar outgoing radiation
boundary condition
\begin{equation}
\partial_t \, \delta u
	\approx
		- c_+ \left[
		      \partial_r \, \delta u + \frac{n-m}{r} \, \delta u
		      \right]
				\qquad
				\text{(at $r = r_\max$)}
					\, \text{.}	
					\label{eqn-outgoing-radiation-outer-BC}
\end{equation}

All our results in this paper use the outer-boundary-condition
parameters given in table~\ref{tab-evolution-outer-BC-pars}, and
take the background slice to be an Eddington-Finkelstein slice of
a mass-$m_\total$ Schwarzschild spacetime, where $m_\total$ is the
total Misner-Sharp mass of the current slice.
\footnote{
	 $m_\total$, and hence the background field
	 variables, are actually slightly time-dependent,
	 violating one of our assumptions in
	 deriving~\eqref{eqn-outgoing-radiation-outer-BC}.
	 This effect is generally small, and we ignore it.
	 }
{}  We discuss the efficacy of these boundary conditions in
section~\ref{sect-sample-evolutions/outer-bndry-errors}.
%
%
\subsubsection{Coordinate Conditions}
\label{sect-BCs/coord-conds}

Our coordinate conditions discussed in section~\ref{sect-coord-conds}
are algebraic equations in~$\alpha$ and $\beta$, and hence need no
boundary conditions.  However, for consistency with more general
coordinate conditions which we will discuss in a future paper, we
use outer (though not inner) boundary conditions for our coordinate
conditions here.

As discussed by \jtciteprefix\jtcite{York-1980-in-Taub-festschrift}, we
assume $\alpha \to 1$ and $\beta^i \to 0$ at spatial infinity to ensure
that our coordinates are asymptotically Minkowskian.

For the lapse function, we approximate this at our finite outer boundary
radius with the scalar Robin outer boundary condition suggested by
\jtciteprefix\jtcite{York-Piran-1982-in-Schild-lectures},
\begin{equation}
(\del_i \alpha) n^i + \frac{\alpha - 1}{r}
	= O \left( \frac{1}{r^3} \right)
	\approx 0
				\qquad
				\text{(at $r = r_\max$)}
					\, \text{.}	
\end{equation}

\jtciteprefix\jtcite{York-Piran-1982-in-Schild-lectures}
\jtcitenumericvsname{describes}{describe} a vector Robin condition
which properly takes into account the shift vector's tensor character,
but for simplicity, at present we instead just use a scalar Robin
condition independently for each coordinate component of the shift
vector,
\begin{equation}
(\del_i \beta^j) n^i + \frac{\beta^j}{r}
	= O \left( \frac{1}{r^3} \right)
	\approx 0
				\qquad
				\text{(at $r = r_\max$)}
					\, \text{.}	
\end{equation}

In spherical symmetry, these conditions become
\begin{equation}
\frac{\partial_r \alpha}{\sqrt{A}} + \frac{\alpha - 1}{r}
	= O \left( \frac{1}{r^3} \right)
	\approx 0
				\qquad
				\text{(at $r = r_\max$)}
					\label{eqn-lapse-scalar-Robin-outer-BC}
\end{equation}
and
\begin{equation}
\frac{\partial_r \beta}{\sqrt{A}} + \frac{\beta}{r}
	= O \left( \frac{1}{r^3} \right)
	\approx 0
				\qquad
				\text{(at $r = r_\max$)}
					\label{eqn-shift-scalar-Robin-outer-BC}
\end{equation}
respectively.
%
%
\section{Numerical Methods}
\label{sect-numerical-methods}

We now describe our methods for discretizing and numerically solving
the $3\,{+}\,1$ evolution and coordinate equations.

The fundamental assumption underlying all our numerical methods is that
all the $3\,{+}\,1$ field variables are smooth functions of the spacetime
coordinates throughout the computational domain.  That is, our numerical
methods are specifically {\em not\/} designed to handle shocks,
gravitational chaos, self-similar behavior of the type found by
\jtcitenameprefix{Choptuik}
\jtciteprefix\jtcite{Choptuik-1993-self-similarity}
\jtcitenamesuffix{},
or any other phenomenology which generates significant power at
arbitrarily high spatial frequencies.  (This smoothness requirement
is the motivation of our coordinate-choice
criterion~\ref{item-all-field-variables-smooth}
in section~\ref{sect-general-conds-for-coords}.)
%
%
\subsection{Nonuniform Gridding}
\label{sect-nonuniform-gridding}

To allow adequate resolution of the rapidly changing field variables
near the black hole while still extending the grid well out into the
weak-field region, some sort of nonuniform grid spacing is highly
desirable.

An adaptive gridding scheme such as that of
\jtcitesnameprefix{Berger and Oliger}
\jtcitesprefix\jtcite{Gropp-1980-adaptive-gridding,Berger-PhD,
Berger-Oliger-1984-adaptive-gridding,
Berger-Jameson-1985-adaptive-gridding,
Berger-1986-adaptive-gridding-data-structures}
\jtcitesnamesuffix{}
is probably the most efficient way to do this, and
\jtcitesprefix\jtcite{Choptuik-PhD,Choptuik-1989-in-Frontiers,
Choptuik-1992-in-dInverno,Choptuik-1993-self-similarity,
Masso-Seidel-Walker-1995-adaptive-gridding,
Brugmann-1996-adaptive-gridding}
report excellent results in using such schemes in spherically symmetric
$3\,{+}\,1$ black hole evolutions.  However, such adaptive gridding schemes
are quite complicated to implement.

For present purposes we have instead chosen a much simpler non-adaptive
nonuniform gridding scheme, where the grid spacing varies smoothly in
a predetermined (and in our case time-independent) manner across the
computational domain.  For the (fairly limited) class of spacetimes we
deal with here, such a scheme provides a substantial efficiency gain
over a simple uniform grid, but is still easy to implement.

To describe our nonuniform gridding scheme, we introduce a new radial
coordinate $\wr$ such that the grid is uniform in~$\wr$.  Although
it would be possible to use $(\wr,\theta,\varphi)$ coordinates as a
tensor basis as well as for the numerical grid (see, example,
\jtcitesprefix\jtcite{ABHSS-1992-diagonal-metric-BH-evolution,Bernstein-PhD}),
we prefer to retain the natural $(r,\theta,\varphi)$ coordinates for
this purpose, treating the nonuniform gridding as a lower-level
numerical technique transparent to our continuum problem formulation.

We therefore implement nonuniform gridding by rewriting individual
$r$-coordinate partial derivatives in terms of $\wr$-coordinate
partial derivatives, \ie{} we write
\begin{mathletters}
					\label{eqn-nonuniform-gridding-xform}
\begin{eqnarray}
\partial_r	& = &	\J \partial_\wr
									\\[1ex]
\partial_{rr}	& = &	\J^2 \partial_{\wr\wr}  +  \K \partial_\wr
					\, \text{,}	
\end{eqnarray}
\end{mathletters}
where we define the coefficients $\J = d\wr / dr$ and
$\K = d^2\wr / dr^2$.

It remains to specify how the nonuniform grid coordinate $\wr$ should
be chosen.  The logarithmic coordinate $\wr = \log(r/r_0)$ is often used
(\eg{}~\jtcitesprefix
\jtcite{ABHSS-1992-diagonal-metric-BH-evolution,Bernstein-PhD}),
but we find that this offers insufficient control over how the grid
resolution varies as a function of (spatial) position.  In particular,
with such a logarithmic coordinate we find that a grid giving good
resolution near the black hole, will typically have insufficient
resolution to accurately represent asymptotically-constant-width
features in the field variables (\eg{}~outward-propagating scalar
field shells) at large radia.

To obtain finer control over how the grid resolution varies as a
function of position, we instead use the \defn{mixed-210} nonuniform
grid coordinate (\jtcite[section~7.1.7]{Thornburg-PhD}) defined by
\begin{equation}
\frac{d \wr}{d r} = \frac{1}{a (r/r_0)^2} + \frac{1}{b (r/r_0)} + \frac{1}{c}
					\, \text{,}	
						\label{eqn-mixed-210-defn}
\end{equation}
where $r_0$, $a$, $b$, and $c$ are suitable (constant) parameters.
(The name ``mixed-210'' comes from ${d\wr} / {dr}$ being a mixture of
$1 / r^2$, $1 / r^1$, and $1 / r^0$ terms.)  Integrating and taking
$\wr = 0$ at $r = r_0$, we have
\begin{equation}
\wr = \frac{r_0}{a} \left( 1 - \frac{1}{r/r_0} \right)
      + \frac{r_0}{b} \log \left( \frac{r}{r_0} \right)
      + \frac{r_0}{c} \left( \frac{r}{r_0} - 1 \right)
					\, \text{.}	
						\label{eqn-mixed-210-wr(r)}
\end{equation}
We also need the inverse function $r(\wr)$; we compute this
numerically.
\footnote{
	 We use Brent's \code{ZEROIN} subroutine
(\jtcitesprefix\jtcite{Forsythe-Malcolm-Moler,Kahaner-Moler-Nash})
	 to numerically find a zero of the function
	 $\wr(r) - \wr_\ast$, where $\wr_\ast$ is the
	 (given) value of $\wr$ for which $r$ is desired.
	 This technique is easy to program and very
	 robust and accurate, but mildly inefficient
	 as it doesn't exploit the smoothness of $\wr(r)$.
	 However, in practice we only need $r(\wr)$ at
	 grid points, so we simply precompute and store
	 all the $r$ values.
	 }

The physical interpretation of this coordinate is that for suitable
choices of $a$, $b$, and $c$, in the innermost part of the grid, the
first term in~\eqref{eqn-mixed-210-defn} is dominant, so the grid
spacing is roughly $\Delta r = a (r/r_0)^2 \cdot \Delta \wr$.  In the
middle part of the grid, the second term in~\eqref{eqn-mixed-210-defn}
is dominant, so the grid spacing is roughly
$\Delta r = b (r/r_0) \cdot \Delta \wr$.  In the outermost part of
the grid, the third term in~\eqref{eqn-mixed-210-defn} is dominant,
so the grid spacing is roughly $\Delta r = c \cdot \Delta \wr$.

By convention, we always place the inner grid boundary at $\wr = 0$,
\ie{} $r = r_0 = r_\min$.  All our results in this paper use the
parameters $r_0 = 1.5$, $a = \infty$ (effectively omitting the first
term in~\eqref{eqn-mixed-210-defn} and~\eqref{eqn-mixed-210-wr(r)}),
$b = 5$, and $c = 100$.  Figure~\ref{fig-mixed-210-coord} shows
$\wr(r)$ and $\Delta \wr(r)$ for these parameters; $\wr$ qualitatively
resembles a logarithmic radial coordinate in the inner part of the grid,
and a uniform radial coordinate in the outer part of the grid.  In
particular, note that the grid spacing $\Delta r$ asymptotes to a
constant value at large radia, allowing asymptotically--constant-width
features to be accurately evolved even in the outer parts of the grid.
%
%
\subsection{Spatial Finite Differencing}
\label{sect-spatial-FD}

Our spatial finite differencing uses a single non-staggered grid,
uniform in~$\wr$, with grid points located at the fixed spatial
coordinate positions $\wr = \wr_\min (\Delta \wr)\wr_\max$.
We label the grid points with the integer grid coordinates
$\iwr = \iwr_\min(1)\iwr_\max$.

To construct a finite difference approximation to a continuum
equation, we first apply the nonuniform-gridding
transformations~\eqref{eqn-nonuniform-gridding-xform} to replace
all $r$-coordinate partial derivatives with $\wr$-coordinate
partial derivatives.  (Or equivalently, we could
use~\eqref{eqn-nonuniform-gridding-xform} to numerically compute the
$r$-coordinate partial derivatives in terms of the $\wr$-coordinate
ones.)  We then discretely approximate each $\wr$-coordinate partial
derivative with a suitable finite difference molecule.
%
%
\subsubsection{Grid Interior}
\label{sect-spatial-FD/grid-interior}

In the grid interior, we use the usual centered 5~point 4th~order
molecules for most $\partial_\wr$ and $\partial_{\wr\wr}$ terms in
the $3\,{+}\,1$ equations.  However, for the $\partial_\wr$ terms derived
from the shift vector Lie derivative terms in the $3\,{+}\,1$ evolution
equations (these terms are shown \underline{underlined} in the evolution
equations~\eqref{eqn-evolution-generic} and~\eqref{eqn-evolution}), we
use upwind--off-centered molecules (still 5~point 4th~order), with the
direction of the upwinding chosen based on the sign of the radial shift
vector $\beta \equiv \beta^r$.  Tables~\ref{tab-FD-molecule-offsets}
and~\ref{tab-FD-molecule-coeffs} show this in detail.
%
%
\subsubsection{Grid Boundaries}
\label{sect-spatial-FD/grid-bndrys}

Our finite differencing near the grid boundaries may be described in
either of two completely equivalent ways, each useful in different
contexts:

\begin{description}
\item[{\rm Off-centered molecules:}]
	In this viewpoint, we use off-centered molecules for partial
	derivatives near the grid boundaries.  To approximate a given
	$\wr$-coordinate partial derivative (either $\partial_\wr$ or
	$\partial_{\wr\wr}$) near one of the boundaries, we choose
	from within one of the molecule families described below, the
	molecule with the closest possible offset (\ie{}~the closest
	possible centering or off-centering) to that used in the grid
	interior, subject to the constraint that the molecule not
	require data from outside the boundary.  For a 1st~derivative
	$\partial_\wr$ we use one of the off-centered 5~point 4th~order
	molecules shown in part~(a) of table~\ref{tab-FD-molecule-coeffs}.
	For a 2nd~derivative $\partial_{\wr\wr}$ we use either one of
	the off-centered 5~point 3rd~order molecules shown in part~(b)
	of this table, or one of the off-centered 6~point 4th~order
	molecules shown in part~(c).  (We discuss the tradeoffs
	between these latter two choices in
	section~\ref{sect-spatial-FD/3rd-vs-4th-order-dww-mols-near-bndrys}.)

\item[{\rm Extrapolation:}]
	In this viewpoint, we first logically extend the grid to
	include a few fictitious grid points beyond each boundary
	of the continuum problem domain, \ie{}~the fictitious points
	are at coordinates
	$\iwr = \iwr_\min{-}1$, $\iwr_\min{-}2$, $\iwr_\min{-}3$, \dots{}
	inside the inner boundary, and
	$\iwr = \iwr_\max{+}1$, $\iwr_\max{+}2$, $\iwr_\max{+}3$, \dots{}
	outside the outer boundary.  To approximate a given
	$\wr$-coordinate partial derivative ($\partial_\wr$ or
	$\partial_{\wr\wr}$) near one of the boundaries, we then
	use the (assumed) smoothness of the grid function being
	finite differenced, to extrapolate values for the grid
	function at the fictitious grid points.   For a 1st~derivative
	$\partial_\wr$ we use the 5~point 4th~order (Lagrange)
	polynomial extrapolants
\footnote{
	 We define the \defn{order} of a Lagrange
	 polynomial interpolant/extrapolant to be the
	 order of the polynomial implicitly fitted to
	 the known grid function values, the local
	 truncation error of the extrapolated value
	 being one order higher than this.  Note that
	 this definition differs from that used for
	 finite difference molecules, where the order
	 is conventionally defined to be that of the
	 local truncation error itself.
	 }
{}	shown in part~(a) of table~\ref{tab-FD-extrapolation-coeffs}.
	For a 2nd~derivative $\partial_{\wr\wr}$ we use either those
	same extrapolants, or the 6~point 5th~order extrapolants
	shown in part~(b) of this table.  (Again, we discuss the
	tradeoffs between these latter two choices in
	section~\ref{sect-spatial-FD/3rd-vs-4th-order-dww-mols-near-bndrys}.)
	Finally, using the newly-extrapolated values, we apply our
	grid-interior finite differencing scheme (including upwinding
	of Lie-derivative $\partial_\wr$ terms) at all the original
	(non-fictitious) grid points.
\end{description}

By convolving the extrapolants in table~\ref{tab-FD-extrapolation-coeffs}
with the various centered and off-centered molecules in
table~\ref{tab-FD-molecule-coeffs}, it's straightforward to show
that the off-centered molecule and extrapolation viewpoints are in
fact exactly equivalent -- they yield identical finite difference
equations, and hence identical results (modulo floating-point
roundoff errors).

Our numerical code implements both viewpoints, and we find each to be
useful in certain circumstances:  Off-centered molecules are most convenient
when the actual molecule coefficients of a finite difference operator
are desired.  This case occurs, for example, when finite differencing
the left hand side operators of elliptic PDEs such as those of coordinate
conditions
\footnote{
	 The coordinate conditions described in
	 section~\ref{sect-coord-conds} can actually
	 be solved independently at each grid point.
	 However, in anticipation of future generalization
	 to more general (elliptic) coordinate conditions,
	 our numerical code uses a general-purpose elliptic
	 solver to compute $\alpha$ and $\beta$ at each
	 time step.  Each elliptic solution is done by
	 first row scaling the resulting linear system
	 to improve its numerical conditioning, then
	 $\sf LU$~decomposing and solving it via
	 \code{LINPACK} (\jtciteprefix\jtcite{LINPACK-book})
	 band matrix routines.
	 }
{} or the $3\,{+}\,1$ initial data problem (\cf{}~paper~I).  In contrast,
extrapolation is most convenient when a finite difference operator
need only be applied to a grid function, without necessarily explicitly
computing any molecule coefficients.  This case occurs, for example,
when finite differencing the right hand side of the $3\,{+}\,1$ evolution
equations.  
%
%
\subsubsection{3rd versus 4th Order $\partial_{ww}$ Molecules
	       Near the Boundaries}
\label{sect-spatial-FD/3rd-vs-4th-order-dww-mols-near-bndrys}

In section~\ref{sect-spatial-FD/grid-bndrys} we describe two slightly
different finite differencing schemes, one where we use the same
(4th)~order of finite differencing for 2nd~derivatives near the
boundaries as elsewhere, and one where we use one order of accuracy
lower (3rd~order).  Here we discuss the tradeoffs between these two
schemes:

\begin{description}
\item[{\rm Overall Accuracy:}]
	One would expect that when using only 3rd~order accurate
	molecules for some terms in the $3\,{+}\,1$ equations near
	the boundaries, the overall accuracy of our results would
	correspondingly be lowered to 3rd~order.  Near the outer
	boundary this is indeed the case.  However, as discussed
	in section~\ref{sect-sample-evolutions/outer-bndry-errors},
	the dominant errors in our numerical solutions there are
	essentially continuum effects, not finite differencing ones,
	so lowering the finite differencing accuracy by one order
	makes little difference to the solutions' accuracies there.

	Near the inner boundary the situation is very different:
	By virtue of the black hole exclusion scheme, the light
	cones there point entirely {\em into\/} the black hole,
	so the future domain of dependence of a grid point near
	the inner boundary contains at most only a few other grid
	points.  
\jtcitesprefix\jtcite{Gustafsson-1971-hyperbolic-BC-FD-convergence,
Gustafsson-1975-hyperbolic-BC-FD-convergence,
Gustafsson-1982-hyperbolic-BC-FD-convergence}
	\jtcitenumericvsname{have}{has} shown theoretically, and
\jtcitesprefix\jtcite{Gary-1975-MOL-BC-4th-order,
Gary-1975-4th-order-MOL-in-Vichnevetsky}
	\jtcitenumericvsname{have}{has} confirmed with numerical
	experiments, that for model problems with this ``outflow
	boundary''
\footnote{
	 That is, ``outflow'' with respect to the
	 computational domain.  This terminology
	 is common in the computational fluid dynamic
	 literature,
\cf{}~footnote~\ref{footnote-hydrodynamics-causality-analogy}.
	 }
{}	causality, off-centered molecules 1~order of accuracy lower
	than the interior molecules may be used near the outflow
	boundary (the inner boundary in our system) {\em without\/}
	lowering the overall accuracy of the solutions.  Unfortunately,
	these results hold only for simple model problems, not for
	nonlinear coupled hyperbolic-elliptic tensor PDEs such as
	the $3\,{+}\,1$ evolution and coordinate equations.  However,
	our numerical results in
section~\ref{sect-sample-evolutions/effects-of-3rd-vs-4th-order-dww}
	strongly suggest that these results do indeed generalize
	to the $3\,{+}\,1$ equations: we find that the $3\,{+}\,1$
	field variables, and diagnostics computed from them, remain
	4th~order accurate even when using only 3rd~order molecules
	for 2nd~derivatives near the inner grid boundary.

\item[{\rm Efficiency and Implementation Convenience:}]
	In order to remain 4th~order, the off-centered 2nd~derivative
	molecules in part~(c) of table~\ref{tab-FD-molecule-coeffs}
	must be one point larger (6~points) than the centered and
	1st~derivative molecules (5~points).  When using the off-centered
	molecules viewpoint, we find that accomodating molecules of
	varying sizes generally adds only modest implementation
	complexities and inefficiencies.

	Correspondingly, the 5th~order extrapolants in part~(b) of
	table~\ref{tab-FD-extrapolation-coeffs} must be one point
	larger than the 4th~order extrapolants in part~(a).  This
	extra size doesn't in itself cause any particular problem,
	but using different extrapolants for 1st and 2nd~spatial
	derivatives
\footnote{
	 Using the 6~point 5th~order extrapolants
	 of part~(b) of table~\ref{tab-FD-extrapolation-coeffs}
	 for 1st~derivatives as well as 2nd~derivatives
	 gives a violently unstable differencing scheme.
	 }
{}	does have significant implications for the overall organization
	of a $3\,{+}\,1$ code:  Because two different extrapolants may
	need to be applied to the same grid function(s), we can't just
	extrapolate all the to-be-finite-differenced grid functions
	once \enmasse{} at the start of each time step or evaluation
	of the $3\,{+}\,1$ evolution equations' right hand side.
	Instead, we must perform a separate extrapolation for each
	individual finite differencing operation.
\footnote{
	 This could be optimized to only
	 re-extrapolate when changing from
	 $\partial_\wr$ to $\partial_{\wr\wr}$
	 or vice versa, but the basic issue
	 remains unchanged.
	 }
$^,$
\footnote{
	 If we use the natural ``pointwise'' organization
	 of the computations, where cache locality is
	 optimized by interleaving algebraic and finite
	 differencing operations so that we do as much
	 computation as possible at one grid point before
	 moving on to the next point, then the extrapolation
	 is somewhat more awkward:  For each individual
	 finite differencing operation {\em at each grid point\/},
	 we must now test to see if we're close enough to the
	 (a) boundary for extrapolation to be needed.  This
	 adds a small extra overhead at {\em each\/} grid
	 point, not just at the near-boundary points.
	 }
\end{description}
%
%
\subsection{Time Integration}
\label{sect-time-integration}

Our time integration uses the method of lines.  This technique
is somewhat uncommon in numerical relativity, so we give a brief
introduction to it in appendix~\ref{app-MOL-intro}.

As discussed in section~\ref{sect-nonuniform-gridding}, for
simplicity's sake we don't use any spatial adaptive gridding in this
work.  For the same reason, we use simple fixed-step fixed-order time
integrators.  We have implemented both the classical 4th~order Runge-Kutta
method and a 4th~order Adams-Bashforth-Moulton predictor-corrector
method (\jtciteprefix\jtcite[section 6.4]{Forsythe-Malcolm-Moler}).
Each time step requires \hbox{4\,(2)}~right-hand-side--function
evaluations for the Runge-Kutta (predictor-corrector) integrator
respectively, \ie{} \hbox{4\,(2)}~complete evaluations of the
coordinate conditions~\eqref{eqn-generalized-EF-lapse}
and~\eqref{eqn-constant-area-shift}, the evolution
equations~\eqref{eqn-evolution}, and their respective boundary
conditions~\eqref{eqn-lapse-scalar-Robin-outer-BC},
\eqref{eqn-shift-scalar-Robin-outer-BC},
and~\eqref{eqn-outgoing-radiation-outer-BC}.

We use a fixed time step, chosen to be slightly less than
$\kappa \, (\Delta r)_\min$, where $\kappa$ is an $O(1)$ parameter
and $(\Delta r)_\min = (\Delta \wr) \Big/ (d \wr / d r)_\min$ is the
$r$-coordinate grid spacing on the inner boundary (\ie{} the smallest
$r$-coordinate grid spacing anywhere in the grid).
\footnote{
	 In detail, we choose the time step $\Delta t$
	 as follows:  For compatability with the
	 predictor-corrector integrator, we use the same
	 (fixed) $\Delta t$ for all time steps.  (Changing
	 $\Delta t$ during an evolution is trivial for
	 the Runge-Kutta integrator, but somewhat awkward
	 for the predictor-corrector time integrator.)
	 We don't do any time interpolation, so we (must)
	 choose $\Delta t$ to evenly divide all the coordinate
	 times where field variables are to be output.
	 To this end, for any real $x > 0$ let
	 $\llfloor x \rrfloor$ denote the largest
	 number $\leq x$ satisfying this latter
	 condition.  We then choose the time step to be
	 $\Delta t = \llfloor \kappa \, (\Delta r)_\min \rrfloor$.
	 }
$^,$
\footnote{
	 This is certainly not the only possible way
	 to choose the time step.  Another plausible
	 scheme would be to normally use a time step
	 $\kappa \, (\Delta r)_\min$, except that if
	 the next such step would take us past an output
	 time, then take a single shorter Runge-Kutta
	 step so as to just reach the output time.
	 }

We have made test evolutions with a variety of $\kappa$~values and
grid resolutions to determine our evolution scheme's empirical stability
limit.
\footnote{
	 Since the Einstein equations are nonlinear,
	 the stability limit presumably also depends
	 on the precise spacetime being evolved.
	 In practice, though, this dependence doesn't
	 seem to be very strong.
	 }
{}  For the spacetimes discussed in this paper, this limit is approximately
$\kappa = \hbox{1.5\,(0.5)}$ for the Runge-Kutta (predictor-corrector)
integrator.  (This limit is weakly resolution-dependent in the sense of
lower resolutions having a somewhat larger $\kappa$ limit, up to about
a 20\% increase at the lowest resolutions used here.)

Comparing the two time integrators, the Runge-Kutta method is easier
to program, more flexible (it allows easy changing of the time step),
and when operating at the stability limit, about 50\% more efficient.
(That is, compared to the predictor-corrector integrator, the Runge-Kutta
integrator requires twice as many right-hand-side--function evaluations
per time step, but allows a maximum stable time step about 3~times larger.)
All our results in this paper are from evolutions using the Runge-Kutta
integrator with $\kappa = 1.5$.
%
%
\subsection{Comparison with Other Methods}
\label{sect-numerical-methods-cmp}

Our numerical methods differ in several ways from those usually used
in~$3\,{+}\,1$ numerical relativity.  Here we discuss some of the advantages
and disadvantages of our key design choices:
%
%
\subsubsection{4th~Order Finite Differencing}
\label{sect-numerical-methods-cmp/4th-order-FD}

Given smooth grid functions and a reasonable grid resolution, 4th~order
finite differencing yields dramatically improved accuracy over the
usual 2nd~order, for (we find) only a modest increase in implementation
effort and computational work per grid point.

For time evolution problems, the most difficult part of 4th~order
finite differencing remains the same as that of 2nd~order finite
differencing: constructing (discovering) a {\em stable\/} differencing
scheme.  Contrary to some suggestions elsewhere
(\eg{}~\jtciteprefix\jtcite[section~19.3]{Numerical-Recipes}), we do
{\em not\/} find it any more difficult to construct stable 4th~order
differencing schemes than 2nd~order ones.  In fact, we find that
allowing molecules to be larger than 3~points (which is essential
for any 4th~order scheme except a nonlocal \defn{compact} one
(\jtcitesprefix\jtcite{Ciment-Leventhal-1975-4th-order-compact-FD,
Hirsh-1975-4th-order-compact-FD})) makes it {\em easier\/} to construct
stable finite differencing schemes.  This is because constructing a
stable differencing scheme often requires considerable trial-and-error
experimentation with various candidate schemes
(\eg{}~\jtciteprefix\jtcite[section~IV]{Marsa-Choptuik-1996-sssf}),
and larger molecules allow the formation of a wider variety of such
candidates consistent with the necessary domains of dependence.
(On the other hand, the requirement of 4th~order accuracy disallows
most averaging techniques and ADI-type schemes.)

With 4th~order finite differencing schemes of the type used here,
there's no need to use a staggered grid of the type sometimes used
with 2nd~order finite differencing schemes
(\eg{}~\jtcitesprefix\jtcite{Choptuik-PhD,Hawley-Evans-1989-in-Frontiers,
Choptuik-1991-consistency}).  We consider this a (modest) advantage,
as staggered grids are somewhat awkward both in programming and in
the eventual data analysis.

It's also interesting to note that we haven't needed to use any type
of causal differencing
(\jtciteprefix\jtcite{Alcubierre-PhD,
Alcubierre-Schutz-1994-causal-differencing})
to obtain a stable evolution scheme, even inside the horizon where
the grid points are moving along spacelike world lines.  We haven't
investigated the reasons for this in detail, but we suspect it's due
to the larger numerical domains of dependence inherent in our scheme's
5~and 6~point molecules.
%
%
\subsubsection{Richardson Extrapolation}
\label{sect-numerical-methods-cmp/Richardson-extrap}

Another technique (which we haven't used here) for increasing
the order of accuracy of any well-behaved (stable, convergent)
finite differencing computation, is Richardson extrapolation
from lower order computations at multiple grid resolutions.
(We briefly review this technique in
appendix~\ref{app-convergence-tests+Richardson-extrap}.)
This is a very powerful technique, and has been used successfully
by (\eg{})
\jtciteprefix\jtcite{Choptuik-Goldwirth-Piran-1992-sssf-cmp-3+1-vs-2+2}
to obtain 4th~order accuracy in spherically symmetric scalar field
evolutions using 2nd~order underlying finite differencing.

For some simple model problems, it's in fact straightforward to show
that such a Richardson extrapolation scheme is in fact exactly equivalent
(yields identical finite difference equations) to a 4th~order finite
differencing scheme of the type we use here.  For example, for the
flat-space linear scalar wave equation and 2nd~order leapfrog finite
differencing, Richardson extrapolating from results at a \hbox{2:1}~ratio
of grid spacings is precisely equivalent to the centered 4th~order
molecules given in table~\ref{tab-FD-molecule-coeffs}.  However, it's
not clear that this equivalence also holds for more complicated systems
of PDEs like the $3+1$ equations.

Though still useful, Richardson extrapolation offers lesser benefits
if parts of the differencing scheme are off-centered:  As explained
in appendix~\ref{app-convergence-tests+Richardson-extrap}, given a
pair of computations at differing grid resolutions (say with a
\hbox{2:1}~ratio of resolutions), in general a single Richardson
extrapolation raises the order of accuracy by~2 if the finite
differencing scheme is fully centered, but only by~1 if the scheme
is even partially off-centered.  Notably, in the presence of
off-centering, Richardson-extrapolating a pair of 2nd~order computations
in general only yields a 3rd~order result, or alternately {\em three\/}
2nd~order computations with differing grid resolutions (say with a
\hbox{4:2:1}~ratio of resolutions) would be needed to obtain a 4th~order
result by Richardson extrapolation.

Similarly, Richardson-extrapolating a \hbox{2:1}~pair of our
(partially off-centered) 4th~order computations would yield ``only''
5th~order results.  This might well be a useful scheme, but we haven't
investigated it as yet.

We should note one possible drawback to such a
Richardson-extrapolation scheme: In practice, our numerical results
in section~\ref{sect-sample-evolutions/convergence-tests} suggest
that although our our 4th~order evolution scheme yields qualitatively
very accurate results at even modest grid resolutions, quite high
grid resolutions are needed before quantitative 4th~order convergence
is fully attained, \ie{}~before the field variables accurately satisfy
the convergence relationship~\eqref{eqn-convergence}, or equivalently
the Richardson error expansion~\eqref{eqn-Richardson-expansion}.

For example, in
figure~\ref{fig-400o10+lower-res.pqw5-Diff-gK-magnitude-conv[t=100]},
notice that the $\text{100o10} - \text{200o10}$ and
$\text{200o10} - \text{400o10}$ curves are still significantly
different in shape for $r \ltsim 70$.  This means that Richardson
extrapolating the 100o10 and 200o10 field variables would quite likely
{\em not\/} yield significantly increased accuracy in this range of~$r$;
in fact it might well even {\em decrease\/} the accuracy there.
The resolution would need to be increased by another factor of~2
(Richardson extrapolating the 200o10 and 400o10 field variables)
before we could be confident that Richardson extrapolation would
work well (yield a significant accuracy improvement over the
higher-resolution pair).
%
%
\subsubsection{Method of Lines}
\label{sect-numerical-methods-cmp/MOL}

We use the method of lines (\defn{MOL}) to time-integrate the evolution
equations.  In comparison to the more common \defn{space and time together}
(SATT) finite differencing schemes, MOL offers both advantages and
disadvantages:

Many SATT schemes in the numerical analysis literature are presented
(only) for simple model problems, and are difficult to generalize to
complicated and ``messy'' problems like the $3\,{+}\,1$ evolution equations.
In particular, with SATT schemes it's often difficult to handle the
2nd~spatial derivatives on the right hand side of the $3\,{+}\,1$ evolution
equations, and schemes using substeps often encounter difficulty with
obtaining suitable boundary conditions and/or treating the coordinate
equations.  In contrast, MOL schemes usually generalize easily to
``messy'' problems.

In the $3\,{+}\,1$ context, MOL decouples the time integration from the
detailed form of the evolution and coordinate equations.  This
decoupling is an advantage in that it isolates the complexity of the
$3\,{+}\,1$ equations in the spatial finite differencing, simplifying the
programming, debugging, and stability analysis of a $3\,{+}\,1$ code.  On
the other hand, this decoupling may also be somewhat of a disadvantage,
in that MOL may give up an $O(1)$ factor in efficiency relative to
a equal-order SATT scheme.  MOL also offers somewhat less flexibility
to tailor the finite differencing of individual terms in the $3\,{+}\,1$
equations to their functional form for (say) improved stability
and/or accuracy.

MOL makes it relatively easy to use finite differencing with higher
than 2nd~order accuracy in both space and time.  Although 4th~order
SATT schemes do exist (see, \eg{}, 
\jtcitesprefix\jtcite{Kreiss-Oliger-1972-4th-order-FD,
Kreiss-Oliger-1973-GARP-report,Oliger-1974-4th-order-FD,
Ciment-Leventhal-1975-4th-order-compact-FD,
Hirsh-1975-4th-order-compact-FD,
Turkel-Abarbanel-Gottlieb-1976-4th-order-FD,Kreiss-1978-Montreal-book}),
they generally don't apply to the $3\,{+}\,1$ equations, due to the
``messy problem'' difficulties noted above.  In contrast, constructing
4th~order MOL schemes is straightforward, even for the full $3\,{+}\,1$
equations.

For almost all time evolution problems, and certainly for the $3\,{+}\,1$
Einstein equations, the most difficult part of both SATT and MOL finite
differencing is the construction of {\em stable\/} differencing schemes.
We find this to be about equally difficult for SATT and MOL; neither
method seems to have a clear advantage here.

We find (only) two significant drawbacks to MOL:  First, MOL schemes
are less widely used than SATT schemes, and the MOL literature is still
somewhat sparse.  Second, the development of MOL adaptive gridding
schemes analogous to the Berger-and-Oliger--type SATT adaptive gridding
schemes noted in section~\ref{sect-nonuniform-gridding}, remains an
open research problem.
%
%
\section{Diagnostics}
\label{sect-diagnostics}

%
%
\subsection{The Energy Constraint}
\label{sect-diagnostics/energy-constraint}

One of our major diagnostics of the accuracy of our results is the
numerically computed energy constraint $C$.
\footnote{
	 $C$ is a much more sensitive diagnostic for
	 this purpose than the momentum constraint
	 $C^i$, because $C$ depends on {\em 2nd\/}~spatial
	 derivatives of the 3-metric components,
	 whereas $C^i$ only depends on 1st~spatial
	 derivatives.
	 }
{}  $C$ has dimensions of $(\text{length})^{-2}$, but there are at least
two different plausible ways to normalize it (make it dimensionless):

Some authors (\eg{}~\cite{2BH-grand-challenge-alliance-1998-moving-BH})
normalize $C$ by taking the absolute value of each of its tensor-level
terms, \ie{} since $C$ is given
by~\eqref{eqn-energy-constraint-generic}, they define
\begin{equation}
C_\abst \equiv |R| + |K\ij K\upij| + |K^2| + |16 \pi \rho|
\end{equation}
so that $C_\relt \equiv C / C_\abst$ is a dimensionless measure
of the extent to which the individual tensor-level terms
in~\eqref{eqn-energy-constraint-generic} cancel out to yield
(ideally) a zero sum~$C$.

However, each of these tensor-level terms is in fact computed as the sum
of several individual terms involving partial derivatives of the state
vector, and these ``inner sums'' may themselves involve nontrivial
cancellations.  For example, for an Eddington-Finkelstein slice of a
mass-$m$ Schwarzschild spacetime, at large~$r$ the computation of $R$
by equation~(B5) of paper~I involves 2~orders in~$r$ of cancellations:
$R = O(m^2/r^4)$ but some of the individual terms in that equation are
$O(1/r^2)$.

We can better measure {\em all\/} the cancellations involved in computing
$C$ by an alternative normalization, where we take the absolute value
of each individual partial-derivative term.  That is, by equations~(B5)
and~(B15) of paper~I, we have
\begin{eqnarray}
C
	& = &
		- 2 \frac{\partial_{rr} B}{A B}
		+ \dhalf \frac{(\partial_r B)^2}{A B^2}
		+ \frac{(\partial_r A)(\partial_r B)}{A^2 B}
		+ \frac{2}{B}
						\nonumber	\\
	&   &
		+ 2 \frac{Y^2}{B^2}
		+ 4 \frac{X Y}{A B}
		- 2 \frac{P^2}{A}
		- 2 Q^2
					\, \text{.}	
				\label{eqn-energy-constraint-partial-derivs}
\end{eqnarray}
We thus define
\begin{eqnarray}
C_\absp
	& \equiv &
		  \left| 2 \frac{\partial_{rr} B}{A B} \right|
		+ \left| \dhalf \frac{(\partial_r B)^2}{A B^2} \right|
		+ \left|
		  \frac{(\partial_r A)(\partial_r B)}{A^2 B}
		  \right|
		+ \left| \frac{2}{B} \right|
						\nonumber	\\
	&   &
		+ \left| 2 \frac{Y^2}{B^2} \right|
		+ \left| 4 \frac{X Y}{A B} \right|
		+ \left| 2 \frac{P^2}{A} \right|
		+ \left| 2 Q^2 \right|
\end{eqnarray}
so that $C_\relp \equiv C / C_\absp$ is a dimensionless measure
of the extent to which the individual partial-derivative terms
in~\eqref{eqn-energy-constraint-partial-derivs} cancel out to yield
(ideally) a zero sum~$C$.

As outlined above, $C_\relt$ and $C_\relp$ measure different things.
For an Eddington-Finkelstein slice of a mass-$m$ Schwarzschild spacetime,
$C_\abst = O(m^2/r^4)$, while $C_\absp = O(1/r^2)$, so
$C_\relt / C_\relp = O(r^2/m^2)$.  That is, requiring $C_\relt$ to be
below a given tolerance at large~$r$ is a much stricter criterion than
requiring $C_\relp$ to be below that same tolerance.

Since $C_\relp$ measures the extent of cancellation in the actual
expression we use to compute~$C$, we think it's a more realistic
diagnostic of the numerical quality of our solutions.  However, the
definition of $C_\relp$ depends on the details of our computational
scheme: it would be difficult to unambiguously compare $C_\relp$ values
for different schemes, or between (say) our spherically symmetric scheme
and a generalization to axisymmetric or fully 3-dimensional spacetimes.
In contrast, $C_\relt$ has a simple, elegant, and
computational-scheme--independent definition.

In describing our numerical results we generally give both diagnostics.
Much of our discussion of them is common to both, so we often refer to
$C_\rel$, meaning either $C_\relt$ or $C_\relp$.
%
%
\subsection{Other Diagnostics}

Two other important diagnostic of the accuracy of our results are the
{\em changes\/} in the numerically computed 3-metric and extrinsic
curvature components when the grid resolution is doubled in a convergence
test.  (We describe our convergence-test methodology in detail
in~appendix~\ref{app-convergence-tests+Richardson-extrap}.)  That
is, defining $Z[\Delta \wr]$ to be the numerically computed value of
the field variable $Z$ using the grid spacing $\Delta \wr$, we use
$\Delta g\ij \equiv g\ij[\Delta \wr] - g\ij[\Delta \wr/2]$ and
$\Delta K\ij \equiv K\ij[\Delta \wr] - K\ij[\Delta \wr/2]$ as
diagnostics of the accuracy of $g\ij$ and $K\ij$ respectively.  For
convenience we normally work with pointwise tensor ``magnitude''
norms of these diagnostics, $\magnitude{\Delta g\ij}$ and
$\magnitude{\Delta K\ij}$.
\footnote{
	 Recall that for a symmetric rank~2 covariant
	 tensor $T\ij$, we define 
	 $\magnitude{T\ij} \equiv \sqrt{T\ij T\upij}$.
	 }

Our remaining diagnostics are mostly described in detail in paper~I
(section~VIII and appendices~B and~C), so we only briefly summarize
them here:  To study the scalar field we use its 3-energy density
$\rho$, its radial 3-energy and 3-momentum densities $4 \pi B \rho$
and $4 \pi B j^r$ respectively, the 4-Ricci scalar
$\four\! R \equiv 8 \pi (\rho - T)$ (which we usually normalize
as $\thalf B \, \four\! R \equiv 4 \pi B (\rho - T)$ to match
$4 \pi B \rho$ and $4 \pi B j^r$), and the (Misner-Sharp)
\xdefn{mass function}~$m$ 
(\jtcitesprefix\jtcite{Hayward-1994-mass-functions,
Hayward-1996-Misner-Sharp-mass-function}).
\footnote{
	 We actually use two different forms of the
	 Misner-Sharp mass function: one computed directly
	 from $g\ij$ and $K\ij$ (we call this form $m_\MS$;
	 \cf{}~appendix~C of paper~I), and the other
	 computed by integrating the local mass density of
\jtciteprefix\jtcite{Guven-OMurchadha-1995-constraints-in-spherical-symmetry-I}
	 (we call this form $m_\mu$; \cf{}~section~VIII of
	 paper~I).  As discussed in appendix~B3 of paper~I,
	 at large~$r$ $m_\MS$ is quite sensitive to small
	 numerical errors in~$g\ij$ and $K\ij$, so we
	 generally define $m \equiv m_\mu$, and refer to
	 this as ``the'' mass function.
	 }
$^,$
\footnote{
	 When measuring the total mass of a slice, to
	 reduce boundary--finite-differencing errors we
	 actually measure the mass 3~grid points inside
	 the outer grid boundary, \ie{} we define
	 $m_\total \equiv m(\iwr=\iwr_\max{-}3)$.
	 }

To study the black hole's growth and mass, we use the areal radius
$h$ of the apparent horizon(s) in each slice, and $m_\bh$, the
mass function interpolated to the outermost apparent horizon's
position.  In a slight abuse of terminology we refer to $m_\bh$ as
the (time-dependent) \defn{mass of the black hole}.  As discussed in
appendix~B of paper~I, in spherical symmetry any apparent horizon
must have \hbox{$r_\areal$ ($\equiv h$ in our coordinates) $= 2m$}.
For any diagnostic $Z$, $Z_h$ denotes the value of $Z$ at
(interpolated to) the or an apparent horizon position.
%
%
\section{Sample Evolutions}
\label{sect-sample-evolutions}

We have made a number of test evolutions to investigate our code's
stability, accuracy, and performance, and to study some of the spacetimes
numerically generated by it.  These evolutions include both vacuum
(Schwarzschild) spacetimes, and dynamic spacetimes containing black
holes surrounded by scalar field shells, but in the interests of brevity
we only discuss the latter here.

We discuss our initial data computation for these evolutions in
detail in paper~I.  Briefly, we begin with an Eddington-Finkelstein
slice of the unit-mass Schwarzschild spacetime, add a Gaussian to
one of the field variable components (always $P$ for the examples
considered here), numerically solve the full 4-vector form of the
York initial data algorithm to project the field variables back into
the constraint hypersurface, and finally numerically coordinate-transform
back to an areal radial coordinate.  For the parameters used here,
this algorithm results in a roughly Eddington-Finkelstein slice
(with $K$ generically nonzero and spatially variable everywhere in
the slice) containing a black hole surrounded at some moderate
radius by a roughly Gaussian scalar field shell.

Table~\ref{tab-test-evolutions-pars} summarizes various parameters
of the evolutions discussed here.  Notice that each model's name
encodes the key numerical parameters (grid resolution and outer
boundary position) and the basic physics (the input perturbation for
the initial data solver).  Most of our discussion of these models is
is independent of their numerical parameters, so we often refer
generically to the families of evolutions sharing common numerical
parameters, and plot results for only one representative set of
numerical parameters within each family.

All the evolutions discussed here use an areal radial coordinate $r$,
with the grid covering the range from $\wr_\min = 0$ ($r_\min = 1.5$)
through one of $\wr_\max = 4$, $10$, or $30$ ($r_\max \approx 248$,
$813$, or $2780$).  The numerical calculations use IEEE double
(64~bit) precision for all floating-point arithmetic.
%
%
\subsection{Scalar Field Phenomenology}
\label{sect-sample-evolutions/scalar-field-phenomenology}

We first consider the pqw5 evolutions, which show a black hole accreting
a relatively thick and moderate-mass scalar field shell.  On the initial
slice the shell's radial density profile $4 \pi B \rho$ has a standard
deviation in~$r$ approximately $1.8$~times the black hole radius, and
the shell has approximately $0.66$~times the black hole mass, \ie{}~$40\%$
of the slice's total mass.

Figure~\ref{fig-200.pqw5-h+scalar-field+mass[t=0(5)50]} shows the
early-time evolution of the scalar field's radial 3-energy and 3-momentum
densities $4 \pi B \rho$ and $4 \pi B j^r$, the normalized 4-Ricci
scalar $\thalf B \, \four\! R \equiv 4 \pi B (\rho - T)$, and the
mass function $m$, for the 200.pqw5 evolution.  (The other pqw5
evolutions don't differ noticeably at the scale and times of this
figure.)

It's clear that the initial slice's scalar field shell is actually a
superposition of two momentarily-coincident shells, one propagating
inward ($j^r < 0$), the other propagating outward ($j^r > 0$).  The
two shells have quite different masses, approximately 0.47~and 0.18~times
the black hole mass ($28\%$~and $11\%$~of the slice's total mass)
respectively.
\footnote{
	 Given the nature of our initial data algorithm
	 (\cf{}~paper~I) and the fact that the pqw5
	 models' initial perturbation is to $P$ only,
	 with $Q$ remaining identically zero
	 (\cf{}~table~\ref{tab-test-evolutions-pars}),
	 the general form of a superposition of ingoing
	 and outgoing shells is as expected.  However,
	 we don't know why the two shells' masses differ
	 so much.
	 }

As the evolution proceeds, the outgoing scalar field shell propagates
outward relatively intact, while the ingoing scalar field shell is
partly captured by the black hole and partly scattered off the
strong-field region's spacetime curvature.  The scattered scalar field
(visible in figure~\ref{fig-200.pqw5-h+scalar-field+mass[t=0(5)50]},
as, for example, the positive pulse in~$4 \pi B j^r$ at $t = 15$
near $r = 20$) then undergoes further scattering off the
strong-field curvature.  The net result of this
\defn{quasinormal-mode ringing} process is the formation of a
sequence of successively-weaker outgoing scalar field shells following
the original one.  The outgoing scalar field shells all propagate
out along asymptotically null geodesics, with asymptotically constant
amplitudes in~$4 \pi B \rho$ and $4 \pi B j^r$, corresponding to
${\sim}\, 1/r^2$~falloffs in~$\rho$ and $j^r$.

These features can be seen in
figure~\ref{fig-200o10+200.pqw5-four-pi-B-rho+mass[t=0(50)500]}, which
shows the later-time evolution of the radial scalar field density and
mass function for the 200o10.pqw5 and 200.pqw5 models.
\footnote{
	 Note that
	 the sharp downward cusps visible in~$4 \pi B \rho$
	 in this figure do {\em not\/} represent zeros,
	 just relatively narrow nonzero minima.
	 }
{}  (For the moment we consider only the 200o10.pqw5 results; as
discussed in section~\ref{sect-sample-evolutions/outer-bndry-errors},
the 200.pqw5 results are contaminated by outer-boundary errors.)
For example (summarizing the scalar field shells' profiles by their
approximate $\text{mean positions} \pm \text{standard deviations}$
in~$4 \pi B \rho$), at $t = 200$ the original outgoing scalar
field shell is at $r \approx 200 \,{\pm}\, 10$, while the
quasinormal-mode--ringing shells are at $r \approx 180 \,{\pm}\, 20$,
$150 \,{\pm}\, 20$, $120 \,{\pm}\, 20$, and $80 \,{\pm}\, 10$.
%
%
\subsection{Quasinormal-Mode Ringing}
\label{sect-sample-evolutions/QNM-ringing}

During the quasinormal-mode ringing process, at any fixed radius
near the black hole, the field undergoes a sequence of damped
(temporal) oscillations, whose frequency is theoretically predicted
to be characteristic of the black hole mass.
Figure~\ref{fig-400o10.pqw5-scalar-field.horizon[t=0-200]} shows
these oscillations for the 400o10.pqw5 evolution.  About 3~oscillation
cycles are visible before the field settles down into a smooth decay
at later times (discussed in detail in
section~\ref{sect-sample-evolutions/late-time-decay-of-SF-near-BH}).
Examining the $\rho_h$ minima/maxima times gives an oscillation period
of about $\tau = 32.5 \,{\pm}\, 1$, where the error estimate is dominated
by the deviations of the oscillation from exact periodicity.

From our results in
section~\ref{sect-sample-evolutions/BH-growth+apparent-horizon-motion},
for this evolution the black hole mass is very nearly constant at
$m = 1.183$ for $t \gtsim 45$ (where most of the oscillations occur),
so our measured oscillation period corresponds to
$\tau = (27.5 \,{\pm}\, 1) \, m_\bh$, in excellent agreement with the
theoretical prediction (\jtciteprefix\jtcite[section~4.4]{Marsa-PhD})
of $\tau = 28.44 \, m_\bh$.
%
%
\subsection{Black Hole Growth and Apparent Horizon Motion}
\label{sect-sample-evolutions/BH-growth+apparent-horizon-motion}

Because the black hole captures some of the scalar field, the black hole
grows during the evolution.  Figure~\ref{fig-200.pqw5-h+m(h)[t=0-50]}
shows the time evolution of the horizon position~$h$ and the contained
(black hole) mass $m(h)$ for the 200.pqw5 model.  Because the scalar
field shell is relatively thick and of moderate mass compared to the
black hole, the black hole grows relatively slowly and smoothly, and
there's always only a single smoothly-moving apparent horizon.

In contrast, consider the 400.pqw1 model, in which the scalar field
shell is relatively thin and massive.  (On the initial slice, the
shell's radial 3-energy density profile $4 \pi B \rho$ has a standard
deviation in~$r$ of only about $0.43$~times the black hole radius,
and the shell has approximately $3$~times the black hole mass,
\ie{}~$75\%$ of the slice's total mass.)

The general phenomenology of this model's scalar field evolution is
fairly similar to that of the 200.pqw5 model.  However, as shown in
figure~\ref{fig-400.pqw1-h+m(h)[t=0-50+19-20]}, the time evolution
of the apparent horizon position $h$ and contained (black hole) mass
$m(h)$ is quite different.  When the thin and massive scalar field
shell falls through the horizon, the black hole grows very rapidly
during a short time interval.  In particular, during the time interval
$19.13 \ltsim t \ltsim 19.71$ there are three distinct apparent horizons
present.  \jtciteprefix\jtcite{Marsa-PhD,Marsa-Choptuik-1996-sssf}
have reported seeing similar behavior in their
spherically-symmetric-scalar-field numerical evolutions,
though with somewhat different initial data.  As discussed in
section~\ref{sect-multiple-apparent-horizons}, this type of transient
multiple--apparent-horizon behavior has important implications for how
a black-hole--exclusion computational scheme should handle the inner
boundary.  

In spherical symmetry, apparent horizons are (occur at positions
given by) precisely the zeros of the horizon function $H$ defined
by~\eqref{eqn-horizon-generic}.  Generically, $H$'s zeros (apparent
horizons) always appear or disappear in pairs, corresponding to local
minima or maxima in~$H$ just touching the $r$ axis.  This behavior
is illustrated in figure~\ref{fig-400.pqw1-H[t=19(0.1)20]}, which
shows $H$ for the 400.pqw1 evolution at times $t = 19(0.1)20$.  
Moreover, at the moment a local minima or maxima just touches the
$r$ axis, the $H$ curve is generically tangent to the axis there, so
at the moment of appearance or disappearance apparent horizons are
generically moving with infinite coordinate speed $dh/dt$.  (This
is visible in figure~\ref{fig-400.pqw1-h+m(h)[t=0-50+19-20]} as the
$h(t)$ curve becoming vertical at the appearance and disappearance
times $t \approx 19.13$ and $t \approx 19.71$.)  As discussed in
section~\ref{sect-multiple-apparent-horizons}, this very fast movement
of the apparent horizon also has important implications for how a
black-hole--exclusion computational scheme should handle the inner
boundary.

It would be interesting to numerically compute the event horizon
for this spacetime (presumably using the methods of
\jtcitesprefix\jtcite{ABBLMSSSW-1995-numerical-event-horizons,
LMSSW-1996-numerical-event-horizons})
to compare it to the apparent horizons, but we haven't done this.
%
%
\subsection{Convergence Tests}
\label{sect-sample-evolutions/convergence-tests}

To assess the overall accuracy level of our numerical evolution
scheme, we begin by considering the magnitude of the energy
constraint, normalized as both $C_\relt$ and $C_\relp$
(\cf{}~section~\ref{sect-diagnostics/energy-constraint}).
Figure~\ref{fig-800o10+lower-res.pqw5-C-rel-conv[t=100]} shows
$|C_\relt|$ and $|C_\relp|$ in the inner part of the grid for the
various-resolution o10.pqw5 evolutions, at the time $t = 100$.
(We discuss the outer part of the grid in
section~\ref{sect-sample-evolutions/outer-bndry-errors}.)

Consider first the 100o10.pqw5, 200o10.pqw5, and 400o10.pqw5 evolutions.
For $r \gtsim 130$ the $C_\rel$ values are very small and are dominated
by floating-point roundoff noise.
\footnote{
	 Roundoff noise is easily recognizable by 
	 its extreme nonsmoothness (order-of-magnitude
	 variations from one grid point to the next).
	 }
{}  At smaller radia (\ie{} in part of the slice with nontrivial dynamics),
the $C_\rel$ values are still quite small:  Even for our lowest resolution
(the 100o10.pqw5 evolution), $C_\relt$ is still ${\ltsim}\, 10^{-2}$
in magnitude throughout the grid, and the more realistically normalized
$C_\relp$ (\cf{}~section~\ref{sect-diagnostics/energy-constraint}) is
${\ltsim}\, 3 \,{\times}\, 10^{-5}$.  Moreover, we generally have
excellent 4th~order convergence of the $|C_\rel|$ values:  In particular,
notice the very nice factor-of-16 ratio between the 200o10.pqw5 and
400o10.pqw5 $|C_\rel|$ values.

However, the $|C_\rel|$ values for the 800o10.pqw5 evolution don't
fully show the expected convergence: in regions where the scalar field
is small but not negligible (\eg{}~$r \ltsim 70$), $|C_\rel|$ (and the
underlying $|C|$ from which $C_\rel$ is computed) for this evolution
shows significant non-smoothness (rapid variations from one grid point to
the next), and is generally only a factor of ${\approx}\, 10$ lower than
for the 400o10.pqw5 evolution.  Although we don't have conclusive proof,
we very strongly believe that these effects are both due entirely to
floating-point roundoff errors.  In support of this hypothesis, we note
that the underlying $|C|$ values in question are very small (typically
${\ltsim}\, 10^{-12}$), and floating-point roundoff effects in the
$3\,{+}\,1$ equations generically grow as $1 / (\Delta \wr)^2$ as the
grid is refined (since the equations contain spatial derivatives up
to 2nd~order).  Furthermore, we note that the 800o10.pqw5 evolution
{\em does\/} show excellent 4th~order convergence near the peak of
the outgoing scalar field shell ($100 \ltsim r \ltsim 130$), where
finite differencing errors from the scalar field shell make $|C|$
somewhat larger (and hence we expect roundoff effects to be less
severe in relative terms).

In an attempt to better estimate the effects of roundoff errors
on our numerical evolutions, we modified our numerical scheme to
(optionally) simulate the effects of somewhat increasing these errors,
by adding a tiny amount of white noise (the noise amplitude is 6~times
the worst-case roundoff error of a single floating-point arithmetic
operation) to the field variables several times per time step.
\footnote{
	 We describe the noise-addition algorithm
	 in detail in appendix~\ref{app-adding-noise}.
	 }
{}  However, this turned out to not always offer reliable estimates
of the roundoff effects
\footnote{
	 In hindsight this isn't unexpected:
	 Actual roundoff errors are often highly
	 discrete and/or highly correlated
(\jtciteprefix\jtcite{Kahan-1996-probabilistic-error-analysis-x}),
	 so (continuous and statistically independent)
	 uniform white noise may not model them
	 well, and in fact may systematically
	 misestimate their actual effects (see also
\jtciteprefix\jtcite{Kahan-1999-correlation-vs-low-probability-events}).
	 }
{} (\cf{}~our discussion in the next 2~paragraphs), and at least in
its present incarnation, we don't recommend this technique to other
researchers.

The 800o10n3.pqw5 evolution is identical to the 800o10.pqw5 evolution
except for having noise added in this manner, and
figure~\ref{fig-800o10+lower-res.pqw5-C-rel-conv[t=100]} also shows this
evolution's $|C_\rel|$ values.  As expected, the extra noise (which is
uncorrelated from one grid point to the next) makes the 800o10n3.pqw5
$|C_\rel|$ values (and the underlying $|C|$ values) much less smooth
than the 800o10.pqw5 values.  This supports our hypothesis that the
(lesser) non-smoothness of the latter $|C|$ values is due to ``normal''
floating-point roundoff effects.

However, in some regions of the grid, the 800o10n3.pqw5 $|C_\rel|$
values (and the underlying $|C|$ values) are consistently {\em smaller\/}
-- by up to an order of magnitude -- than the 800o10.pqw5 values!
This is particularly prominent for $20 \ltsim r \ltsim 47$ and
$r_\min \ltsim r \ltsim 19$, and is also noticable for
$52 \ltsim r \ltsim 60$.  We don't understand why adding noise would
make $|C_\rel|$ (and $|C|$) {\em smaller\/}, though we suspect a
dithering effect may be involved here.

Another way to quantitatively assess the accuracy of our numerical
evolution scheme is to consider 3-grid convergence tests
(\cf{}~appendix~\ref{app-convergence-tests+Richardson-extrap})
for the individual $g\ij$ and $K\ij$ coordinate components.
Figure~\ref{fig-400o10+lower-res.pqw5-Diff-gK-magnitude-conv[t=100]}
shows the results of some tests of this type, for the same o10.pqw5
evolutions, again at the time $t = 100$.  Here we have done a
3-grid convergence test on each $g\ij$ and $K\ij$ component, but for
conciseness we plot only the tensor magnitudes of the $g\ij$ and
$K\ij$ differences between the different-resolution models,
$\magnitude{\Delta g\ij}$ and $\magnitude{\Delta K\ij}$.
\footnote{
	 Recall that for a symmetric rank~2 covariant
	 tensor $T\ij$, we define 
	 $\magnitude{T\ij} \equiv \sqrt{T\ij T\upij}$.
	 }

The $\magnitude{\Delta g\ij}$ and $\magnitude{\Delta K\ij}$ values all
show excellent 4th~order convergence, except for some noise in the
highest-resolution $\magnitude{\Delta K\ij}$ values.  This noise has
the same characteristics described above for the 800o10.pqw5 $|C_\rel|$
values in figure~\ref{fig-800o10+lower-res.pqw5-C-rel-conv[t=100]},
and from its qualitative form is almost certainly due to floating-point
roundoff errors.  As additional evidence for this, note that this noise
increases just as we would expect when we add extra noise to the field
variables, \ie{} going from the
$\magnitude{K\ij[\text{400o10.pqw5}] - K\ij[\text{800o10.pqw5}]}$
values to the
$\magnitude{K\ij[\text{400o10.pqw5}] - K\ij[\text{800o10n3.pqw5}]}$
ones.  There's no sign here of anything like the anomalous behavior
seen in the $|C_\rel|$ results.
%
%
\subsection{The Effects of 3rd versus 4th Order $\partial_{ww}$
	    Molecules Near the Inner Boundary}
\label{sect-sample-evolutions/effects-of-3rd-vs-4th-order-dww}

Our spatial finite differencing is generally 4th~order
(\cf{} section~\ref{sect-spatial-FD}), \ie{} we approximate spatial
partial derivatives by 4th~order finite difference molecules.  In
particular, we approximate spatial 2nd~partial derivatives by the
4th~order molecules of part~(c) of table~\ref{tab-FD-molecule-coeffs},
or equivalently we use the quintic extrapolants of part~(b) of
table~\ref{tab-FD-extrapolation-coeffs}.

However, as discussed in
section~\ref{sect-spatial-FD/3rd-vs-4th-order-dww-mols-near-bndrys},
there are some efficiency and implementation-convenience advantages
to approximating spatial 2nd~partial derivatives near the grid
boundaries by the 3rd~order finite difference molecules of part~(b)
of table~\ref{tab-FD-molecule-coeffs}, or equivalently the quartic
extrapolants of part~(a) of table~\ref{tab-FD-extrapolation-coeffs}.
The 100e4.pqw5, 200e4.pqw5, and 400e4.pqw5 evolutions use this
alternative finite differencing scheme, but are otherwise identical
to the 100.pqw5, 200.pqw5, and 400.pqw5 evolutions (respectively).

In general, the results of the e4.pqw5 evolutions are very similar
to those of the corresponding pqw5 evolutions.  Compared to the pqw5
evolutions, the e4.pqw5 evolutions' errors (measured either by the
constraints, or by 3-grid convergence tests on the $g\ij$ and $K\ij$
components) are substantially larger at the innermost few grid points
on the initial slice, but these larger errors damp out within a few
time steps, and thereafter the e4.pqw5 evolutions have similar
accuracies to the pqw5 evolutions.  For example,
figure~\ref{fig-200e4+200.pqw5-C-relp.inner[t=0+various+100]}
shows $|C_\relp|$ for the innermost part of the grid for the
200.pqw5 and 200e4.pqw5 evolutions.  Notice that on the initial slice
($t = 0$), at the innermost grid point the 200e4.pqw5 evolution's
error is about 3~orders of magnitude larger than that of the
200.pqw5 evolution.  However, even by $t = 0.01$ the errors are
within an order of magnitude of each other, and by $t = 0.1$
(3~normal-sized time steps for this evolution), or at larger radia,
they're essentially identical.

More generally, we find (details omitted for brevity) that the
e4.pqw5 evolutions show full 4th~order convergence of $|C_\rel|$
and other accuracy diagnostics.  This extends the the model-problem
analytical results of
\jtcitesprefix\jtcite{Gustafsson-1971-hyperbolic-BC-FD-convergence,
Gustafsson-1975-hyperbolic-BC-FD-convergence,
Gustafsson-1982-hyperbolic-BC-FD-convergence}
and the numerical experiments of
\jtcitesprefix\jtcite{Gary-1975-MOL-BC-4th-order,
Gary-1975-4th-order-MOL-in-Vichnevetsky},
to the full $3+1$ initial data and evolution equations.
%
%
\subsection{Outer-Boundary Errors}
\label{sect-sample-evolutions/outer-bndry-errors}

As discussed in section~\ref{sect-BCs}, our computational domain only
extends out to a finite outer boundary radius $r = r_\max$.  Conceptually,
there are three distinct ways in which this introduces errors in our
solutions:
\begin{itemize}
\item	Because our (continuum) outer boundary conditions only
	approximate the true $r \geq r_\max$ physics of the
	$3\,{+}\,1$ equations, ``true'' infinite-domain solutions
	of the equations won't exactly satisfy our (continuum)
	boundary conditions at $r = r_\max$.  In a slight abuse
	of language, we refer to this effect as the ``inaccuracy''
	of our (continuum) outer boundary conditions.
\item	Finite differencing errors occur when implementing our
	(continuum) outer boundary conditions.
\item	Floating point roundoff errors occur when implementing the
	finite differenced outer boundary conditions.
\end{itemize}
For the spacetimes considered in this paper, the first of these processes
-- the inaccuracy of our outer boundary conditions -- is always the
dominant error source.

Independent of what causes them, it's useful to (further) subdivide the
outer boundary errors into \defn{static mismatching errors} occurring
at all times, and \defn{dynamic reflection errors} occurring only when
propagating waves in the field variables (\eg{} those accompanying an
outward-moving scalar field shell) reach the outer boundary.

Static mismatching errors are caused by the falloff rates of the
field variables at large $r$ not precisely matching those we assumed
in deriving our our outer boundary conditions
(\cf{}~section~\ref{sect-BCs} and table~\ref{tab-evolution-outer-BC-pars}).
The result of these mismatches is that our outer boundary conditions
slightly perturb the field variables at the outermost grid point.  In
general this perturbation also violates the constraints.  In a time
evolution these errors spread throughout their domain of dependence,
so the net result of static mismatching errors is a wave of (small)
field variable errors and constraint violation, originating at the
outer boundary and propagating inward at the speed of light.  

Figure~\ref{fig-200o10.pqw5-C-rel-outer[t=0(50)200]} illustrates
static mismatching errors.  This shows the magnitude $|C_\rel|$
of the energy constraint in the outer part of the grid for the
200o10.pqw5 evolution, for times $t = 0(50)200$.  The
inward-propagating constraint-violation wave is clearly visible
above the floating-point roundoff error ``noise floor''.  The
constraint-violation wave has amplitude
${\approx}\, 0.3$ in~$|C_\relt|$, or
${\approx}\, 3 \,{\times}\, 10^{-6}$ in~$|C_\relp|$.

As a continuum effect, static mismatching errors are essentially
independent of the finite difference grid resolution,
\footnote{
	 We have verified this (details omitted for brevity)
	 by comparing the errors in the different-resolution
	 o10.pqw5 evolutions.
	 }
{} and can only be made smaller by using more sophisticated outer
boundary conditions or by moving the outer boundary radius
$r = r_\max$ farther out.  The effects of varying $r_\max$ can
be seen by comparing the constraint-violation--wave amplitudes in
(for example) the 200.pqw5, 200o10.pqw5, and 200o30.pqw5 evolutions.
Comparing these (details omitted for brevity), for this data the
constraint violation near the outer boundary scales approximately
as $|C_\relt| \,{\sim}\, (r_\max)^{0.25 \pm 0.05}$
or $|C_\relp| \,{\sim}\, (r_\max)^{-1.75 \pm 0.25}$.  There's a
slight Gibbs-type oscillation at the crest of the inward-propagating
constraint-violation wavefront, but this doesn't grow significantly
with time, and there's no sign of any finite differencing instability.
The constraint violation stays approximately constant in amplitude
in both $|C_\relt|$ and $|C_\relp|$ as it propagates inward.

When outgoing waves in the field variables reach the outer boundary,
then dynamic reflection errors also occur: one can think of these as
being due to the outward-propagating waves not precisely matching our
simple Sommerfeld-type outgoing-radiation
condition~\eqref{eqn-scaled-delta-u-outgoing-radiation}.  This
mismatching causes some of the outgoing waves in both the geometry
and matter fields to effectively reflect off the outer boundary and
propagate back inwards.

Dynamic reflection errors can be seen in 
figure~\ref{fig-200o10+200.pqw5-four-pi-B-rho+mass[t=0(50)500]}
as the difference in~$4 \pi B \rho$ between the 200o10.pqw5 and
200.pqw5 evolutions at times $t \gtsim 350$:  The 200o10.pqw5
evolution's outer boundary (at $r_\max \approx 813$) is far enough
out so the matter hasn't yet reached it at these times, while the
200.pqw5 evolution's outer boundary (at $r_\max \approx 248$) is
relatively close in, so the matter encounters it relatively early
in the evolution (at $t \gtsim 225$), resulting in the spurious
inward-propagating scalar field shell visible at (for example)
\footnote{
	 Here we're again summarizing the shell's profiles by their
	 approximate $\text{mean positions} \pm \text{standard deviations}$
	 in~$4 \pi B \rho$), as in
	 section~\ref{sect-sample-evolutions/scalar-field-phenomenology}.
	 }
{} $r \approx 165 \,{\pm}\, 15$ at $t = 350$,
$r \approx 115 \,{\pm}\, 15$ at $t = 400$, and
$r \approx 65 \,{\pm}\, 15$ at $t = 450$.  Comparing the amplitudes
of the outgoing and reflected scalar field shells for the 200.pqw5
evolution, it's clear that only a very small fraction of the
outgoing scalar field is reflected: the outer boundary's reflection
coefficient (defined as the ratio of the reflected to the incident
peak amplitudes), is approximately~$10^{-6}$ in~$4 \pi B \rho$, or
$10^{-3}$ in the scalar field amplitudes $P$ and $Q$.

Like static mismatching errors, dynamic reflection errors are
essentially a continuum effect, not a finite differencing artifact,
and can only be made smaller by using more sophisticated outer
boundary conditions or by moving the outer boundary radius
$r = r_\max$ farther out.  The effects of increasing $r_\max$ can
again be seen by comparing the outer-boundary reflection coefficient
in (for example) the 200.pqw5, 200o10.pqw5, and 200o30.pqw5 evolutions.
Comparing these (details again omitted for brevity), the outer-boundary
reflection coefficient scales approximately as ${\sim}\, 1/(r_\max)^4$
in~$4 \pi B \rho$, or ${\sim}\, 1/(r_\max)^2$ in~$P$ or~$Q$.

The dynamic reflections in the geometry variables (as measured, say,
by $|C_\rel|$) show quite different behavior: in these variables
the reflection coefficient is approximately unity, \ie{} the $|C_\rel|$
values associated with the ingoing reflected matter are approximately
the same as those associated with the outgoing wave prior to the
reflection.  Fortunately, these values are generally very small,
so we don't consider this to be a serious problem.
%
%
\subsection{Late-Time Decay of the Scalar Field near the Black Hole}
\label{sect-sample-evolutions/late-time-decay-of-SF-near-BH}

After the quasinormal-mode ringing dies out, the scalar field near the
black hole continues to decay.  In the limit of very late times, both
analytical perturbation theory
(\eg{}~\jtcitesprefix\jtcite{
Nollert-Schmidt-1992-Schw-quasinormal-modes,
Gundlach-Price-Pullin-1994-late-time-fields-I,
Gomez-Winicour-Schmidt-1994-sssf-late-time})
and explicit numerical calculations
(\eg{}~\jtcitesprefix\jtcite{
Gundlach-Price-Pullin-1994-late-time-fields-II,Marsa-PhD,
Marsa-Choptuik-1996-sssf}) show that the scalar field $\phi$ at either
any fixed coordinate radius $r$, or at the horizon position $h(t)$,
should decay as ${\sim}\, 1/t^3$.  In our computational scheme $\phi$
itself isn't readily accessible, but it's clear that $P$ and $Q$
should both display this same decay rate, and $\four\! R$ and
$4 \pi B \rho$ should decay at twice this rate, \ie{}~as
${\sim}\, 1/t^6$.

Figure~
\hbox{\ref{fig-400o30+400o10.pqw5-scalar-field.horizon[t=150-5000]}(a)}
%
%
\addtocounter{footnote}{1}
\footnotetext{
\label{footnote-400o10==400o30-in-overlap-region}
	     Within the overlap region $t \leq 500$, the
	     $4 \pi B \rho_h$ and $\four\! R_h$ values
	     are identical between the two evolutions to
	     within ${<}\, 20$~parts per million.
	     }
{} shows the late-time behavior of $\four\! R_h$ and $4 \pi B \rho_h$ 
for the 400o30.pqw5 evolution.  Both diagnostics decay roughly as
${\sim}\, 1/t^{12}$ for $200 \ltsim t \ltsim 500$, then the decay
rate steepens somewhat through $t \approx 700$.  Both diagnostics
then reach reach (nonzero) minima at $t \approx 750$, increase again
to maxima at $t \approx 950$, and thereafter decay at later times,
eventually asymptoting roughly to ${\sim}\, 1/t^{7.5}$.
(As discussed below, the sudden jump visible at $t = 5000$ is due to
dynamic reflections off the outer boundary.)  A closer examination
shows that there is also considerable further structure in these
diagnostics: as shown in figure~
\hbox{\ref{fig-400o30+400o10.pqw5-scalar-field.horizon[t=150-5000]}(b)},
they both vary systematically by factors of~$2$ or so in addition
to the average decay rates just described.  In other words, the
diagnostics' actual decay rates aren't constant.

This behavior clearly disagrees with both the perturbation-theory
predictions, and previous numerical results.  We have not been able
to identify a numerical artifact or error source which would account
for any of the behavior we see:
\begin{itemize}
\item	Reflections from the outer boundary don't arrive back
	at the horizon until $t \gtsim 5000$,
\footnote{
	 That is, since our initial data has compact support,
	 with only negligible scalar field for $r \gg 20$
	 (\eg{}~$4 \pi B \rho < 10^{-50}$ for $r > 75$),
	 we don't expect significant dynamic reflections
	 from the outer boundary until the outgoing matter
	 propagates out to $r_\max$, and then the reflections
	 must propagate back in to the horizon, for a total
	 of $t \gtsim 2 r_\max + O(\ln r) \approx 5500$\,
%
%
\footnotemark
{}	 until reflection effects should be visible at the
	 horizon.  It's thus somewhat peculiar that the
	 reflections are already visible at $t = 5000$ in
 figure~\hbox{\ref{fig-400o30+400o10.pqw5-scalar-field.horizon[t=150-5000]}(a)}.
	 From examination of the detailed field variable
	 profiles, the explanation for this is that as
	 the reflected matter propagates back inwards, a
	 series of low-amplitude precursor waves develops
	 somewhat ahead of the main ingoing matter shell;
%
%
\footnotemark
{}	 the $t = 5000$ point in
 figure~\hbox{\ref{fig-400o30+400o10.pqw5-scalar-field.horizon[t=150-5000]}(a)}
	 is caused by the arrival of the first of these
	 precursor waves at the horizon.
	 }
%
%
\addtocounter{footnote}{-1}
\footnotetext{
	     The $O(\ln r)$ term corrects for the fact
	     that at large $r$, the outgoing speed of light
	     is $c_+ = 1 - O(1/r)$ in our (asymptotically
	     Eddington-Finkelstein) coordinates.
	     }
\addtocounter{footnote}{1}
\footnotetext{
	     The development of these precursor waves does
	     seem to represent a type of instability in our
	     evolution scheme, but we don't consider this to
	     be a significant problem, because the waves are
	     very weak (their peak amplitude in~$4 \pi B \rho$
	     is only ${\approx}\, 10^{-18}$ at $t = 5000$), they
	     seem to grow only for {\em inward\/}-propagating
	     matter shells, and even then their growth is
	     quite slow (their doubling time is roughly
	     $500m$).
	     }
	{} so they shouldn't be relevant here.
\item	Finite differencing errors don't appear to be responsible:
	A convergence test between the 100o30.pqw5, 200o30.pqw5,
	and 400o30.pqw5 evolutions shows that finite differencing
	errors in the 200o30.pqw5 $4 \pi B \rho_h$ and
	$|\four\! R_h|$ values are generally ${\ltsim}\, 0.1\%$
	for $t \leq 2500$,
\footnote{
	 The convergence test does show much larger errors
	 in the narrow time range $700 \ltsim t \ltsim 800$,
	 but even there the peak $\text{200o30.pqw5} - \text{400o30.pqw5}$
	 difference is only ${\approx}\, 1.5\%$.  We haven't
	 investigated the reasons for this narrow time range
	 having an order of magnitude larger finite differencing
	 error than other times, but it's interesting to note
	 that this is the same time range as the minima in
	 both the matter diagnostics themselves.
	 }
{}	rising to (only) ${\approx}\, 0.3\%$ by $t = 5000$, and the
	400o30.pqw5 errors should be another factor of~16 smaller
	still.
\item	Floating-point roundoff errors don't appear to be responsible:
	Comparing the 200o30.pqw5 and 200o30n3.pqw5 evolutions shows
	that adding extra noise to the field variables changes both
	diagnostics by only ${\ltsim}\, 1\%$ for $t \leq 3000$, rising
	to (only) ${\approx}\, 10\%$ by $t = 4500$.  However, as
	noted in section~\ref{sect-sample-evolutions/convergence-tests},
	our noise-addition technique may sometimes systematically
	misestimate the effects of floating-point roundoff errors
(\jtciteprefix\jtcite{Kahan-1996-probabilistic-error-analysis-x,
Kahan-1999-correlation-vs-low-probability-events}),
	so we can't be certain these aren't significant error
	sources.
\end{itemize}

Even though we haven't identified any cause for these results, we don't
think they represent genuine features of the true continuum physics.
Rather, we suspect that they represent some as-yet-undetermined artifact
in our numerical evolution scheme and/or our data analysis techniques.
%
%
\section{Conclusions and Directions for Further Research}
\label{sect-conclusions}

In this paper we present a number of results on the $3+1$ numerical
evolution of dynamic black hole spacetimes, using the ``black hole
exclusion'' technique to avoid singularities.

We first discuss the various design tradeoffs for the inner boundary
placement in a black-hole--exclusion computational scheme:  We argue
that the (an) apparent horizon provides a natural reference point for
placing the inner boundary, and we discuss the causality of such
black-hole--excluded evolutions.  We identify two main options for
the inner boundary location: it may be coincident with the (an)
apparent horizon, or it may be somewhat inside.  We discuss the
advantages and disadvantages of both options, concluding that at
present it's not clear which is preferable.

We then discuss the further complications which arise when our slices
may contain multiple apparent horizons:  In practice, these are often
present only for a limited time interval in an evolution, and in particular
the apparent horizon which a black-hole--exclusion evolution scheme
has been using as a reference for its inner boundary placement, may
at some point disappear.  Assuming the slicing still contains another
apparent horizons or horizons containing the disappearing one, we
discuss how the evolution scheme might transition its inner boundary
reference to the new ``farther-out'' apparent horizon.  We again
outline several possible design options, and conclude that further
research is needed to investigate these and other possible ways of
handling the disappearing--apparent-horizon problem.

We then present a set of general criteria for choosing the
coordinates in a black-hole--exclusion computational scheme.
Like a number of other researchers
(\eg{}~\jtcitesprefix\jtcite{Marsa-PhD,Marsa-Choptuik-1996-sssf}),
we argue that Eddington-Finkelstein coordinates and their
generalizations are ideally suited for use with the
black-hole--exclusion technique.  Unfortunately, although it's
easy to generalize these coordinates for dynamic spherically
symmetric spacetimes, they have no geometrically-preferred
generalization to generic multi--black-hole spacetimes with
no Killing vectors.

We then turn to a detailed presentation of our black-hole--exclusion
computational scheme itself.  To summarize our scheme's key properties:
\begin{itemize}
\item	We require that a black hole be present in all our slices,
	including the initial data slice.  As discussed in detail in
	paper~I, we use the usual York algorithm to construct our
	initial data, solving the full 4-vector form of the York
	equations numerically because our slices have $K$ nonzero
	and spatially variable.
\item	Our formulation of the $3\,{+}\,1$ equations uses a free evolution,
	where the constraints are not used explicitly in the evolution,
	but serve solely as diagnostics of the code's accuracy.  We
	discuss two different ways to normalize the constraints.
\item	We place the inner boundary somewhat inside the apparent horizon
	(typically at ${\approx}\, 75\%$ of the initial slice's horizon
	radius), rather than on the horizon as is sometimes done.  At
	present we maintain the inner boundary at a fixed spatial coordinate
	position throughout the evolution, though we plan to generalize
	this in the future.
\item	We have experimented with a number of different slicing and
	spatial coordinate conditions, but in this paper we discuss
	only a single choice for these, the combination of a generalized
	Eddington-Finkelstein slicing and a constant-area radial
	coordinate.  (Our initial data is such that this radial
	coordinate is in fact an areal one.)
\item	We use finite differencing to discretize the $3\,{+}\,1$
	equations.  However, the details of our finite differencing
	scheme are quite different from those commonly used
	in~$3\,{+}\,1$ computations:
	\begin{itemize}
	\item	For spatial finite differencing, in the grid
		interior we use centered 5~point 4th~order
		molecules, except for upwind--off-centered
		5~point 4th~order molecules for the Lie derivative
		(shift-vector advection) terms in the $3\,{+}\,1$
		evolution equations.  Near the grid boundaries
		we use off-centered \hbox{5(6)}~point 4th~order
		molecules for \hbox{1st(2nd)}~derivatives.
	\item	We have also experimented with an alternate
		finite differencing scheme, in which we use
		5~point 3rd~order molecules for 2nd~spatial
		derivatives near the grid boundaries.  This has
		some performance and implementation-convenience
		advantages, and still gives overall 4th~order
		accuracy for the numerical solutions.
	\item	Contrary to some suggestions elsewhere
(\eg{}~\jtciteprefix\jtcite[section~19.3]{Numerical-Recipes}),
		we do {\em not\/} find it any more difficult to
		construct stable 4th~order differencing schemes
		than 2nd~order ones.  In fact, we find that in
		some ways it's easier, since allowing molecules
		to be larger than 3~points opens up a wider
		variety of possibilities for different
		candidate finite differencing schemes
(\cf{}~\jtciteprefix\jtcite[section~IV]{Marsa-Choptuik-1996-sssf})
		consistent with the necessary domains of dependence.
	\item	Our time integration uses the method of lines,
		with a fully-explicit fixed--time-step 4th~order
		Runge-Kutta time integrator.
	\item	We use a single smoothly-graded nonuniform
		spatial grid, with the grid points at fixed
		(time-independent) coordinate positions.  Our
		nonuniform gridding qualitatively resembles
		using a logarithmic radial coordinates, but
		offers finer control over how the grid
		resolution varies with position; normally we
		adjust the parameters so the grid has relatively
		high resolution in the strong-field region,
		dropping to a moderate (asymptotically constant)
		resolution at large~$r$.
	\item	We don't use any type of staggered grid.
	\item	We don't use any type of ``causal differencing''.
	\end{itemize}
\end{itemize}

Our numerical results show that this evolution scheme is stable
and can run ``forever'': we have evolved for over $t = 4000m$
with no difficulties.

Even at the lowest resolution presented here ($\Delta \wr = 1/100$,
yielding $\Delta r/r \approx 3\%$ at~$r = 3m$ and $2\%$ at~$r = 20m$),
our evolution scheme is still very accurate.  For example, after
evolving the 100o10.pqw5 initial data to $t = 100$ with nontrivial
dynamics, $|C_\relt| \,{\ltsim}\, 10^{-2}$ throughout the grid,
and the more realistically normalized $|C_\relp|$
(\cf{}~section~\ref{sect-diagnostics/energy-constraint}) is
${\ltsim}\, 3 \,{\times}\, 10^{-5}$.  The actual errors in the $g\ij$
and $K\ij$ components themselves, as determined by 3-grid convergence
tests, are also very small: the errors in \hbox{$g\ij$\,($K\ij$)} are
${\ltsim}\, 10^{-5}$\,($3 \,{\times}\, 10^{-7}$) in most of the grid,
rising to
${\approx}\, 3 \,{\times}\, 10^{-5}$\,($3 \,{\times}\, 10^{-5}$)
near the horizon, and
${\approx}\, 10^{-4}$\,($10^{-3}$) at the inner
grid boundary.

As expected from our scheme's 4th~order accuracy, at higher resolutions
all these error measures decrease very rapidly:  Each doubling of the
grid resolution reduces the errors by a factor of~16.  We do see some
floating-point-roundoff-error effects at very high resolutions
($\Delta \wr = 1/800$, yielding $\Delta r/r \approx 0.4\%$ at~$r = 3m$
and $0.25\%$ at~$r = 20m$).  For the resolutions we've used, roundoff
effects are never serious, generally just slight increases and/or
noisiness in~$|C|$ and the 3-grid convergence tests' $K\ij$ error
estimates.
\footnote{
	 In one test evolution we do see a very peculiar
	 effect where artificially adding additional
	 white noise to the field variables several times
	 per time step throughout an evolution, causes
	 $|C|$ to {\em decrease\/} by up to an order
	 of magnitude in some parts of the grid!  We
	 suspect this may be a dithering effect, but
	 we don't understand the details of this.
	 }

We use a simple pseudo-scalar Sommerfeld-type outgoing radiation
outer boundary condition for the $3+1$ evolution equations.  This
works quite well for the outgoing matter, probably because our
nonuniform spatial grid allows us to uwe quite large outer boundary
radia.  For example, even at our innermost outer boundary position,
$r_\max \approx 248$, only about $10^{-3}$ of the outgoing scalar
field amplitude (\ie{} about $10^{-6}$ of the outgoing energy
density) incident on the boundary is reflected back inwards, and this
reflection coefficient falls approximately as ${\sim}\, 1/(r_\max)^4$
($1/(r_\max)^2$) as the outer boundary is moved farther out.

We have made a number of test evolutions to investigate our code's
stability, accuracy, and performance.  These evolutions include both
vacuum (Schwarzschild) spacetimes, and dynamic spacetimes containing
black holes surrounded by scalar field shells.  Our results from these
tests are generally consistent with theoretical predictions and past
numerical studies, but there are a few surprises:

When a relatively thin and massive scalar shell accretes into the
black hole, for a short time 3~distinct apparent horizons are present.
We present general arguments, and show a numerical example demonstrating,
that apparent horizons should generically appear and disappear in
pairs, and at their moment of appearance/disappearance they generically
move with {\em infinite\/} coordinate speed (\ie{} horizontally in
the usual spacetime diagram).  Such highly superluminal motion might
cause difficulties for a black-hole--exclusion evolution scheme which
attempts track the apparent horizon's location and use it as a
reference point for the scheme's inner boundary placement.

Although our measurements of quasinormal-mode ringing in our numerical
evolutions show excellent agreement with both black hole perturbation
theory and previous numerical calculations, our measurements of the
late-time power-law tail of the field near the horizon strongly
disagree with both perturbation theory and previous numerical results.
We think this discrepancy is due to a numerical artifact of some sort
in our computational scheme and/or data analysis techniques, but we
don't have any specific explanation for it at this time.

Our numerical code implementing the computational scheme described
here and in paper~I is available from the author's web site.
\footnote{
	 \hbox{\protect\tt http://www.thp.univie.ac.at/$\boldmath\sim$jthorn}
	 }
{}  The code is written in ISO~C (about 35K~lines, although this
includes considerable code for additional computations not described
either here or in paper~I), and should be easily portable to any modern
computer system.  It may be freely modified and/or redistributed under
the terms of the GNU General Public License.
%
%
\section*{Acknowledgments}

We thank
M.~Choptuik and R.~A.~Matzner for helpful comments on an
   earlier version of our computational scheme, and
S.~A.~Major for many valuable comments on this paper's exposition.
We thank 
R.~A.~Matzner,
   the University of Texas at Austin Center for Relativity,
W.~G.~Unruh, R.~Parachoniak, M.-A.~Potts,
   the University of British Columbia Physics Department,
P.~C.~Aichelburg,
   and the \Universitat{} Wien Institut \fur{} Theoretische Physik
for their hospitality and the use of their research facilities
at various times during the course of this research.
We thank
J.~Wolfgang,
R.~Parachoniak,
P.~Luckham, J.~Thorn,
M.~P\"{u}rrer, and W.~Simon
for assistance with setting up and maintaining computer facilities,
and
G.~Rodgers, J.~Thorn,
the (American) National Science Foundation (grant PHY~8806567),
the (American) Texas Advanced Research Program (grant TARP-085),
and the (Austrian) Fonds zur F\"{o}rderung der wissenschaftlichen Forschung
  (project P12754-PHY) for financial support.
%
%
\appendix
%
%
\section{Evolution Equations for the Spherically Symmetric Scalar Field System}
\label{app-sssf-equations}

We have tabulated the detailed $3\,{+}\,1$ equations for the
constraints and all our diagnostics, written explicitly in terms of
our state variables $A$, $B$, $X$, $Y$, $P$, and $Q$, in appendix~B
of paper~I and section~\ref{sect-diagnostics/energy-constraint} of
this paper.  Here we tabulate the corresponding equations for the
time evolution of the state variables:
\begin{mathletters}
							\label{eqn-evolution}
\begin{eqnarray}
\partial_t A
	& = &	{}
		- 2 \alpha X
		+ \beta \U{\partial_r A}
		+ 2 (\U{\partial_r \beta}) A
									\\
\partial_t B
	& = &	{}
		- 2 \alpha Y
		+ \beta \U{\partial_r B}
									\\
\partial_t X
	& = &	{}
		- \partial_{rr} \alpha
		+ \thalf (\partial_r \alpha) \frac{\partial_r A}{A}
							\nonumber	\\
	&   &	{}
		- \alpha \frac{\partial_{rr} B}{B}
		+ \thalf \alpha {\left( \frac{\partial_r B}{B} \right)}^2
		+ \thalf \alpha \frac{(\partial_r A) (\partial_r B)}{A B}
		+ 2 \alpha \frac{X Y}{B}
		- \alpha \frac{X^2}{A}
							\nonumber	\\
	&   &	{}
		+ \beta \U{\partial_r X}
		+ 2 (\U{\partial_r \beta}) X
							\nonumber	\\
	&   &	{}
		- 2 \alpha P^2
									\\
\partial_t Y
	& = &	{}
		- \thalf (\partial_r \alpha) \frac{\partial_r B}{A}
							\nonumber	\\
	&   &	{}
		- \thalf \alpha \frac{\partial_{rr} B}{A}
		+ \tquarter \alpha \frac{(\partial_r A) (\partial_r B)}{A^2}
		+ \alpha
		+ \alpha \frac{X Y}{A}
							\nonumber	\\
	&   &	{}
		+ \beta \U{\partial_r Y}
									\\
\partial_t P
	& = &	(\partial_r \alpha) Q
		+ \alpha \partial_r Q			\nonumber	\\
	&   &	{}
		+ (\U{\partial_r \beta}) P
		+ \beta \U{\partial_r P}
									\\
\partial_t Q
	& = &	(\partial_r \alpha) \frac{P}{A}
		- \thalf \alpha \frac{\partial_r A}{A} \frac{P}{A}
		+ \alpha \frac{\partial_r B}{B} \frac{P}{A}
		+ \alpha \frac{\partial_r P}{A}		\nonumber	\\
	&   &	{}
		+ \alpha \frac{X}{A} Q
		+ 2 \alpha \frac{Y}{B} Q		\nonumber	\\
	&   &	{}
		+ \U{\beta \partial_r Q}
					\, \text{,}	
\end{eqnarray}
\end{mathletters}
where the partial derivatives in the shift vector Lie derivative terms
are \underline{underlined} for reasons discussed in
section~\ref{sect-spatial-FD}.
%
%
\section{A Brief Introduction to the Method of Lines}
\label{app-MOL-intro}

In this appendix we give a brief introduction to the
\xdefn{method of lines} (\defn{MOL}) for parabolic or hyperbolic PDEs.
This technique has a number of desirable features, but it's not widely
known and the literature on it is quite sparse.

The best general MOL references we're aware of are
\jtcitesprefix\jtcite{Hyman-1976-Courant-MOL-report,Hyman-1979-MOL}.  Other
very good MOL references are
\jtcitesprefix\jtcite{Madsen-Sincovec-1973-LLNL-MOL-report,
Madsen-Sincovec-1974-MOL-in-Oden-etal,Vemuri-Karplus-1981-MOL-section,
Hyman-1989-MOL-in-Buchler}.  Unfortunately, all of these except
\jtciteprefix\jtcite{Vemuri-Karplus-1981-MOL-section} are somewhat
inaccessible, and none go into the level of detail necessary for a
full understanding of the design choices in developing a method of
lines code.  We have discussed MOL at this latter level of detail in
\jtciteprefix\jtcite[sections~7.3.2--7.3.9]{Thornburg-PhD}.  Other useful
MOL references include
\jtciteprefix\jtcite{Liskovets-1965-MOL-review}
	(reviews much of the earlier Soviet work on this topic),
\jtciteprefix\jtcite{Carver-1976-MOL-intro},
\jtciteprefix\jtcite{Schiesser-1991-MOL}
	(very elementary),
and \jtcitesprefix\jtcite{Gustafsson-1971-hyperbolic-BC-FD-convergence,
Gustafsson-1975-hyperbolic-BC-FD-convergence,
Gustafsson-1982-hyperbolic-BC-FD-convergence,
Gary-1975-MOL-BC-4th-order,Gary-1975-4th-order-MOL-in-Vichnevetsky}
	(technical discussions of finite differencing techniques for
	boundary conditions).

The basic idea of MOL is simple:  We initially finite difference
\footnote{
	 The method of lines can also be used
	 with finite element or pseudospectral
	 discretizations, but for simplicity we
	 only discuss finite difference methods
	 here.
	 }
{} only the spatial derivatives in the PDE, keeping the time derivatives
continuous.  This yields a set of coupled ODEs
\footnote{
	 The coupling is due to the spatial finite
	 differencing; the example below (starting
	 with~\eqref{eqn-scalar-wave}) should clarify
	 this.
	 }
{} for the time dependence of the field variables at the spatial grid
points.  (In the terminology of relativity, these ODEs give the time
dependence of the field variables along the spatial-grid-point world
lines.)  A suitable ODE integrator is then used to time-integrate these
ODEs.

A simple example should help to clarify this:  Consider the flat-space
linear scalar advection equation written in 1st~order form,
\begin{equation}
\partial_t u = - c \, \partial_x u
							\label{eqn-scalar-wave}
\end{equation}
on the domain $(t,x) \in [0,\infty) \times [0,1)$, with the (smooth)
coefficient function $c$ being everywhere positive (so that propagation
is solely rightward), subject to the periodic boundary condition
\begin{equation}
u(t,x{=}0) = u(t,x{=}1)
					\, \text{.}	
					\label{eqn-scalar-wave-periodic-BC}
\end{equation}

To treat this problem by MOL, we first discretize the spatial
dimension with the uniform grid $x_k = k \, \Delta x$ for
\hbox{$k = 0$, $1$, $2$, \dots, $K{-}1$}, where $\Delta x \equiv 1/K$
for some integer $K$.  (For the moment we keep the time dimension
continuous.)  Introducing the usual notation $u_k \equiv u(x_k)$
and $c_k \equiv c(x_k)$, and approximating the spatial derivative
in the evolution equation~\eqref{eqn-scalar-wave} with the usual
centered 2nd~order finite differencing molecule, we obtain the
coupled system of ODEs
\begin{equation}
\partial_t u_k
	= - c_k \frac{u_{k+1 \bmod K} - u_{k-1 \bmod K}}
		     {2 \, \Delta x}
				\qquad
				\text{(for $k = 0$, $1$, $2$, \dots, $K{-}1$)}
				      \label{eqn-scalar-wave-periodic-MOL-ODEs}
\end{equation}
for the time dependence of the state variables $\{ u_k \}$.  These
ODEs can then be integrated using any suitable ODE integrator.

Although the time integration of ODE systems is generally most easily,
accurately, and efficiently done using modern adaptive methods,
\footnote{
	 These are discussed in detail
	 \jtcitenumericvsname{in}{by} (\eg{})
\jtcitesprefix\jtcite{Gear-ODE-book,Shampine-Gordon-ODE-book,
Gear-1981-ODE-review,Gupta-SacksDavis-Tischer-1985-ODE-review,
Byrne-Hindmarsh-1987-stiff-ODE-review},
	 and are implemented in numerical codes such as
	 (\eg{})
	 \code{DE/STEP}
	    (\jtciteprefix\jtcite{Shampine-Gordon-ODE-book})
	 \code{ODEPACK/LSODE}
	    (\jtciteprefix\jtcite{Hindmarsh-1983-ODEPACK-codes}),
	 and \code{RKSUITE}
	    (\jtciteprefix\jtcite{Brankin-Gladwell-Shampine-1992-RKSUITE-codes}).
	 However, such methods and codes don't yet
	 handle the varying-size ODE systems which
	 would result from applying MOL to a
	 Berger-and-Oliger--type adaptive gridding scheme
	 (\cf{}~section~\ref{sect-nonuniform-gridding}).
	 More generally, the integration of such
	 adaptive gridding schemes into MOL remains
	 an open problem.
	 }
{} for MOL ODE systems simpler methods are often appropriate.  In
particular, in MOL there's normally little point in making the time
integration much more accurate than the spatial finite differencing.
Assuming a PDE whose solutions vary roughly equally rapidly in space
and time, then for the (common) case where no spatial adaptive gridding
is done and the spatial finite differencing is of fixed order, this
suggests using a correspondingly simple fixed-step fixed-order time
integrator.  \jtciteprefix\jtcite{Hyman-1989-MOL-in-Buchler} and
\jtciteprefix\jtcite[section~7.3.9]{Thornburg-PhD} discuss the choice
of MOL time integrators in more detail.

Despite its superficial differences, MOL is actually closely connected
to the more common ``space and time together'' (\xdefn{SATT}) finite
difference methods for PDEs:  Essentially all ODE integrators use
finite differencing in the time dimension, so the overall result of
applying such an integrator to a system of MOL ODEs, is to (implicitly)
construct and solve a (complicated) system of finite difference equations
in space and time.

For example, if we were to use the leapfrog time integrator
\begin{equation}
\u(t + \Delta t)
	= \u(t - \Delta t)
	  + 2 \, \Delta t \, (\partial_t \u)(t)
	  + O \bigl( (\Delta t)^2 \bigr)
					\label{eqn-leapfrog-time-integrator}
\end{equation}
(where $\u$ denotes the state vector $\{u_k\}$) to time integrate
the MOL ODEs~\eqref{eqn-scalar-wave-periodic-MOL-ODEs}, it's easy
to see that at each time level $t_n \equiv n \, \Delta t$ we would
obtain the finite difference equations
\begin{equation}
\frac{{_{n+1} u _k} - {_{n-1} u _k}}
     {2 \, {\Delta t}}
	= - c_k \frac{{_n u _{k+1 \bmod K}} - {_n u _{k-1 \bmod K}}}
		     {2 \, \Delta x}
			\qquad
			\text{(for $k = 0$, $1$, $2$, \dots, $K{-}1$)}
			       \label{eqn-scalar-wave-periodic-MOL-leapfrog-FD}
\end{equation}
where we use the notation ${_n u _k} \equiv u(t{=}t_n, x{=}x_k)$.
These are precisely the equations for the standard 2nd~order
centered-in-time centered-in-space (\defn{CTCS}) leapfrog SATT
finite differencing scheme, applied to the original
PDE~\eqref{eqn-scalar-wave}.

Corresponding to the equivalence of a MOL scheme to a suitable
SATT scheme, the usual CFL stability limit of an explicit SATT scheme
has an exact analog in MOL:  When time integrating a system of MOL
ODEs with an explicit ODE integrator, the maximum time step will
be limited by stability considerations
(\jtciteprefix\jtcite[section~7.3.9]{Thornburg-PhD}) to precisely the
same value as the equivalent SATT scheme's CFL limit.  For example,
the scalar-wave-equation finite difference
equations~\eqref{eqn-scalar-wave-periodic-MOL-leapfrog-FD} have
the stability limit $|c \, \Delta t| \leq \Delta x$, which may be
viewed either as the usual CFL limit for the CTCS leapfrog SATT
scheme, or the ODE-integration stability limit for the leapfrog time
integrator~\eqref{eqn-leapfrog-time-integrator} applied to the system
of ODEs~\eqref{eqn-scalar-wave-periodic-MOL-ODEs}.

Now consider a variant form of our example problem, where we extend
the spatial domain from $[0,1)$ to $[0,1]$, and introduce the left
boundary condition
\begin{equation}
u(t, x{=}0) = \sin t
					\, \text{.}	
						\label{eqn-scalar-wave-left-BC}
\end{equation}
(There is no right boundary condition:  The system~\eqref{eqn-scalar-wave}
and~\eqref{eqn-scalar-wave-left-BC} is well-posed without one,
since no part of the problem domain causally depends on the right
boundary.)

To treat this variant by the MOL, we first extend the grid to
include $x_K = 1$.  We then time-differentiate the left boundary
condition~\eqref{eqn-scalar-wave-left-BC} and combine it with the
interior evolution equation~\eqref{eqn-scalar-wave} to obtain the
\defn{merged} evolution equation
\begin{equation}
\partial_t u
	= \left\{
	  \begin{tabular}{l@{\qquad}l}
	  $\cos t$			& (if $x = 0$)		\\
	  $- c \, \partial_x u$		& (if $x > 0$)		
	  \end{tabular}
	  \right.
					\, \text{.}	
					   \label{eqn-scalar-wave-with-left-BC}
\end{equation}

Finally, we spatially finite difference this in the same manner as
before for the grid interior, but using the off-centered finite
difference molecule
\begin{equation}
\partial_x =
	\frac{1}{2 \, \Delta x}
	\Bigl[
	\begin{array}{c@{\quad}c@{\quad}c}
	+1	& -4	& \U{+3}	
	\end{array}
	\Bigr]
		+ O \bigl( (\Delta x)^2 \bigr)
\end{equation}
at the right boundary grid point, to obtain the MOL ODEs
\begin{mathletters}
\begin{eqnarray}
\partial_t u_0	& = &	\cos t
									\\
\partial_t u_k	& = &	- c_k \frac{u_{k+1} - u_{k-1}}
				   {2 \, \Delta x}
				\qquad
				\text{(for $k = 1$,~$2$,~$3$,~\dots,~$K{-}1$)}
									\\
\partial_t u_K	& = &	- c_K \frac{u_{K-2} - 4 u_{K-1} + 3 u_K}
				   {2 \, \Delta x}
					\, \text{.}	
\end{eqnarray}
\end{mathletters}
These can again be integrated using any suitable ODE integrator.
(Note that the leapfrog ODE integrator~\eqref{eqn-leapfrog-time-integrator}
would {\em not\/} be suitable here, in fact it would be unconditionally
unstable for this problem.  However, this is ``just'' a problem with
this time integrator; these ODEs can easily be stably integrated by
using a time integration scheme with better stability properties.  For
example, any of the integrators discussed in
section~\ref{sect-time-integration} would suffice here.)
%
%
\section{Convergence Tests and Richardson Extrapolation}
\label{app-convergence-tests+Richardson-extrap}

In this appendix we briefly review our methodology for convergence
testing and also briefly discuss the related technique of Richardson
extrapolation.  We have previously discussed some of this material
in more detail in appendix~E of paper~I.
%
%
\subsection{Convergence Tests}

As has been forcefully emphasized by
\jtcitenameprefix{Choptuik}
\jtcitesprefix\jtcite{Choptuik-PhD,Choptuik-1991-consistency,
Choptuik-Goldwirth-Piran-1992-sssf-cmp-3+1-vs-2+2}
\jtcitenamesuffix{},
a careful comparison of the errors in approximating the same physical
system with the same algorithm, but at different grid resolutions, can
yield a great deal of information about, and very stringent tests of,
a computational scheme's numerical performance and correctness.

In the present context, consider some diagnostic grid function $Z$,
and consider first the case where its true (continuum) value $Z^\ast$
is known a~priori.  (For example, in
section~\ref{sect-sample-evolutions/convergence-tests} we take $Z$ to be
$|C|$ at time $t = 100$ for the o10.pqw5 evolutions, so $Z^\ast = 0$.)
Suppose we have a pair of numerical computations of $Z$, identical except
for having a \hbox{2:1}~ratio of grid spacings $\Delta \wr$.  As discussed
in detail by \jtciteprefix\jtcite{Choptuik-1991-consistency}, if all the
field variables are smooth and the code's numerical errors are dominated
by truncation errors from $n$th~order finite differencing, then the
numerically computed $Z$ values must satisfy the Richardson expansion
\begin{mathletters}
					\label{eqn-Richardson-expansion}
\begin{eqnarray}
Z[\Delta \wr]
	& = &	Z^\ast
		+ (\Delta \wr)^n f
		+ O \bigl( (\Delta \wr)^{n+2} \bigr)
									\\
Z[\Delta \wr/2]
	& = &	Z^\ast
		+ (\Delta \wr/2)^n f
		+ O \bigl( (\Delta \wr)^{n+2} \bigr)
\end{eqnarray}
\end{mathletters}
at each grid point, where $Z[\Delta \wr]$ denotes the numerically
computed $Z$ using the grid spacing $\Delta \wr$, and $f$ is an
$O(1)$ smooth function depending on various high order derivatives
of $Z^\ast$ and the field variables, but {\em not\/} on the grid
resolution.  [We're assuming centered finite differencing here in
writing the higher order terms as $O \bigl( (\Delta \wr)^{n+2} \bigr)$,
otherwise they would only be $O \bigl( (\Delta \wr)^{n+1} \bigr)$.]
Neglecting the higher order terms, i.e.~in the limit of small
$\Delta \wr$, we can eliminate $f$ to obtain the \defn{2-grid}
convergence relationship
\begin{equation}
Z[\Delta \wr/2] - Z^\ast
	= \frac{1}{2^n} \Bigl( Z[\Delta \wr] - Z^\ast \Bigr)
					\, \text{,}	
							\label{eqn-convergence}
\end{equation}
which must hold at each grid point common to the two grids.

If $Z^\ast$ isn't known ahead of time (for example, in
section~\ref{sect-sample-evolutions/convergence-tests} we also take
$Z$ to be an individual coordinate component of $g\ij$ or $K\ij$ at
time $t = 100$ for the o10.pqw5 evolutions), then we consider a triplet
of numerical computations of $Z$, identical except for having a
\hbox{4:2:1}~ratio of grid spacings $\Delta \wr$.  Analogously to
the previous case, we now have the Richardson expansion
\begin{mathletters}
					\label{eqn-Richardson-expansion-3grid}
\begin{eqnarray}
Z[\Delta \wr]
	& = &	Z^\ast
		+ (\Delta \wr)^n f
		+ O \bigl( (\Delta \wr)^{n+2} \bigr)
									\\
Z[\Delta \wr/2]
	& = &	Z^\ast
		+ (\Delta \wr/2)^n f
		+ O \bigl( (\Delta \wr)^{n+2} \bigr)
									\\
Z[\Delta \wr/4]
	& = &	Z^\ast
		+ (\Delta \wr/4)^n f
		+ O \bigl( (\Delta \wr)^{n+2} \bigr)
\end{eqnarray}
\end{mathletters}
at each grid point.  Again neglecting the higher order terms, i.e.~in
the limit of small $\Delta \wr$, we can eliminate both $f$ and $Z^\ast$
to obtain the \defn{3-grid} convergence relationship
\begin{equation}
Z[\Delta \wr/2] - Z[\Delta \wr/4]
	= \frac{1}{2^n} \Bigl( Z[\Delta \wr] - Z[\Delta \wr/2] \Bigr)
					\, \text{,}	
						\label{eqn-convergence-3grid}
\end{equation}
which must hold at each grid point common to all three grids.

To assess how well numerical data satisfy one of the convergence
relationships~\eqref{eqn-convergence} or~\eqref{eqn-convergence-3grid},
we plot the differences
$Z[\Delta \wr] - Z^\ast$ and $Z[\Delta \wr] - Z^\ast$ (2-grid)
or $Z[\Delta \wr/2] - Z[\Delta \wr/4]$
   and $Z[\Delta \wr] - Z[\Delta \wr/2]$ (3-grid)
as a function of position, on a common logarithmic scale.  If
(and only if) the data satisfy the convergence criterion, the two
plots will be identical except for a vertical offset corresponding
to a factor of~$2^n$.

It's important to note that we apply both the 2-grid and 3-grid
convergence tests {\em pointwise\/}, \ie{} independently at each grid
point common to the different-resolution grids.  This makes these
analyses much more sensitive than a simple comparison of gridwise norms,
which would tend to ``wash out'' any convergence problems occurring
at only a small subset of the grid points (\eg{} near a boundary).

Notice that for both the 2-grid and the 3-grid convergence tests, the
parameter $n$, the order of the convergence, is (or at least should be)
known in advance from the form of the finite differencing scheme.  (For
our computational scheme $n = 4$.)  Thus the factor of $2^n$ convergence
ratio in~\eqref{eqn-convergence} or~\eqref{eqn-convergence-3grid} isn't
fitted to the data points, but is rather an a~priori prediction with
{\em no\/} adjustable parameters.  A convergence test of either type
is thus a very strong test of the validity of the overall finite
differencing scheme and the Richardson
expansion~\eqref{eqn-Richardson-expansion}
or~\eqref{eqn-Richardson-expansion-3grid}.
%
%
\subsection{Richardson Extrapolation}

So far we have described how to use the Richardson
error expansions~\eqref{eqn-Richardson-expansion}
and~\eqref{eqn-Richardson-expansion-3grid} to {\em assess\/} the
accuracy of our numerical results.  Alternatively, we can use these
same expansions to (in many circumstances) {\em improve\/} this
accuracy, by the technique of \defn{Richardson extrapolation}:
Suppose we once again have a pair of numerical computations of
$Z$, identical except for having a \hbox{2:1}~ratio of grid spacings
$\Delta \wr$, but now suppose further that we don't know the true
(continuum) value $Z^\ast$.  If we assume that the Richardson error
expansion~\eqref{eqn-Richardson-expansion} does indeed hold, then
we can solve directly for $Z^\ast$:
\begin{mathletters}
						\label{eqn-Richardson-extrap}
\begin{eqnarray}
Z^\ast
	& = &	Z[\Delta \wr]
		- \frac{2^n}{2^n - 1}
		  \Bigl( Z[\Delta \wr] - Z[\Delta \wr/2] \Bigr)
		+ O \bigl( (\Delta \wr)^{n+2} \bigr)
									\\
	& = &	Z[\Delta \wr/2]
		- \frac{1}{2^n - 1}
		  \Bigl( Z[\Delta \wr] - Z[\Delta \wr/2] \Bigr)
		+ O \bigl( (\Delta \wr)^{n+2} \bigr)
\end{eqnarray}
\end{mathletters}

This is a remarkable result: the computed $Z^\ast$ is 2~orders more
accurate than any of the numerical computations used to calculate it!
[If the differencing scheme isn't fully centered, then the computed
$Z^\ast$ is only accurate to $O \bigl( (\Delta \wr)^{n+1} \bigr)$.]

Notice that when doing Richardson extrapolation, we must {\em assume\/}
that the Richardson error expansion~\eqref{eqn-Richardson-expansion}
does indeed hold.  However, in practice, although we can't verify
this directly if $Z^\ast$ is unknown (the only case where Richardson
extrapolation is interesting), we can verify it indirectly by doing a
2-grid convergence test on (say) the energy constraint $C$ for the
same \hbox{2:1}~pair of computations: if this shows good convergence
at the expected order, then it's a fairly safe assumption that other
diagnostics like $Z$ (from the same numerical computation) also show
similar convergence.

However, there is one drawback to Richardson extrapolation:
there's no longer any way to estimate the remaining
$O \bigl( (\Delta \wr)^{n+2} \bigr)$ error
[$O \bigl( (\Delta \wr)^{n+1} \bigr)$ for the non-centered case]
in the Richardson-extrapolated $Z^\ast$ result.
We discuss the use of Richardson extrapolation further in
section~\ref{sect-numerical-methods-cmp/Richardson-extrap}.
%
%
\section{Adding Noise to the Evolution}
\label{app-adding-noise}

Although the effects of floating-point roundoff errors are fairly easy
to model for each individual floating-point arithmetic operation
(\jtciteprefix\jtcite[section~4.2.2]{Knuth-v2},
\jtciteprefix\jtcite{Goldberg-1991-FP-article}),
it's not easy to estimate the net (cumulative) effects of these errors
on an entire computational scheme such as ours.

In the spirit of convergence testing
(\cf{}~appendix~\ref{app-convergence-tests+Richardson-extrap}), the
obvious way to try to estimate the effects of floating-point roundoff
errors would be to repeat our computations with a higher floating-point
precision, and see how much various diagnostics change.  Unfortunately,
this would be somewhat awkward to do, since we already use IEEE double
(64~bit) precision for all floating-point arithmetic, and this is the
highest precision generally available.  Instead, to estimate roundoff
effects we repeat our computations with a simulated lower precision.
The obvious choice of IEEE single (32~bit) precision is much too low
(results with this precision would be quite inaccurate), so we use a
software simulation of a slight lowering of the precision, more precisely
a slight increase in floating-point roundoff errors.

In detail, we simulate increased floating-point rounding errors
as follows:  Each time our time stepping algorithm calls for a
right-hand-side function evaluation (\ie{}~4~times per time step,
\cf{}~section~\ref{sect-time-integration}), we add a tiny amount of
uniformly distributed white noise
\footnote{
	 We generate the white noise via the Unix
	 \hbox{\tt random(3)} random number generator.
	 This is a high-quality generator using a
	 nonlinear additive feedback algorithm
	 (\jtcite[section~3.2.2]{Knuth-v2}); it
	 does {not\/} suffer from the usual
	 approximate--linear-dependence or
	 low-order-bit--nonrandomness problems which
	 afflict most linear-congruential random number
	 generators (\jtcite[section~3.3.4]{Knuth-v2}).
	 }
{} to each component of the time-integration state vector
($A$, $B$, $X$, $Y$, $P$, and $Q$).  We then use the added-noise
state vector as usual in computing the Einstein equations and
integrating them ahead to the next time level.  Note that this
means that the added noise is cumulative: after $N$ time steps we
have added a total of $4N$ white-noise samples (interspersed
throughout the time evolution) to each state-vector component at
each grid point.

Each time we add noise, we scale its amplitude independently
at each grid point to that of the corresponding state-vector
component there:  We scale the noise to be uniform over the range
$\pm 3$~\ulp{} for that component, where a \defn{\ulp} denotes a
``\hbox{{\bf u}nit} in the \hbox{{\bf l}ast} \hbox{{\bf p}lace}''
(\jtcite[section~4.2.2]{Knuth-v2}, \jtcite{Goldberg-1991-FP-article}),
the value of the least significant fraction bit of a binary
floating-point number.  For comparison, floating-point roundoff errors
are often modelled as adding white noise uniform over the range
$\pm \thalf$~\ulp{} for each floating-point arithmetic operation
(\jtciteprefix\jtcite[section~4.2.2]{Knuth-v2},
\jtciteprefix\jtcite{Goldberg-1991-FP-article}).
[However, it's important to realise that such a model has
serious flaws: actual floating-point roundoff errors are often
highly discrete and/or highly correlated
(\jtciteprefix\jtcite{Kahan-1996-probabilistic-error-analysis-x}),
so naive statistical analyses based on modelling them as uncorrelated
continuous random variables can be highly misleading (see also
\jtciteprefix\jtcite{Kahan-1999-correlation-vs-low-probability-events}).]
%

%
\clearpage
%
%
\begin{figure}
%
%
\caption[Eddington-Finkelstein Slicing of Schwarzschild Spacetime]
	{
	This figure shows the light cones and the Eddington-Finkelstein
	slicing of Schwarzschild spacetime.  Part~(a) is plotted
	in Kruskal-Szekeres coordinates $(u,v)$, and shows the
	$r = 1.5(0.5)3.5$ surfaces
\footnotemark[0\ref{footnote-x(delta)y}]
{}	and the Eddington-Finkelstein $t = -8(1)4$ slices, with
	the $t = -4(1)4$ slices labelled.  Part~(b) is plotted in
	Eddington-Finkelstein coordinates $(r,t)$, and shows the
	Eddington-Finkelstein $t = -1(0.5)3$ slices and the
	$r = 1.5$ and $r = 2$ (horizon) surfaces.  The legend
	is common to both parts of the figure.
	}
\label{fig-Schw/EF-slicing}
%
%
\end{figure}
%
\begin{figure}
%
%
\caption[$\wr$ Nonuniform Gridding Coordinate]
	{
	This figure shows the behavior of our nonuniform gridding
	coordinate $\wr$.  Part~(a) shows the actual $\wr$ coordinate,
	while part~(b) shows the grid spacing $\Delta r$ for grids
	with resolutions of (from top to bottom)
	$\Delta \wr = 0.01$, $0.005$, $0.0025$, and $0.00125$
	(we refer to these as ``100'', ``200'', ``400'', and ``800''
	evolutions respectively).  The diagonal dashed lines labeled
	along the top and right of part~(b) show lines of constant
	relative grid spacing $\Delta r/r$.  
	In both parts of the figure the vertical dashed lines
	show the outer grid boundaries for $\wr_\max = 4$, $10$,
	and~$30$.
	}
\label{fig-mixed-210-coord}
%
%
\end{figure}
%
\begin{figure}
%
%
\caption[Scalar Field and Mass for 200.pqw5 Evolution at Times -- Early Times]
	{
	This figure shows the scalar field's radial 3-energy
	and 3-momentum densities $4 \pi B \rho$ and $4 \pi B j^r$,
	the normalized 4-Ricci scalar
	$\thalf B \, \four\! R \equiv 4 \pi B (\rho - T)$, and
	the mass function $m$, for the 200.pqw5 evolution at times
	$t = 0(5)50$.  The vertical dashed line near the left side
	of each subplot shows the apparent horizon position~$h$.
	Notice that the left vertical scale changes by a factor
	of~3 between the $t = 0$\,--\,$25$ and $t = 25$\,--\,$50$
	subfigures; for comparison the $t = 25$ subfigure is
	shown twice, once at each scale.
	}
\label{fig-200.pqw5-h+scalar-field+mass[t=0(5)50]}
%
%
\end{figure}
%
\begin{figure}
%
%
\caption[Scalar Field and Mass for 200o10.pqw5 and 200.pqw5 Evolutions
	 -- Later Times]
	{
	This figure shows the scalar field's radial density
	profile $4 \pi B \rho$ (on a logarithmic scale) and the
	$m$ function (on a linear scale) for the 200o10.pqw5 and
	200.pqw5 evolutions at times $t = 0(50)500$.  Notice the change
	in scales for $r$ and $4 \pi B \rho$ (but not $m$) compared
	with figure~\ref{fig-200.pqw5-h+scalar-field+mass[t=0(5)50]}:
	the box in the $t = 0$ and $t = 50$ subplots shows the
	area plotted in that figure.  The 200.pqw5 $4 \pi B \rho$
	curves in the $t \leq 300$ subfigures, and the 200.pqw5 $m$
	curves for all the subfigures, are at this scale indistinguishable
	from, and hence hidden under, the 200o10.pqw5 curves.  The
	200.pqw5 $4 \pi B \rho$ curves in the $t \geq 350$ subplots
	show significant outer boundary reflections; we discuss these
	in section~\ref{sect-sample-evolutions/outer-bndry-errors}.
	}
\label{fig-200o10+200.pqw5-four-pi-B-rho+mass[t=0(50)500]}
%
%
\end{figure}
%
\begin{figure}
%
%
\caption[Early-Time Behavior of the Scalar Field at the Horizon]
	{
	This figure shows the quasinormal-mode oscillations
	of the 4-Ricci scalar at the horizon, $\four\! R_h$
	(upper curve),
	and the energy density at the horizon, $\rho_h$ (lower curve),
	for the 400o10.pqw5 evolution at times $0 \leq t \leq 200$.  
	(For visual clarity the plot actually shows $\rho_h/100$
	instead of $\rho_h$.)
	Note that while $\four\! R_h$ changes sign (shown by
	the different plot symbols) in the oscillations,
	$\rho_h$ is by definition always positive semi-definite,
	and is in fact strictly positive at all times in this
	evolution.
	}
\label{fig-400o10.pqw5-scalar-field.horizon[t=0-200]}
%
%
\end{figure}
%
\begin{figure}
%
%
\caption[Apparent Horizon Position and Mass for 200.pqw5 Model]
	{
	This figure shows the apparent horizon's position $h$
	and contained mass $m(h)$ for the 200.pqw5 evolution
	at times $0 \le t \le 50$.  (At later times the black hole
	doesn't grow significantly.)
	}
\label{fig-200.pqw5-h+m(h)[t=0-50]}
%
%
\end{figure}
%
\begin{figure}
%
%
\caption[Apparent Horizon Position and Mass for 400.pqw1 Model]
	{
	This figure shows the apparent horizon's position $h$
	and contained mass $m(h)$ for the 400.pqw1 evolution,
	(a)~for the time range $0 \le t \le 50$, and
	(b)~at an expanded scale for the time range
	$19 \le t \le 20$.  (The uneven density of points is
	to ensure good resolution of the near-vertical parts
	of the curve in part~(b).)
	}
\label{fig-400.pqw1-h+m(h)[t=0-50+19-20]}
%
%
\end{figure}
%
\begin{figure}
%
%
\caption[Horizon Function $H$ for 400.pqw1 Model for Times $t = 19(0.1)20$]
	{
	This figure shows the left hand side of the
	apparent horizon equation, the horizon function $H$
	defined by~\eqref{eqn-horizon-generic},
	for the 400.pqw1 evolution at times $t = 19(0.1)20$.
	As discussed in the text
(section~\ref{sect-sample-evolutions/BH-growth+apparent-horizon-motion}),
	apparent horizons correspond to zeros of $H$.
	}
\label{fig-400.pqw1-H[t=19(0.1)20]}
%
%
\end{figure}
%
\begin{figure}
%
%
\caption[$|C_\rel|$ Convergence for o10.pqw5 Evolutions]
	{
	This figure shows the energy constraint for the
	o10.pqw5 evolutions, at time $t = 100$.  Part~(a)
	shows the energy constraint normalized as $C_\relt$,
	while part~(b) shows it normalized as $C_\relp$;
	the vertical scale is the same in both parts.
	For reference, both subfigures also show the scalar field's
	scaled radial density $4 \pi B \rho$ (which at this scale
	doesn't differ appreciably between the different-resolution
	evolutions), at the same time.  The vertical dashed
	line near the left side of each subfigure shows the
	apparent horizon position~$h$.  The scale bars show a
	factor of~$16$ in $|C_\rel|$, for comparison with the
	vertical spacings between the energy constraint plots
	at the different resolutions.
	}
\label{fig-800o10+lower-res.pqw5-C-rel-conv[t=100]}
%
%
\end{figure}
%
\begin{figure}
%
%
\caption[$\magnitude{\Delta g\ij}$ and $\magnitude{\Delta K\ij}$
	 Convergence for o10.pqw5 Evolutions]
	{
	This figure shows the tensor magnitudes of the differences
	in the 3-metric and extrinsic curvature,
	$\magnitude{\Delta g\ij}$ and $\magnitude{\Delta K\ij}$
	(parts~(a) and~(b) respectively), between pairs of o10.pqw5
	evolutions which are identical except for a factor of~2
	difference in grid resolution, all at time $t = 100$.
	For reference, it also shows the scalar field's scaled
	radial density $4 \pi B \rho$ (which at this scale doesn't differ
	appreciably between the different-resolution evolutions), at
	the same time.  The vertical dashed line near the left side
	of each subfigure shows the apparent horizon position~$h$.
	The scale bars show a factor of~$16$ in $\magnitude{\Delta g\ij}$
	and $\magnitude{\Delta K\ij}$, for comparison with the
	vertical spacings between the plots at the different
	resolution pairs.
	}
\label{fig-400o10+lower-res.pqw5-Diff-gK-magnitude-conv[t=100]}
%
%
\end{figure}
%
\begin{figure}
%
%
\caption[Difference between 3rd and 4th Order Finite Difference Molecules
	 Near the Inner Boundary]
	{
	This figure shows the normalized energy constraint $|C_\relp|$
	for the innermost part of the grid in the 200.pqw5 and 200e4.pqw5
	evolutions, at times $t = 0$, $0.01$, $0.1$, $1$, $10$, and $100$.
	The vertical dashed line near the right side of each subplot
	except the $t = 100$ one shows the apparent horizon position~$h$;
	for $t = 100$ the horizon has expanded to include
	the entire subplot.
	}
\label{fig-200e4+200.pqw5-C-relp.inner[t=0+various+100]}
%
%
\end{figure}
%
\begin{figure}
%
%
\caption[Static Mismatching Errors at the Outer Boundary]
	{
	This figure shows the effects of static mismatching errors
	in the outer boundary conditions.  It shows the energy
	constraint in the outer part of the grid at times
	$t = 0(50)200$, for the 200o10.pqw5 evolution.
	Part~(a) shows the energy constraint normalized as $C_\relt$,
	while part~(b) shows it normalized as $C_\relp$; the
	legend and the axis scales are common to both parts.
	}
\label{fig-200o10.pqw5-C-rel-outer[t=0(50)200]}
%
%
\end{figure}
%
\begin{figure}
%
%
\caption[Late-Time Decay of the Scalar Field at the Horizon]
	{
	This figure shows the decay of the scalar field at the
	apparent horizon in the 400o30.pqw5 and 400o10.pqw5 evolutions,
	for times $150 \leq t \leq 5000$.  (At earlier times the
	field is dominated by quasinormal-mode ringing,
	\cf{}~section~\ref{sect-sample-evolutions/QNM-ringing} and
	figure~\ref{fig-400o10.pqw5-scalar-field.horizon[t=0-200]}.)
	Part~(a) shows the energy density at the horizon, $\rho_h$,
	and the magnitude of the 4-Ricci scalar at the horizon,
	$|\four\! R_h|$, along with ${\sim}\, 1/t^{12}$ and
	${\sim}\, 1/t^{7}$ falloff lines.
	Part~(b) shows the ratios of the same data points
	to the falloff lines plotted in part~(a), together
	with similarly-ratioed falloffs ${\sim}\, t^{\pm 1/2}$
	and ${\sim}\, t^{\pm 1}$ faster/slower to give an
	indication of the uncertainties in the nominal
	falloff rates.
	Part~(a) shows data from the 400o30.pqw5 evolution;
	part~(b) shows both this same data, and also the (much
	more densely sampled) $t \leq 500$ data from the 400o10.pqw5
	evolution.
\footnotemark[0\ref{footnote-400o10==400o30-in-overlap-region}]
	}
\label{fig-400o30+400o10.pqw5-scalar-field.horizon[t=150-5000]}
%
%
\end{figure}
%
\clearpage
%
\begin{table}
\begin{center}
$$
\begin{array}{c@{\qquad}c@{\qquad}c}
\text{Field Variable}	& m	& n	\\
\hline 
A			& 0	& 2	\\
B			& 2	& 1	\\
X			& 0	& 3	\\
Y			& 0	& 2	\\
P			& 0	& 1	\\
Q			& 0	& 1	
\end{array}
$$
\end{center}
%
%
\caption[Evolution Outer Boundary Condition Parameters]
	{
	This table gives the parameters $m$ and $n$
	used for the evolution outer boundary
	condition~\eqref{eqn-outgoing-radiation-outer-BC}.
	}
\label{tab-evolution-outer-BC-pars}
%
%
\end{table}
%
\begin{table}
\begin{center}
\def\X{\hbox to 4.0\arraycolsep{}}
$$
\begin{array}{lcrccrccrcccrc}
		& \multicolumn{9}{c}{\partial_\wr}
		&
		& \multicolumn{3}{c}{\partial_{\wr\wr}}
									\\
\cline{2-10}\cline{12-14}
		& \multicolumn{3}{c}{\text{most terms}}
		& \multicolumn{6}{c}{\text{Lie derivative terms}}
		&
		& \multicolumn{3}{c}{\text{All terms}}
									\\
\cline{5-10}
\iwr		&  &    &
		& \multicolumn{3}{c}{\beta \ge 0}
		& \multicolumn{3}{c}{\beta < 0}
		&                                   &  &    &
									\\
\hline 
%
\iwr_\min	&  & +2 &  &  & +2 &  &  & +2 &  &  &  & +2 & 		\\
\iwr_\min + 1	&  & +1 &  &  & +1 &  &  & +1 &  &  &  & +1 & 		\\
\iwr_\min + 2	&  &  0 &  &  & +1 &  &  &  0 &  &  &  &  0 & 		\\
\text{interior}	&\X&  0 &\X&\X& +1 &\X&\X& -1 &\X&  &\X&  0 &\X		\\
\iwr_\max - 2	&  &  0 &  &  &  0 &  &  & -1 &  &  &  &  0 & 		\\
\iwr_\max - 1	&  & -1 &  &  & -1 &  &  & -1 &  &  &  & -1 & 		\\
\iwr_\max	&  & -2 &  &  & -2 &  &  & -2 &  &  &  & -2 & 		
\end{array}
$$
\end{center}
%
%
\caption[Finite Difference Molecule Offsets]
	{
	This table gives the offsets of the finite difference
	molecules we use in various parts of the grid.  For the
	shift vector Lie derivative terms in the evolution equations
	(these terms are shown \underline{underlined} in the evolution
	equations~\eqref{eqn-evolution-generic} and~\eqref{eqn-evolution}),
	the molecule offsets also depend on the sign of the
	shift vector.  The molecules themselves are given in
	table~\ref{tab-FD-molecule-coeffs}.
	}
\label{tab-FD-molecule-offsets}
%
%
\end{table}
%
\begin{table}
%
%
%
%
\mathsurround=0em
\def\arraystretch{1.5}
\begin{flushleft}
\vspace*{-3ex}
%
Part~(a):\\
\centerline{
$
\begin{array}{crcccrccrccrccrccrccrccrccrccrc}
 &    &	&
	& \multicolumn{27}{c}{
		\displaystyle
		\partial_\wr
		=
		\frac{1}{12 \, \Delta\wr} \, \Bigl[ \quad \dots \quad \Bigr]
		+ O \bigl( (\Delta\wr)^4 \bigr)
			     }
								      \\[1.5ex]
\cline{5-31} 
\multicolumn{3}{c}{\text{Offset}}
	&
	& \multicolumn{3}{c}{i{-}4}
	& \multicolumn{3}{c}{i{-}3}
	& \multicolumn{3}{c}{i{-}2}
	& \multicolumn{3}{c}{i{-}1}
	& \multicolumn{3}{c}{\underline{\,i\,}}
	& \multicolumn{3}{c}{i{+}1}
	& \multicolumn{3}{c}{i{+}2}
	& \multicolumn{3}{c}{i{+}3}
	& \multicolumn{3}{c}{i{+}4}
									\\
\hline 
 & -2 &	&
	& & \llap{$\Bigl[\,$}{+3} &
			& & -16 &	& & +36 &	& & -48 &
	& & \underline{+25}\rlap{$\,\Bigr]$} &
	& &    &	& &    &	& &    &	& &    &
									\\
 & -1 &	&
	& &    &	& & \llap{$\Bigl[\,$}{-1} &
					& & +6 &	& & -18 &
	& & \underline{+10} &
	& & +3\rlap{$\,\Bigr]$} &
			& &    &	& &    &	& &    &
									\\
 &  0 &	&
	& &    &	& &    &	& & \llap{$\Bigl[\,$}{+1} &
							& & -8 &
	& & \underline{\,\,0} &
	& & +8 &	& & -1\rlap{$\,\Bigr]$} &
					& &    &	& &    &
									\\
 & +1 &	&
	& &    &	& &    &	& &    &	& &
							\llap{$\Bigl[\,$}{-3} &
	& & \underline{-10} &
	& & +18 &	& & -6 &	& & +1\rlap{$\,\Bigr]$} &
							& &    &
									\\
 & +2 &	&
	& &    &	& &    &	& &    &	& &    &
	& & \llap{$\Bigl[\,$}\underline{-25} &
	& & +48 &	& & -36 &	& & +16	&	& &
							-3\rlap{$\,\Bigr]$} &
%
\end{array}
$
}
%
\vspace{2.5ex}
%
Part~(b):\\
\centerline{
$
\begin{array}{crcccrccrccrccrccrccrccrccrccrc}
 &    &	&
	& \multicolumn{27}{c}{
		\displaystyle
		\partial_{\wr\wr}
		=
		\frac{1}{12 \, (\Delta\wr)^2}
			\, \Bigl[ \quad \dots \quad \Bigr]
		+ O \bigl( (\Delta\wr)^4 \text{~or~} (\Delta\wr)^3 \bigr)
			     }
								      \\[1.5ex]
\cline{5-31} 
\multicolumn{3}{c}{\text{Offset}}
	&
	& \multicolumn{3}{c}{i{-}4}
	& \multicolumn{3}{c}{i{-}3}
	& \multicolumn{3}{c}{i{-}2}
	& \multicolumn{3}{c}{i{-}1}
	& \multicolumn{3}{c}{\underline{\,i\,}}
	& \multicolumn{3}{c}{i{+}1}
	& \multicolumn{3}{c}{i{+}2}
	& \multicolumn{3}{c}{i{+}3}
	& \multicolumn{3}{c}{i{+}4}
									\\
\hline 
 & -2 &	&
	& & \llap{$\Bigl[\,$}{+11} &
			& &  -56 &	& & +114 &	& & -104 &
	& & \underline{+35}\rlap{$\,\Bigr]$} &
	& &      &	& &      &	& &      &	& &      &
									\\
 & -1 &	&
	& &      &	& & \llap{$\Bigl[\,$}{-1} &
					& &   +4 &	& &   +6 &
	& & \underline{-20} &
	& & +11\rlap{$\,\Bigr]$} &
			& &      &	& &      &	& &      &
									\\
 &  0 &	&
	& &      &	& &      &	& & \llap{$\Bigl[\,$}{-1} &
							& &  +16 &
	& & \underline{-30} &
	& & +16  &	& & -1\rlap{$\,\Bigr]$} &
					& &      &	& &      &
									\\
 & +1 &	&
	& &      &	& &      &	& &      &	& &
							\llap{$\Bigl[\,$}{+11} &
	& & \underline{-20} &
	& &   +6 &	& &   +4 &	& & -1\rlap{$\,\Bigr]$} &
							& &      &
									\\
 & +2 &	&
	& &      &	& &      &	& &      &	& &      &
	& & \llap{$\Bigl[\,$}\underline{+35} &
	& & -104 &	& & +114 &	& &  -56 &	& &
							+11\rlap{$\,\Bigr]$} &
%
\end{array}
$
}
%
\vspace{2.5ex}
%
Part~(c):\\
\centerline{
$
\begin{array}{crcccrccrccrccrccrccrccrccrccrccrccrc}
 &    &	&
	& \multicolumn{33}{c}{
		\displaystyle
		\partial_{\wr\wr}
		=
		\frac{1}{12 \, (\Delta\wr)^2}
			\, \Bigl[ \quad \dots \quad \Bigr]
		+ O \bigl( (\Delta\wr)^4 \bigr)
			     }
								      \\[1.5ex]
\cline{5-37} 
\multicolumn{3}{c}{\text{Offset}}
	&
	& \multicolumn{3}{c}{i{-}5}
	& \multicolumn{3}{c}{i{-}4}
	& \multicolumn{3}{c}{i{-}3}
	& \multicolumn{3}{c}{i{-}2}
	& \multicolumn{3}{c}{i{-}1}
	& \multicolumn{3}{c}{\underline{\,i\,}}
	& \multicolumn{3}{c}{i{+}1}
	& \multicolumn{3}{c}{i{+}2}
	& \multicolumn{3}{c}{i{+}3}
	& \multicolumn{3}{c}{i{+}4}
	& \multicolumn{3}{c}{i{+}5}
									\\
\hline 
 & -2 &	&
	& & \llap{$\Bigl[\,$}{-10} &
	& & +61	&
	& & -156 &
	& & +214 &
	& & -154 &
	& & \underline{+45}\rlap{$\,\Bigr]$} &
	& &    &
	& &    &
	& &    &
	& &    &
	& &    &
									\\
 & -1 &	&
	& &    &
	& & \llap{$\Bigl[\,$}{+1} &
	& & -6 &
	& & +14 &
	& & -4 &
	& & \underline{-15} &
	& & +10\rlap{$\,\Bigr]$} &
	& &    &
	& &    &
	& &    &
	& &    &
									\\
 &  0 &	&
	& &    &
	& &    &
	& &    &
	& & \llap{$\Bigl[\,$}{-1} &
	& & +16 &
	& & \underline{-30} &
	& & +16 &
	& & -1\rlap{$\,\Bigr]$} &
	& &    &
	& &    &
	& &    &
									\\
 & +1 &	&
	& &    &
	& &    &
	& &    &
	& &    &
	& & \llap{$\Bigl[\,$}{+10} &
	& & \underline{-15} &
	& & -4 &
	& & +14 &
	& & -6 &
	& & +1\rlap{$\,\Bigr]$} &
	& &    &
									\\
 & +2 &	&
	& &    &
	& &    &
	& &    &
	& &    &
	& &    &
	& & \llap{$\Bigl[\,$}\underline{+45} &
	& & -154 &
	& & +214 &
	& & -156 &
	& & +61 &
	& & -10\rlap{$\,\Bigr]$} &
%
\end{array}
$
}
%
\end{flushleft}
%
%
\caption[Finite Difference Molecule Coefficients]
	{
	This table gives the coefficients of all of our finite
	difference molecules as a function of their offsets,
	(a)~for $\partial_\wr$, (b)~and (c)~for $\partial_{\wr\wr}$.
	(As discussed in the text, parts~(b) and~(c) correspond to
	using the extrapolants of parts~(a) and~(b) of
	table~\ref{tab-FD-extrapolation-coeffs}, respectively.)
	The $i{\pm}k$ column headings denote grid points for
	molecules applied at grid point~$i$.  (The application
	point is also denoted by underlining.)  In part~(b), the
	centered molecule has an error of $O \bigl( (\Delta\wr)^4 \bigr)$,
	while the off-centered molecules have errors of
	$O \bigl( (\Delta\wr)^3 \bigr)$.  In part~(c), notice that
	the off-centered 2nd~derivative molecules must be 1~point
	larger than the centered one (6~points instad of 5)
	in order to still have 4th~order local truncation error.
	}
\label{tab-FD-molecule-coeffs}
%
%
\end{table}
%
\begin{table}
%
%
\mathsurround=0pt
\setbox0=\hbox{$\;$}\arraycolsep=0.5\wd0	
\def\arraystretch{1.5}
%
%
\begin{flushleft}
Part~(a): 5~point 4th~order (quartic) extrapolants:\\
$$
\begin{array}{cccr@{}ccr@{}ccr@{}ccr@{}ccr@{}ccc}
f_{i{\mp}1} & =
	& + &   5&f_i
	& - &  10&f_{i{\pm}1}
	& + &  10&f_{i{\pm}2}
	& - &   5&f_{i{\pm}3}
	& + &    &f_{i{\pm}4}
	& + & \displaystyle O \bigl( (\Delta \wr)^5 \bigr)
							\\
f_{i{\mp}2} & =
	& + &  15&f_i
	& - &  40&f_{i{\pm}1}
	& + &  45&f_{i{\pm}2}
	& - &  24&f_{i{\pm}3}
	& + &   5&f_{i{\pm}4}
	& + & \displaystyle O \bigl( (\Delta \wr)^5 \bigr)
							\\
f_{i{\mp}3} & =
	& + &  35&f_i
	& - & 105&f_{i{\pm}1}
	& + & 126&f_{i{\pm}2}
	& - &  70&f_{i{\pm}3}
	& + &  15&f_{i{\pm}4}
	& + & \displaystyle O \bigl( (\Delta \wr)^5 \bigr)
\end{array}
$$
\end{flushleft}
%
%
\begin{flushleft}
Part~(b): 6~point 5th~order (quintic) extrapolants:\\
$$
\begin{array}{cccr@{}ccr@{}ccr@{}ccr@{}ccr@{}ccr@{}ccc}
f_{i{\mp}1} & =
	& + &   6&f_i
	& - &  15&f_{i{\pm}1}
	& + &  20&f_{i{\pm}2}
	& - &  15&f_{i{\pm}3}
	& + &   6&f_{i{\pm}4}
	& - &    &f_{i{\pm}5}
	& + & \displaystyle O \bigl( (\Delta \wr)^6 \bigr)
							\\
f_{i{\mp}2} & =
	& + &  21&f_i
	& - &  70&f_{i{\pm}1}
	& + & 105&f_{i{\pm}2}
	& - &  84&f_{i{\pm}3}
	& + &  35&f_{i{\pm}4}
	& - &   6&f_{i{\pm}5}
	& + & \displaystyle O \bigl( (\Delta \wr)^6 \bigr)
\end{array}
$$
\end{flushleft}
%
%
\caption[Finite Difference Extrapolation Coefficients]
	{
	This table gives the coefficients of our Lagrange (polynomial)
	extrapolants for use in finite differencing.  As discussed in
	the text, when computing $\partial_\wr$ we always use the
	extrapolants in part~(a); when computing $\partial_{\wr\wr}$
	we may use the extrapolants of either part~(a) or~(b),
	corresponding to using the molecules of part~(b) or~(c)
	of table~\ref{tab-FD-molecule-coeffs}, respectively.
	}
\label{tab-FD-extrapolation-coeffs}
%
%
\end{table}
%
\begin{table}
\begin{center}
%
%
\squeezetable
\mathsurround=0em
\begin{tabular}{lclclcclcclcccrcrrcc@{\,}c@{\,}c@{\,}c@{\,}c@{}c@{}c@{}c@{}c@{}c@{\quad}c@{}c@{}c@{}c@{}c@{}c@{\quad}cccccc}
%
\nop{Name}
& \nop{Notes}
	& \nop{Delta wr}	& \multicolumn{6}{c}{$\Delta r/r$}
				& \multicolumn{3}{c}{$\Delta r$ at}
			& \nop{\}}
	& \multicolumn{3}{c}{\nop{wr_max}}
				& \nop{r_max}
				& \nop{t_max}
			& \nop{\}}
			& \multicolumn{10}{c}{Input Perturbation}
				& \multicolumn{6}{c}{Approximate $4 \pi B \rho$ Profile}
					& \multicolumn{5}{c}{Initial Mass}
						& Final
									\\
%
%
\cline{4-9}
\cline{36-40}
%
%
Name
& Notes
	& $\Delta \wr$		& \multicolumn{3}{c}{at $r{=}3$}
				& \multicolumn{3}{c}{at $r{=}20$}
				& \multicolumn{3}{c}{large $r$}
			& \nop{\}}
	& \multicolumn{3}{c}{$\wr_\max$}
				& $r_\max$
				& $t_\max$
			& \nop{\}}
			& \multicolumn{10}{c}{for Initial Data Solver}
				& \multicolumn{6}{c}{of $t = 0$ Scalar Field Shell}
					& BH
					& $+$	& SF
					& $=$	& Total	& BH Mass
									\\
%
%
\hline 
%
%
\begin{tabular}{@{}l@{}}
100.pqw5	\\
200.pqw5	\\
400.pqw5	\\
100e4.pqw5	\\
200e4.pqw5	\\
400e4.pqw5	\\
100o10.pqw5	\\
200o10.pqw5	\\
400o10.pqw5	\\
800o10.pqw5	\\
800o10n3.pqw5	\\
100o30.pqw5	\\
200o30.pqw5	\\
200o30n3.pqw5	\\
400o30.pqw5	
\end{tabular}
& \begin{tabular}{@{}l@{}}
  \nop{100.pqw5}		\\
  \nop{200.pqw5}		\\
  \nop{400.pqw5}		\\
  \nop{100e4.pqw5}(a)		\\
  \nop{100e4.pqw5}(a)		\\
  \nop{100e4.pqw5}(a)		\\
  \nop{100o10.pqw5}		\\
  \nop{200o10.pqw5}		\\
  \nop{400o10.pqw5}		\\
  \nop{800o10.pqw5}		\\
  \nop{800o10n3.pqw5}(b)	\\
  \nop{100o30.pqw5}		\\
  \nop{200o30.pqw5}		\\
  \nop{200o30n3.pqw5}(b)	\\
  \nop{400o30.pqw5}		
  \end{tabular}
	& \begin{tabular}{@{}l@{}}
	  0.01		\\
	  0.005		\\
	  0.0025	\\
	  0.01		\\
	  0.005		\\
	  0.0025	\\
	  0.01		\\
	  0.005		\\
	  0.0025	\\
	  0.00125	\\
	  0.00125	\\
	  0.01		\\
	  0.005		\\
	  0.005		\\
	  0.0025	
	  \end{tabular}		&
				& \begin{tabular}{@{}l@{}}
				  0.030		\\
				  0.015		\\
				  0.0076	\\
				  0.030		\\
				  0.015		\\
				  0.0076	\\
				  0.030		\\
				  0.015		\\
				  0.0076	\\
				  0.0038	\\
				  0.0038	\\
				  0.030		\\
				  0.015		\\
				  0.015		\\
				  0.0076	
				  \end{tabular}
				&
					&
					& \begin{tabular}{@{}l@{}}
					  0.020		\\
					  0.010		\\
					  0.0050	\\
					  0.020		\\
					  0.010		\\
					  0.0050	\\
					  0.020		\\
					  0.010		\\
					  0.0050	\\
					  0.0025	\\
					  0.0025	\\
					  0.020		\\
					  0.010		\\
					  0.010		\\
					  0.0050	
					  \end{tabular}
					&
						&
						& \begin{tabular}{@{}l@{}}
						  0.96		\\
						  0.48		\\
						  0.24		\\
						  0.96		\\
						  0.48		\\
						  0.24		\\
						  0.96		\\
						  0.48		\\
						  0.24		\\
						  0.12		\\
						  0.12		\\
						  0.96		\\
						  0.48		\\
						  0.48		\\
						  0.24		
						  \end{tabular}
						&
			& \begin{tabular}{@{}c@{}}
			  \raise3.25ex\hbox{$
			  \left.
			  \begin{tabular}{@{}c@{}}
			  \nop{100}	\\
			  \nop{200}	\\[-1ex]
			  \nop{400}	
			  \end{tabular}
			  \right\}
			  		    $}
						\\
			  \raise3.25ex\hbox{$
			  \left.
			  \begin{tabular}{@{}c@{}}
			  \nop{100}	\\
			  \nop{200}	\\[-1ex]
			  \nop{400}	
			  \end{tabular}
			  \right\}
			  		    $}
						\\
			  \raise6.25ex\hbox{$
			  \left.
			  \begin{tabular}{@{}c@{}}
			  \nop{100o10}	\\
			  \nop{200o10}	\\
			  \nop{400o10}	\\[-1ex]
			  \nop{800o10}	\\
			  \nop{800o10n3}
			  \end{tabular}
			  \right\}
			  		    $}
						\\
			  \raise4.0ex\hbox{$
			  \left.
			  \begin{tabular}{@{}c@{}}
			  \nop{100o30}	\\
			  \nop{200o30}	\\[-1.5ex]
			  \nop{200o30n3}\\
			  \nop{400o30}	
			  \end{tabular}
			  \right\}
			  		    $}
			  \end{tabular}
	&
	& \begin{tabular}{@{}r@{}}
	  \nop{100}		\\
	  \nop{200}	 4	\\
	  \nop{400}		\\
	  \nop{100}		\\
	  \nop{200}	 4	\\
	  \nop{400}		\\
	  \nop{100o10}		\\
	  \nop{200o10}		\\
	  \nop{400o10}	10	\\
	  \nop{800o10}		\\
	  \nop{800o10n3}	\\
	  \nop{100o30}		\\
	  \nop{200o30}	\lower1.5ex\hbox{30}	\\
	  \nop{200o30n3}	\\
	  \nop{400o30}		
	  \end{tabular}
	&			& \begin{tabular}{@{}r@{}}
				  \nop{100}		\\
				  \nop{200}	248	\\
				  \nop{400}		\\
				  \nop{100}		\\
				  \nop{200}	248	\\
				  \nop{400}		\\
				  \nop{100o10}		\\
				  \nop{200o10}		\\
				  \nop{400o10}	813	\\
				  \nop{800o10}		\\
				  \nop{800o10n3}	\\
				  \nop{100o30}		\\
				  \nop{200o30}	\lower1.5ex\hbox{2780}	\\
				  \nop{400o30}		\\
				  \nop{200o30n3}	
				  \end{tabular}
					& \begin{tabular}{@{}r@{}}
					  \nop{100}		\\
					  \nop{200}	500	\\
					  \nop{200}		\\
					  \nop{100}		\\
					  \nop{200}	500	\\
					  \nop{200}		\\
					  \nop{100o10}		\\
					  \nop{200o10}		\\
					  \nop{400o10}	2000	\\
					  \nop{800o10}		\\
					  \nop{800o10n3}	\\
					  \nop{100o30}		\\
					  \nop{200o30} \lower1.5ex\hbox{5000} \\
					  \nop{200o30n3}	\\
					  \nop{400o30}		
					  \end{tabular}
			& $
			  \left.
			  \begin{tabular}{@{}c@{}}
			  \nop{100}	\\
			  \nop{200}	\\
			  \nop{400}	\\
			  \nop{100}	\\
			  \nop{200}	\\
			  \nop{400}	\\
			  \nop{100o10}	\\
			  \nop{200o10}	\\
			  \nop{400o10}	\\
			  \nop{800o10}	\\
			  \nop{800o10n3}\\
			  \nop{100o30}	\\
			  \nop{200o30}	\\
			  \nop{200o30n3}\\
			  \nop{400o30}	
			  \end{tabular}
			  \right\}
			  $
			& $P$	& $\rightarrow$	& $P$	& $+$
			& 0.02	& ${\times} \Gaussian(r_\init{=}$
			& 20	& $, \sigma{=}$
			& 5	& $)$
				& 0.0735	& ${\times} \Gaussian(r{=}$
				& 21.8	& $, \sigma{=}$
				& 3.5	& $)$
					& 0.976
					& $+$	& 0.641
					& $=$	& 1.617	& 1.183
									\\
%
%
\hline 
%
%
\begin{tabular}{@{}l@{}}
100.pqw1	\\
200.pqw1	\\
400.pqw1	\\
800.pqw1	\\
\end{tabular}
& \nop{}
	& \begin{tabular}{@{}l@{}}
	  0.01		\\
	  0.005		\\
	  0.0025	\\
	  0.00125	
	  \end{tabular}		&
				& \begin{tabular}{@{}l@{}}
				  0.030		\\
				  0.015		\\
				  0.0076	\\
				  0.0038	
				  \end{tabular}
				&
					&
					& \begin{tabular}{@{}l@{}}
					  0.020		\\
					  0.010		\\
					  0.0050	\\
					  0.0025	
					  \end{tabular}
					&
						&
						& \begin{tabular}{@{}l@{}}
						  0.96		\\
						  0.48		\\
						  0.24		\\
						  0.12		
						  \end{tabular}
						&
			& $
			  \left.
			  \begin{tabular}{@{}c@{}}
			  \nop{100}	\\
			  \nop{200}	\\
			  \nop{400}	\\[-1ex]
			  \nop{800}	
			  \end{tabular}
			  \right\}
			  $
	&
	& 4
	&			& 248
	  			& \phantom{0}200
			&
			& $P$	& $\rightarrow$	& $P$	& $+$
			& 0.1\phantom{0}
			 	& ${\times} \Gaussian(r_\init{=}$
			& 20	& $, \sigma{=}$
			& 1	& $)$
				& 1.42\phantom{00}
					& ${\times} \Gaussian(r{=}$
				& 22.71	& $, \sigma{=}$
				& 0.76	& $)$
					& 0.892
					& $+$	& 2.692
					& $=$	& 3.584	& 2.447
%
%
\end{tabular}
\end{center}
\tablenotetext[1]
   {
   These evolutions use an alternate finite differencing
   scheme near the grid boundaries, as discussed in
   sections~\ref{sect-spatial-FD/3rd-vs-4th-order-dww-mols-near-bndrys}
   and~\ref{sect-sample-evolutions/effects-of-3rd-vs-4th-order-dww}.
   }
\tablenotetext[2]
   {
   These evolutions have a small amount of white noise
   added to the state vector each time the right-hand-side
   function in the evolution equations is evaluated, as
   discussed in appendix~\ref{app-adding-noise}.
   }
%
%
\caption[Test Evolutions]
	{
	This table gives the initial data and other parameters for
	our test computations.  For each model this table shows a
	descriptive name, the spatial grid resolution, the spatial grid
	size, the coordinate time for which we have evolved the system,
	the input perturbation for our initial data solver
	(\cf{}~paper~I), a qualitative description of the
	resulting initial data, a summary of the initial slice's
	mass distribution (broken down into black hole, scalar field,
	and total mass), and the final black hole mass at the
	end of the evolution.  
	}
\label{tab-test-evolutions-pars}
%
%
\end{table}
%
\end{document}